\documentclass[prd,showpacs,floatfix,amsmath,amssymb,floatfix]{revtex4}
\usepackage{graphicx,dcolumn,booktabs,bm}
\usepackage{longtable,lscape}
\usepackage{txfonts}
\usepackage{overpic}
\usepackage{multirow}
\usepackage{amssymb}
\usepackage{indentfirst}
\usepackage{feynmf}   
\usepackage{slashed}  
\usepackage{cases}
\usepackage{color,ulem}
\usepackage{graphicx}
\usepackage{epsfig}
\usepackage{subfigure}
\usepackage{calligra}
\usepackage{mathtools}
\usepackage{dcolumn}

\graphicspath{{Figures/}}

\newcommand{\defeq}{\vcentcolon=}

\begin{document}


\title{Tri-meson bound state $BBB^{\ast}$ via delocalized $\pi$~Bond}
\author{Li Ma$^{1}$}\email{ma@hiskp.uni-bonn.de}
\author{Qian Wang$^{1,2}$}\email{wangqian@hiskp.uni-bonn.de}
\author{Ulf-G. Mei{\ss}ner$^{1,3,4}$}\email{meissner@hiskp.uni-bonn.de}

\affiliation{
$^1$Helmholtz-Institut f\"{u}r Strahlen- und Kernphysik and Bethe Center for Theoretical Physics, Universit\"{a}t Bonn, D-53115 Bonn, Germany\\
$^2$Institute of Quantum Matter, South China Normal University, Guangzhou 510006, China\\
  $^3$Institut f\"{u}r Kernphysik, Institute for Advanced Simulation, and J\"{u}lich Center for Hadron Physics, Forschungszentrum J\"{u}lich, D-52425 J\"{u}lich, Germany\\
  $^4$Tbilisi State University, 0186 Tbilisi, Georgia}

\date{\today}

\begin{abstract}

During the last decades, numerous exotic states which cannot be explained by the conventional quark model
 have been observed in experiment. Some of them can be understood as two-body hadronic molecules,
 such as the famous $X(3872)$, analogous to deuteron in nuclear physics. 
 Along the same line, the existence of the triton leaves an open question 
  whether there is a bound state formed by three hadrons.
  Since, for a given potential, a system with large reduced masses is more easier to form a bound state, 
 we study the $BBB^{\ast}$ system with the one-pion exchange  potential as an exploratory
  step by solving the three-body Schr\"odinger Equation. 
  We predict that a tri-meson molecular state for the $BBB^{\ast}$ system is probably
  existent as long as the molecular states of its two-body subsystem $BB^*$ exist.

 \end{abstract}

\pacs{14.40.Rt, 36.10.Gv} \maketitle

\section{Introduction}\label{sec1}

In the last few decades, numerous exotic states named as ``XYZ" as well as charm-strange mesons
beyond the conventional quark model
have been reported by many experimental collaborations. For a review of these exotic states, 
we recommend Refs.~\cite{Klempt:2007cp,Klempt:2009pi,Brambilla:2010cs,Olsen:2014qna,Oset:2016lyh,Chen:2016qju,Chen:2016spr,Esposito:2016noz,Lebed:2016hpi,Hosaka:2016pey,Dong:2017gaw,Guo:2017jvc,Olsen:2017bmm,Francis:2016hui,Karliner:2017qjm,Eichten:2017ffp}.
Some of them can be understood in the hadronic molecular picture~\cite{Guo:2017jvc}
which is an analog of the deuteron as a loosely bound state of a proton and a neutron. 
In their formation, the one-pion exchange potential (OPEP)
plays an important role, e.g. in the formation of
 deuteron and the $X(3872)$~\cite{Machleidt:1987hj,Tornqvist:2003na}, due to its long-range property. 
 Analogously, the existence of the triton arouses interest in the study of the three-hadron system, 
 especially after the large accumulation of experimental data, which might give some hints about
 the existence of this kind of bound states.
 In general, to solve the three-body problem, one should solve the
  Faddeev equations rigorously~\cite{Malfliet:1968tj,Eichmann:2009qa,Ishii:1995bu,Eichmann:2011vu,Huang:1993yd,Ishii:1993np,Ishikawa:2002ti,SanchisAlepuz:2011jn,Elster:2008hn,Eichmann:2008ef,Popovici:2010ph,Fujiwara:2003wr,Stadler:1991zz}. 
  However, for a specific system, one can do some approximation to simplify the problem,
   such as the Fixed Center Approximation (FCA) in the study of the $X(2175)$ as a resonance
   of the $\phi K\bar{K}$ system~\cite{MartinezTorres:2008gy} and the approximation on unitary
    chiral dynamics on the
    $\pi K \bar{K}$ and $\pi\pi\eta$ systems~\cite{MartinezTorres:2011vh}. 
    The FCA has also been widely applied to other systems, such as the $KK\bar{K}$~\cite{Torres:2011jt}, 
    the $D\bar{D}^{\ast}K$ and $\bar{D}D^{\ast}K$~\cite{Ren:2018pcd},
    the $J/\psi K\bar{K}$~\cite{MartinezTorres:2009xb}, 
    the $NDK$, $\bar{K} DN$ and $ND\bar{D}$~\cite{Xiao:2011rc},
    the $N\bar{K}K$~\cite{Xie:2010ig}, the  $BD\bar{D}$ and 
    $BDD$~\cite{Dias:2017miz,Jido:2008kp,MartinezTorres:2008kh} systems.
    There are many other studies that employ FCA method discussed in Refs.~\cite{Bayar:2012rk,Oset:2012gi,MartinezTorres:2010ax,Liang:2013yta,Bayar:2015oea,YamagataSekihara:2010qk,Roca:2010tf,Xie:2011uw,Debastiani:2017vhv}. 
      The isobar familism has also been applied to discuss three-body systems, such as the strange dibaryon resonance in the $\bar{K}NN-\pi\sigma N$ system~\cite{Ikeda:2007nz}, the effect of the $\Delta (1236)$ isobar on the three nucleon bound states~\cite{Hajduk:1979yn}, and other systems~\cite{MartinezTorres:2008kh,Ikeda:2008ub,Gal:2013dca,Dreissigacker:1981az}. 
      The dimer familism is another approximation method for a three-body system where a composite field is intoduced to describe its two-body subsystem when rescattering with a third particle, which has been applied to the three-hadron systems~\cite{Konig:2015aka,Konig:2016yka,Wilbring:2017fwy,Schmidt:2018vvl}. 
    Recently, a series of studies~\cite{Hammer:2017uqm,Hammer:2017kms,Meng:2017jgx} of a three-particle system
    in a finite volume via the dimer field have be proposed to gain a insights about a three-body system in a discretized space such as used in lattice QCD. 
It is worth mentioning that the system $BB^*B^*-B^*B^*B^*$ has been studied recently in Ref.~\cite{Garcilazo:2018rwu} based on the colored interaction for its subsytems.
 By solving exactly the Faddeev equations for the tri-meson system, the authors find a bound state about 90 $\mathrm{MeV}$ below three B mesons threshold. Similar discussions on the $\Omega NN$ and $\Omega \Omega N$ systems can be found in Ref.~\cite{Garcilazo:2019igo}.

As discussed above, the OPEP plays an important role in binding the two-hadron system. 
From another point of view,
one can view it as a pion shared by the two constituents and form a bound state.
 It can be regarded as a bond similar to the $\sigma~bond$ in hydrogen molecules. 
 There is another kind of bond called delocalized $\pi~bond$ universally existing in benzene molecules, 
 which is a pair of electrons shared by the six carbon atoms. 
 A simple extension is replacing the carbon atoms by hadrons. 
 We have studied the role of the delocalized $\pi~bond$ in forming the 
 three-body bound state for the double heavy tri-meson systems,  
i.e. $DD^{\ast}K$, $D\bar{D}^{\ast}K$, $BB^{\ast}\bar{K}$ and 
 $B\bar{B}^{\ast}\bar{K}$~\cite{Ma:2017ery}, based on the sufficient information of it sub two-body system
 and the Born-Oppenheimer Approximation (BOA) which works well for a system with
  several heavy  and light particles \cite{Moroz:2014eba}.
  The crucial idea is to use Born-Oppenheimer (BO) potential for considering the influence
   of the light part on the dynamics of the heavy part.
  Therefore, it is a fascinating idea whether 
 the delocalized $\pi~bond$ and the BOA could be applied to a three-heavy system,
such as the $BBB^{\ast}$ system with large reduced mass. 

 The same three bottomed meson system has been studied in Ref.~ \cite{Wilbring:2017fwy}
 by calculating the scattering amplitudes between the $Z_b(10610)$ or the $Z_b(10650)$
 and the bottomed meson. The universal bound states of three bottomed mesons from the Efimov
 effect has been ruled out.
 As in their calculation, only the contact interaction is included which might be the reason why
 they do not find a bound state.
 After including the long-range OPEP, the case might be different. Thus we solve the 
 three-body Schr\"odinger equation to discuss the $BBB^{\ast}$ system by considering the OPEP.  
 Without an assumption about its two-body subsystem, i.e. the molecular nature of 
 the $Z_b(10610)$ or $Z_b(10650)$, we focus on the three-body bound state as a function of
 the binding energy of its
 subsystem.  Hopefully, the present extensive investigations will be
 useful to deepen our  the understanding of a system made of three heavy particles.

 This paper is organized as follows. After the introduction, the formalism and the inputs for the
 $BBB^{\ast}$ system are presented in Sec.~\ref{sec2}. The dynamics of the two-body subsystem and the
 corresponding BO potential are given in Sec. \ref{sec3} and Sec. \ref{sec4}, respectively.
 By constructing a proper interpolating wave functions for the $BBB^{\ast}$ in Sec.~\ref{sec5},
 we solve the three-body Schr\"odinger equation in Sec.~\ref{sec6}. Numerical results and
 discussions are given in the following section. The summary is presented in the last section.
 Some technicalities are relegated to the appendix.

\section{Formalism and the inputs}\label{sec2}

The BOA has been successfully used in few-body system with several heavy and light
particles~\cite{Braaten:2014qka,Moroz:2014eba}. 
For a three-body system with one light and two heavy mesons, such as the $DD^*K$~\cite{Ma:2017ery} system,
the three-body Schr\"odinger equation is divided into two sub equations, one is the motion of
light meson with two static sources
 and the other one is 
the equation for the two heavy mesons with the BO potential~\cite{Ma:2017ery}, 
which reflects the influence of the light meson on the dynamics of the two heavy mesons. 
If the interaction between the light meson and the heavy meson is attractive, 
it would make the two heavy mesons come closer, thus facilitating the formation of a bound state of
the whole system. 
However, for a three heavy meson system, although the application of BOA is not straightforward,
one can employ the 
underlying permutation symmetry, which means the corresponding
 dynamics is invariant under the interchange of any two constituents,
for the system and continue to use the BOA for the calculation.

\begin{figure}[t!]
  \begin{center}
  \rotatebox{0}{\includegraphics*[width=0.35\textwidth]{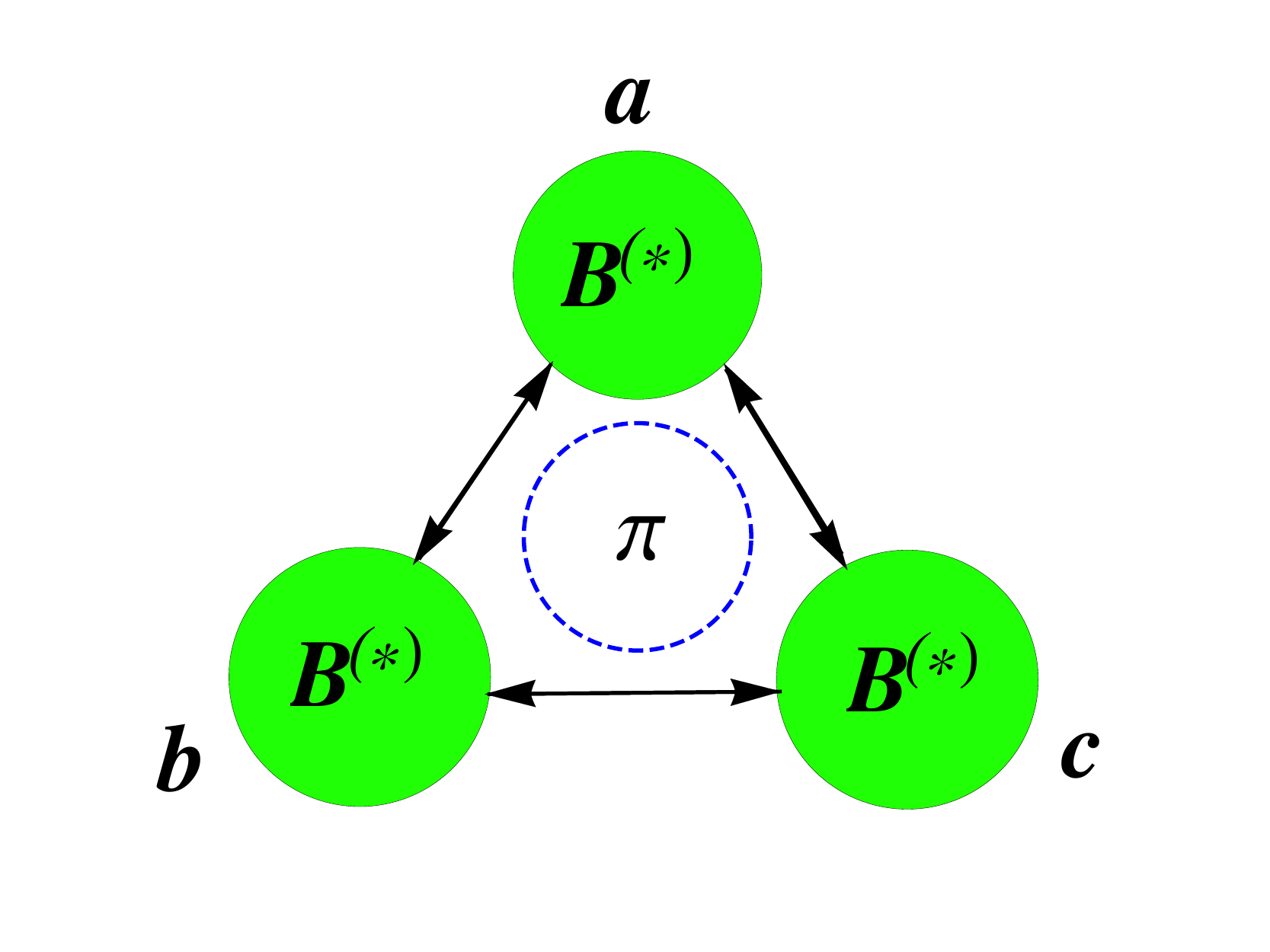}}
  \caption{Dynamical illustration of the $BBB^{\ast}$ system with a circle describing the
    delocalized $\pi~bond$ inside. Since the three constituents have the same probabilities
    to be the $B$ and $B^{\ast}$, one can rewrite the system as $B_a^{(\ast)}B_b^{(\ast)}B_c^{(\ast)}$.}
    \label{BBB}
  \end{center}
\end{figure}

The OPEP indicates that there is only one virtual pion exchanged by any two constituents as shown
in Fig.~\ref{BBB}.
One can use $a$, $b$ and $c$ to label the three mesons in the original channel, i.e. $B_a^{*}B_bB_c$.
It changes into $B_aB_b^{*}B_c$ via one-pion exchange (OPE) between $a$ and $b$, and
the channel $B_aB_b^{*}B_c$ changes into $B_aB_bB_c^{*}$ through the OPE
between $b$ and $c$. When the virtual pion arises between $a$ and $c$, 
it returns back into the original channel $B_a^{*}B_bB_c$. 
Within this scenario, the virtual pion is not localized between any two constituents but
rather shared by the whole system.
It is very similar with the benzene molecule which has a pair of electrons shared by the
six carbon atoms,  which is called $delocalized~\pi~bond$ in molecular physics. 
 Since the three constituents have the same probability to be the $B$ and $B^{\ast}$ mesons,
  one can write the system as $B_a^{(\ast)}B_b^{(\ast)}B_c^{(\ast)}$. 
  Furthermore, the order of the $a$, $b$ and $c$ labels of the three mesons is artificial, 
  as the system is invariant under the interchange of the $a$, $b$ and $c$. 
  This interchange symmetry will help to simplify our calculations. 
  The point is that one can count the influence of each heavy meson on the
  dynamics of the other two mesons one by one.  
   In other words, we can divide the system into three two-body subsystems $a~b$, $b~c$ and $a~c$. 
 In each subsystem, one should add the BO potential from the remaining one. 
 The existence of a negative common eigenvalue for the three subsystems may
  partly answer whether there is three-body bound state for the three heavy system.
  For simplicity, we call this method as $Born-Oppenheimer~potential~method$ ($BO~potential~method$).

Before performing the calculation, we define the isospin wave functions of the 
 $BBB^{\ast}$ systems as $|I_2, I_3, I_{3z}\rangle$ with $I_2$ the isospin of the sub-$BB^{\ast}$ system.
 $I_3$ and $I_{3z}$ represent the total isospin of the three-body system and its $z$ direction, respectively.
  One thus obtains the isospin wave functions of the $BBB^{\ast}$ system, 
\begin{eqnarray*}
\left|1, \frac{3}{2}, \frac{3}{2}\right\rangle &=& | (B^{+}B^{\ast +})B^{+} \rangle,\label{eq1} \\
\left|1, \frac{3}{2}, -\frac{3}{2}\right\rangle &=& | (B^{0}B^{\ast 0})B^{0} \rangle, \\
\left|1, \frac{3}{2}, \frac{1}{2}\right\rangle &=& \frac{1}{\sqrt{3}}[| (B^{+}B^{\ast 0})B^{+} \rangle+| (B^{0}B^{\ast +})B^{+} \rangle+|( B^{+}B^{\ast +})B^{0} \rangle], \\
\left|1, \frac{3}{2}, -\frac{1}{2}\right\rangle &=& \frac{1}{\sqrt{3}}[| (B^{0}B^{\ast 0})B^{+} \rangle+| (B^{+}B^{\ast 0})B^{0} \rangle+| (B^{0}B^{\ast +})B^{0} \rangle], \\
\left|1, \frac{1}{2}, \frac{1}{2}\right\rangle &=& \frac{1}{\sqrt{6}}[2 | (B^{+}B^{\ast +})B^{0} \rangle-| (B^{0}B^{\ast +})B^{+} \rangle-| (B^{+}B^{\ast 0})B^{+} \rangle], \\
\left|1, \frac{1}{2}, -\frac{1}{2}\right\rangle &=& \frac{1}{\sqrt{6}}[| (B^{0}B^{\ast +})B^{0} \rangle+| (B^{+}B^{\ast 0})B^{0} \rangle-2 | (B^{0}B^{\ast 0})B^{+} \rangle], \\
\left|0, \frac{1}{2}, \frac{1}{2}\right\rangle &=& \frac{1}{\sqrt{2}}[| (B^{0}B^{\ast +})B^{+} \rangle-| (B^{+}B^{\ast 0})B^{+} \rangle], \label{D11} \\
\left|0, \frac{1}{2}, -\frac{1}{2}\right\rangle &=& \frac{1}{\sqrt{2}}[| (B^{0}B^{\ast +})B^{0} \rangle-| (B^{+}B^{\ast 0})B^{0} \rangle]. \label{D12}
\end{eqnarray*}
Since $BB^{\ast}$ can couple with $B^{\ast}B^{\ast}$ via OPE, the coupled channel
effect is not negligible. We only consider the next close $BB^{\ast}B^{\ast}$ channel in our calculation.
If we distinguish the specific locations of the constituents as $a$, $b$ and $c$, 
there are six channels in total, i.e. $B_a^{\ast}B_bB_c$, $B_aB_b^{\ast}B_c$, $B_aB_bB_c^{\ast}$,
$B_a^{\ast}B_b^{\ast}B_c$, $B_a^{\ast}B_bB_c^{\ast}$ and $B_aB_b^{\ast}B_c^{\ast}$. 

The Lagrangians with SU(2) chiral symmetry (we only consider OPE) and C-parity conservation read 
\begin{eqnarray}
\mathcal{L}_{P}&=&-i\frac{2g}{f_\pi}\bar{M}
P^{*\mu}_b\partial_{\mu}\phi_{ba}P^{\dag}_{a}+i\frac{2g}{f_\pi}\bar{M} P_b\partial_{\mu}\phi_{ba}P^{*\mu\dag}_{a} \nonumber\\
&-& \frac{g}{f_\pi}
P^{*\mu}_b\partial^{\alpha}\phi_{ba}\partial^{\beta}P^{*\nu\dag}_{a}\epsilon_{\mu\nu\alpha\beta}
+ \frac{g}{f_\pi}
\partial^{\beta}P^{*\mu}_b\partial^{\alpha}\phi_{ba}P^{*\nu\dag}_{a}\epsilon_{\mu\nu\alpha\beta}, \\
\widetilde{\mathcal{L}_{P}}&=&-i\frac{2g}{f_\pi}\bar{M}\widetilde{P^{\dag}_{a}}
\partial_{\mu}\phi_{ab}\widetilde{P^{*\mu}_b}-i\frac{2g}{f_\pi}\bar{M}\widetilde{P^{*\mu\dag}_{a}}\partial_{\mu}\phi_{ab}\widetilde{P_b} \nonumber\\
&+& \frac{g}{f_\pi}
\partial^{\beta}\widetilde{P^{*\mu\dag}_{a}}\partial^{\alpha}\phi_{ab}\widetilde{P^{*\nu}_b}\epsilon_{\mu\nu\alpha\beta}
-\frac{g}{f_\pi}
\widetilde{P^{*\mu\dag}_{a}}\partial^{\alpha}\phi_{ab}\partial^{\beta}\widetilde{P^{*\nu}_b}\epsilon_{\mu\nu\alpha\beta},\label{anti-pseudo-exchange}
\end{eqnarray}
where the heavy flavor meson fields $P$ and $P^*$ represent $P=(B^-, \bar{B}^0)$, $P^*=(B^{*-},
\bar{B}^{*0})$, respectively. Its corresponding heavy anti-meson fields
$\widetilde{P}$ and $\widetilde{P}^*$ represent
$\widetilde{P}=(B^+, B^0)$,
$\widetilde{P}^*=(B^{*+}, B^{*0})$.
$\phi$ is the pion matrix
\begin{eqnarray}
\phi=\left(
         \begin{array}{cc}
           \frac{\pi^0}{\sqrt{2}} & \pi^+ \\
           \pi^- & -\frac{\pi^0}{\sqrt{2}} \\
         \end{array}
       \right).
\end{eqnarray}
We use the pion decay constant $f_{\pi}=132~\mathrm{MeV}$~\cite{Zhao:2014gqa}.
The pionic coupling constant $g\!=\!0.57$ is extracted from the width of $D^{*+}$
 by assuming heavy quark flavor symmetry\cite{Patrignani:2016xqp}. 
 All the parameters and input datas are listed in Table~\ref{tab:coupling-constant}. 
 Here, we neglect isospin breaking effect and use the masses of their 
 charged particles. 

\begin{table}[htbp]
  \caption{The coupling constants and meson masses in our calculation. The meson masses are taken
    from the PDG \cite{Patrignani:2016xqp}}
\label{tab:coupling-constant}
\begin{center}
\begin{tabular}{c | c   }
\hline \hline   {mass(MeV)} & {coupling constants} \\
 \hline
 $m_{\pi}=139$  &  $g=0.57$  \\
 $M_{B}=5279$ &  $f_{\pi}=132.00$ MeV  \\
 $M_{B^*}=5325$ &    \\
 \hline\hline
\end{tabular}
\end{center}
\end{table}

\begin{figure}[ht]
  \begin{center}
  \rotatebox{0}{\includegraphics*[width=1.00\textwidth]{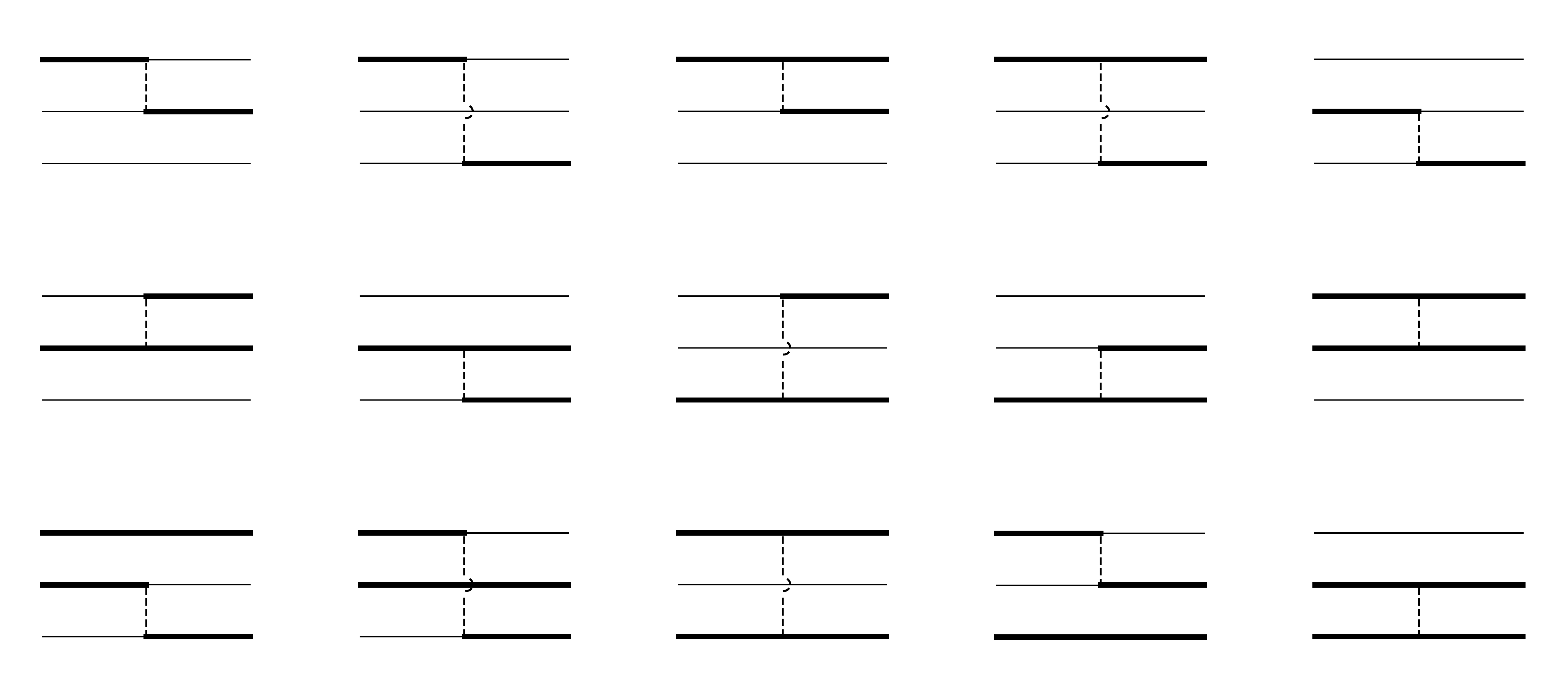}}
  \caption{The leading order OPE diagrams for the transitions among the relevant three-body
    channels, i.e. $B_a^{\ast}B_bB_c$, $B_aB_b^{\ast}B_c$, $B_aB_bB_c^{\ast}$, $B_a^{\ast}B_b^{\ast}B_c$,
    $B_a^{\ast}B_bB_c^{\ast}$ and $B_aB_b^{\ast}B_c^{\ast}$. The solid and bold solid lines represent
    the $B$ and $B^{\ast}$ meson fields, respectively.
    Dotted lines represent pion fields. }
    \label{Feym}
  \end{center}
\end{figure}

Under SU(2) chiral symmetry, the OPE interaction is of order $\mathcal{O}(p^0)$ for the three-body
system. In this paper, we only take into account the OPEP to the $\mathcal{O}(p^0)$ order. 
  Thus, there are four kinds of effective potentials. We use $V_1$ to denote the effective 
  potential for the interaction $BB^{\ast}\to B^{\ast}B$. $V_2$ and $V'_2$ denote
   the process $BB^{\ast}\to B^{\ast}B^{\ast}$ and its reverse,  respectively. 
   $V_3$ represents diagonal process $B^{\ast}B^{\ast}\to B^{\ast}B^{\ast}$. 
   Since the interactions are physical, the effective potentials should be unitary which
   gives $V'_2=(V_2)^{\dag}$. $\vec{r}_{ij}$ is used to denote the relative displacement between
   the $i$-th and $j$-th particles.
   Thus, the effective potentials of the three-body system in the channel space $|\mathcal{BBB}\rangle
   \defeq\{ B_a^{\ast}B_bB_c, B_aB_b^{\ast}B_c, B_aB_bB_c^{\ast}, B_a^{\ast}B_b^{\ast}B_c, B_a^{\ast}B_bB_c^{\ast}, B_aB_b^{\ast}B_c^{\ast} \}$ take the following form, 
\begin{eqnarray}
V_{BBB^{\ast}}=\left(
         \begin{array}{cccccc}
           0 & V_1(\vec{r}_{ab}) &  V_1(\vec{r}_{ac})  &  V_2(\vec{r}_{ab}) & V_2(\vec{r}_{ac}) & 0 \\
          V_1(\vec{r}_{ba}) & 0  &  V_1(\vec{r}_{bc})  & V'_2(\vec{r}_{ba}) & 0 & V_2(\vec{r}_{bc}) \\
          V_1(\vec{r}_{ca}) & V_1(\vec{r}_{cb}) & 0 & 0 & V'_2(\vec{r}_{ac}) & V'_2(\vec{r}_{bc}) \\
          V_2(\vec{r}_{ba}) & V'_2(\vec{r}_{ab}) & 0 & V_3(\vec{r}_{ab}) & V_1(\vec{r}_{bc}) & V_1(\vec{r}_{ac}) \\
          V_2(\vec{r}_{ca}) & 0 & V'_2(\vec{r}_{ca}) & V_1(\vec{r}_{cb}) & V_3(\vec{r}_{ac}) & V_1(\vec{r}_{ab}) \\
          0 & V_2(\vec{r}_{cb}) & V'_2(\vec{r}_{cb}) & V_1(\vec{r}_{ca}) & V_1(\vec{r}_{ba}) & V_3(\vec{r}_{bc}) \\
         \end{array}
       \right). \label{VBBB}
\end{eqnarray}
Its graphical illustration is shown in Fig.~\ref{Feym}.

\section{The break-up state and two-body subsystem}\label{sec3}

For the three bottomed meson system, suppose one of the constituent is infinitely
away from the remaining two mesons. The system can be divided into a two-body subsystem plus a free meson.
A bound state solution of the two-body subsystem indicates
a break-up state for the three-body system, i.e. a two-body bound state plus a free meson.
 In the OPE model, as there is no direct interaction between two $B$  mesons,
one could expect a break-up state with the subsystem $BB^{*}$ with quantum number $J^{P}=1^+$
and a free meson $B$.
We can detach the subsystem $B_{a}B_{b}^{*}$ first, and explore its binding solution. The
Hamiltonian of the subsystem in the channel space $|\mathcal{BBB}\rangle\defeq\{ B_a^{\ast}B_bB_c,
B_aB_b^{\ast}B_c, B_aB_bB_c^{\ast}, B_a^{\ast}B_b^{\ast}B_c, B_a^{\ast}B_bB_c^{\ast}, B_aB_b^{\ast}B_c^{\ast} \}$
reads
\begin{eqnarray}
H_{ab}=\left(
         \begin{array}{cccccc}
           T_{*} & V_1(\vec{r}_{ab}) &  0  &  V_2(\vec{r}_{ab}) & 0 & 0 \\
          V_1(\vec{r}_{ba}) &  T_{*}  &  0  & V'_2(\vec{r}_{ba}) & 0 & 0 \\
          0 & 0 & T & 0 & 0 & 0 \\
          V_2(\vec{r}_{ba}) & V'_2(\vec{r}_{ab}) & 0 & T_{**}+V_3(\vec{r}_{ab})+\delta M & 0 & 0 \\
          0 & 0 & 0 & 0 &  T_{*}+\delta M & V_1(\vec{r}_{ab}) \\
          0 & 0 & 0 & 0 & V_1(\vec{r}_{ba}) &  T_{*}+\delta M \\
         \end{array}
       \right), 
\end{eqnarray}
where the $T_{*}=-({1}/{2\mu_{*}})\nabla_{ab}^2$ and $T_{**}=-({1}/{2\mu_{**}})\nabla_{ab}^2$
are the relative kinetic energy for the $BB^*$ and $B^*B^*$ in their center-of-mass frame,
 respectively, with $\mu_{*}=({M_BM_{B^*}})/({M_B+M_{B^*}})$, $\mu_{**}=({M_{B^*}})/{2}$, $\nabla_{ab}^2
=({1}/{r_{ab}})({d^2}/{dr_{ab}^2})r_{ab}-({\overrightarrow{L_{ab}}^2})/({r_{ab}^2})$. Here
 $\overrightarrow{L_{ab}}$ is the angular momentum operator between meson $a$ and $b$. 
We also have the  mass gap $\delta M=M_{B^*}-M_B$. 
The effective potentials $V_1$, $V_2$, $V'_2$ and $V_3$ depend on the isospin of the specific channels,
thus we rewrite the $BB^*$ wave functions with fixed isospin 
\begin{eqnarray*}
|1, 1\rangle &=& |B^{+} B^{\ast +} \rangle, \\
|1, -1\rangle &=& |B^{0} B^{\ast 0} \rangle, \\
|1, 0\rangle &=& \frac{1}{\sqrt{2}}[|B^+ B^{\ast 0}\rangle+|B^0 B^{\ast +} \rangle], \\
|0, 0\rangle &=& \frac{1}{\sqrt{2}}[|B^+ B^{\ast 0} \rangle-|B^0 B^{\ast +} \rangle].
\end{eqnarray*}

\begin{figure}[ht]
  \begin{center}
  \rotatebox{0}{\includegraphics*[width=0.33\textwidth]{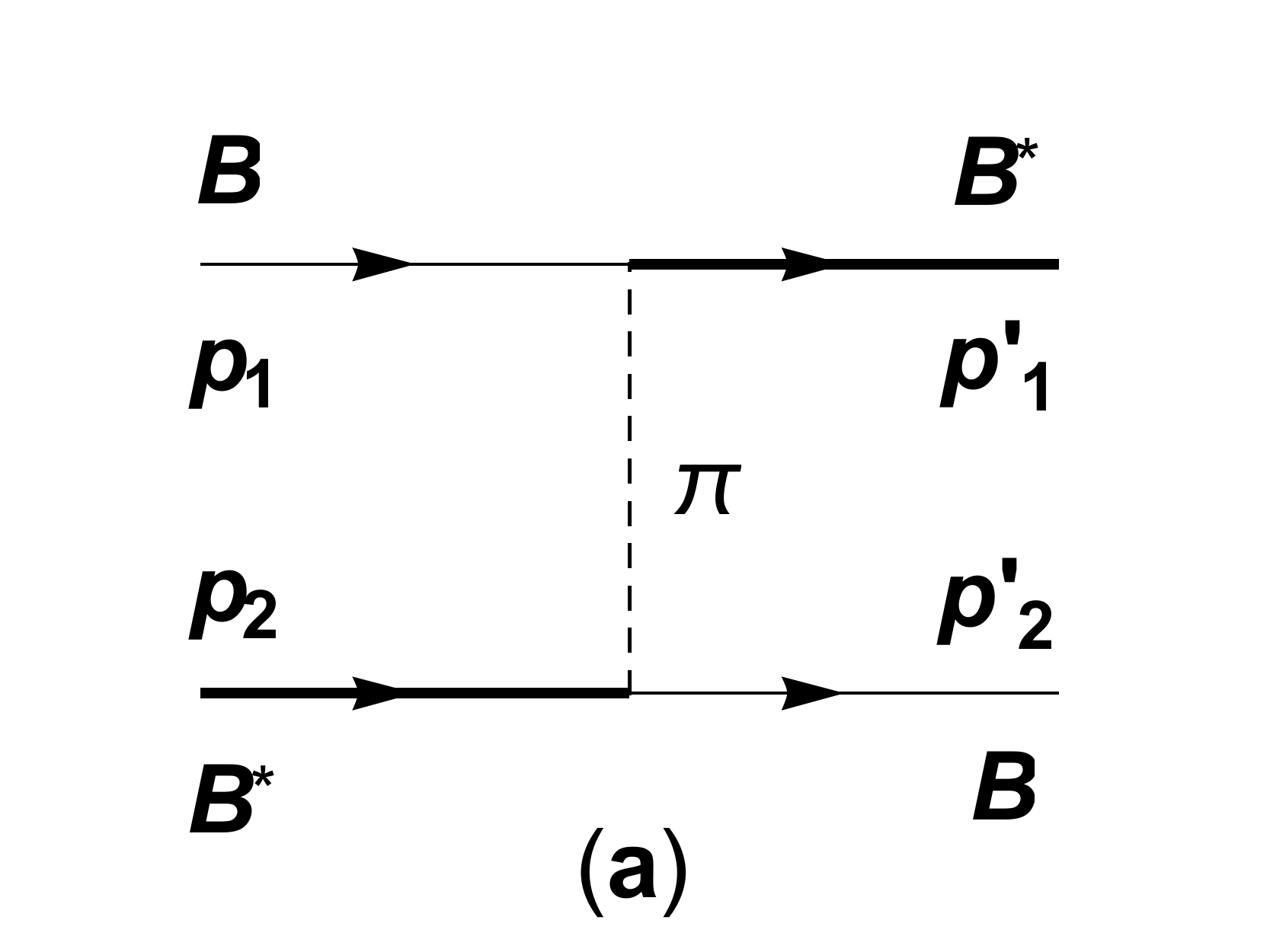}}
  \rotatebox{0}{\includegraphics*[width=0.33\textwidth]{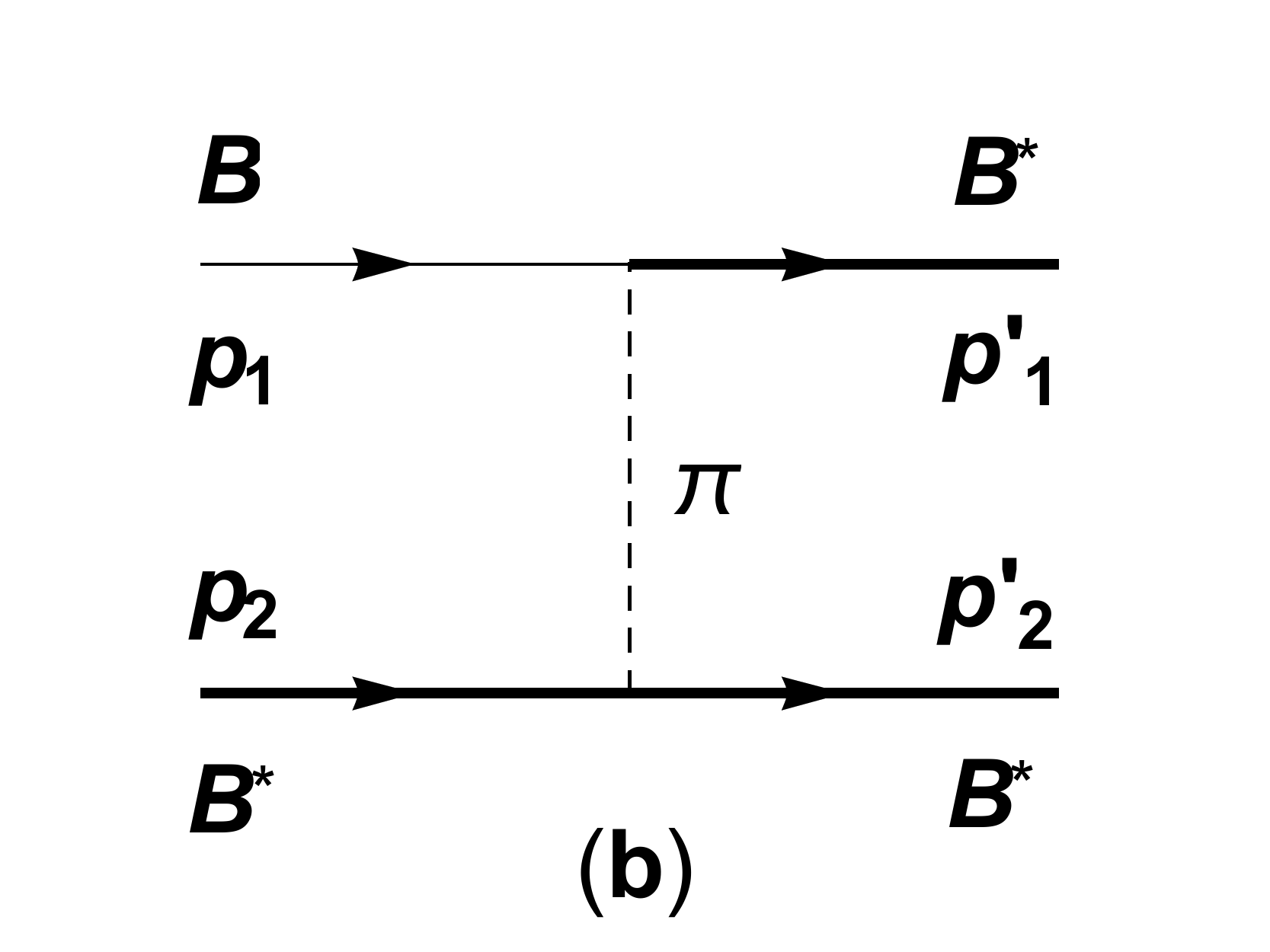}}
  \rotatebox{0}{\includegraphics*[width=0.33\textwidth]{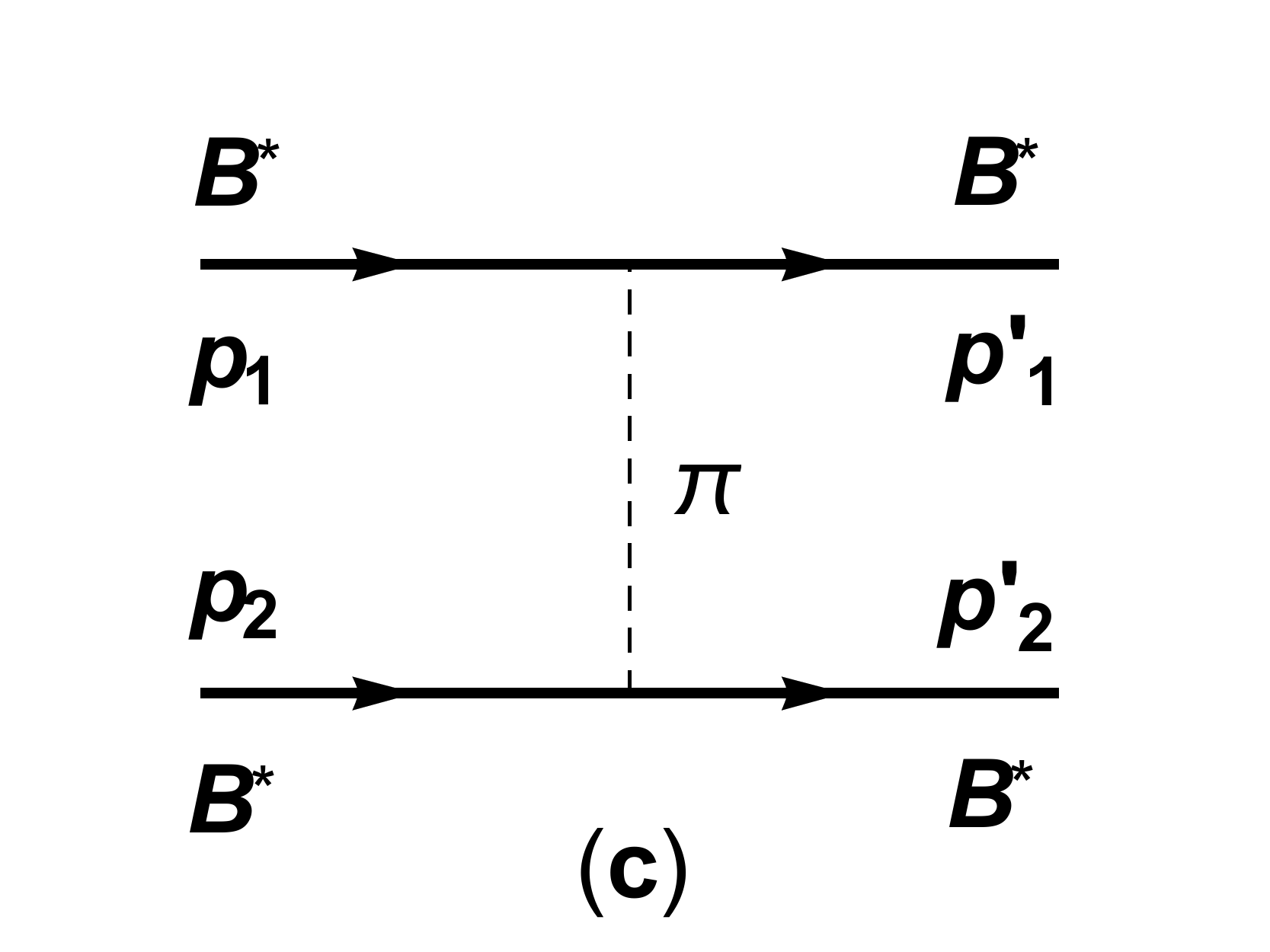}}
      \caption{The u-channel Feynman diagrams  for describing both the
       $BB^{\ast}$ and $B^{\ast}B^{\ast}$ system's interaction at tree level. 
       The regular and bold lines stand for the $B$ and the $B^{\ast}$ fields, 
       respectively. The dotted lines denote the pion fields.}
    \label{Fey}
  \end{center}
\end{figure}
For the specific channels $B_a^*B_b$, $B_aB_b^*$ and $B_a^*B_b^*$, the Schr\"odinger equation in
the channel space $\{ B_a^*B_b, ~B_aB_b^*, ~B_a^*B_b^* \}$ takes the form
\begin{eqnarray*}
\left(
        \begin{array}{ccc}
 T_{*} & V_1(\vec{r}_{ab})  & V_2(\vec{r}_{ab})  \\
 V_1(\vec{r}_{ba}) & T_{*}  & V'_2(\vec{r}_{ba})  \\
 V_2(\vec{r}_{ba}) & V'_2(\vec{r}_{ab})  & T_{**}+V_3(\vec{r}_{ab})+\delta M  
       \end{array}
       \right)
      \left(
\begin{array}{c}
   \frac{1}{\sqrt{2}}\psi(\vec{r}_{ab})  \\
   \frac{1}{\sqrt{2}}\psi(\vec{r}_{ab})  \\
    \psi'(\vec{r}_{ab}) 
\end{array}
\right) = E_b
 \left(
\begin{array}{c}
   \frac{1}{\sqrt{2}}\psi(\vec{r}_{ab})  \\
   \frac{1}{\sqrt{2}}\psi(\vec{r}_{ab})  \\
    \psi'(\vec{r}_{ab}) 
\end{array}
\right).
\end{eqnarray*}

Based on this, we can derive the scattering amplitude at the tree level 
\begin{equation}
\langle f | S | i \rangle = \delta_{fi} + (2\pi)^4\delta^4(p_f-p_i) i M_{fi} =  \delta_{fi} - 2\pi \delta(E_f-E_i) i V_{fi},
\end{equation}
where the $T$-matrix is the interaction part of the $S$-matrix and the $M$ is
defined as the invariant matrix element. In the second equation
we have applied the first order of Born series expansion
on the Lippmann-Schwinger equation with $V_{fi}$ being the effective potential.
The relation between the scattering amplitude $M_{fi}$
and the potential $V_{fi}$ is 
\begin{equation}
V_{fi}=-\frac{M_{fi}}{\sqrt{ \mathop\prod\limits_{f}2{p_f}^0
\mathop\prod\limits_{i} 2{p_i}^0}}\approx -\frac{M_{fi}}
{\sqrt{\mathop\prod\limits_{f} 2{m_f} \mathop\prod\limits_{i}
2{m_i}}},
\end{equation}
where $p_{f(i)}$ and $m_{f(i)}$ denote the four-momentum and the mass of the final (initial)
state.

In the calculation, $p_1(E_1,\vec{p})$ and $p_2(E_2,-\vec{p})$
denote the four-momenta of the initial state particles in the center-of-mass
system shown in Fig.~\ref{Fey}, while $p'_1(E'_1,\vec{p'})$ and
 $p'_2(E'_2,-\vec{p'})$ denote the four-momenta of the final state particles, respectively.
$q=p'_1-p_1=(E'_1-E_1,\vec{p'}-\vec{p})=(E_2-E'_2,\vec{q}) $
is the transferred four-momentum. For convenience, we always use
$
\vec{q}=\vec{p'}_1-\vec{p}_1
$
and
$
\vec{k}=(\vec{p'}_1+\vec{p}_1)/2
$
instead of $\vec{p'}$ and $\vec{p}$ in the calculations.
The effective potential in coordinate space can be derived by Fourier transformation
\begin{equation*}
V(\vec{r})=\frac{1}{(2\pi)^3}\int d^3\vec{q}\, e^{i\vec{q}\cdot\vec{r}}\, V(\vec{q}\,F^{2}(\vec{q})~.
\end{equation*}
To take into account in a rough way the substructure of each vertex,
a monopole form factor 
\begin{equation}
F_i(q)=\frac{\Lambda^2-m_{\pi}^2}{\Lambda^2-q_i^2}=\frac{\Lambda^2-m_{\pi}^2}{{\tilde{\Lambda}}_i^2+\vec{q}_i^2},
\end{equation}
with $m_\pi$ the pion mass and 
\begin{equation}
\tilde{\Lambda}^{(\prime)2}=\Lambda^2-(\Delta M^{(\prime)})^2, 
\end{equation}
is used to suppress the contribution from UV energies.
Here, $\Delta M=M_B^{\ast}-M_B$ and $\Delta M'=(M_B^{\ast}-M_B)/2$.
As the parameter $\Lambda$ is related to  non-perturbative QCD,
it cannot be well determined. Here we only explore its effect on the binding energy
of the $BB^*$ with the quantum number $J^P=1^+$ system.    
To solve the time-independent
Schr\"{o}dinger equation in coordinate space, the potential $V(\vec{q},\vec{k})$ in momentum space
can be transformed in to that in coordinate space as shown in the Appendix.

\begin{figure}[ht]
  \begin{center}
  \rotatebox{0}{\includegraphics*[width=0.45\textwidth]{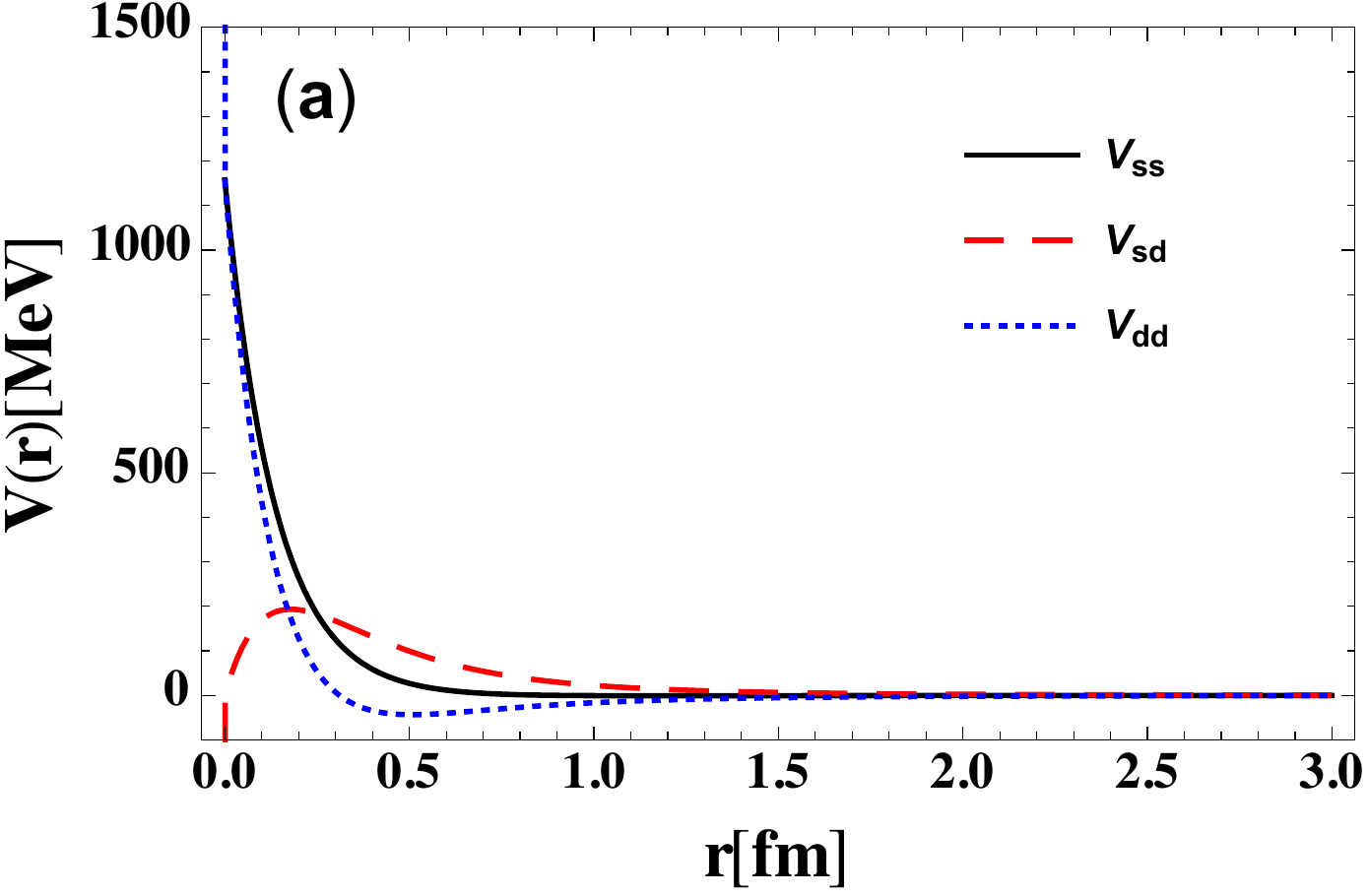}}
    \rotatebox{0}{\includegraphics*[width=0.45\textwidth]{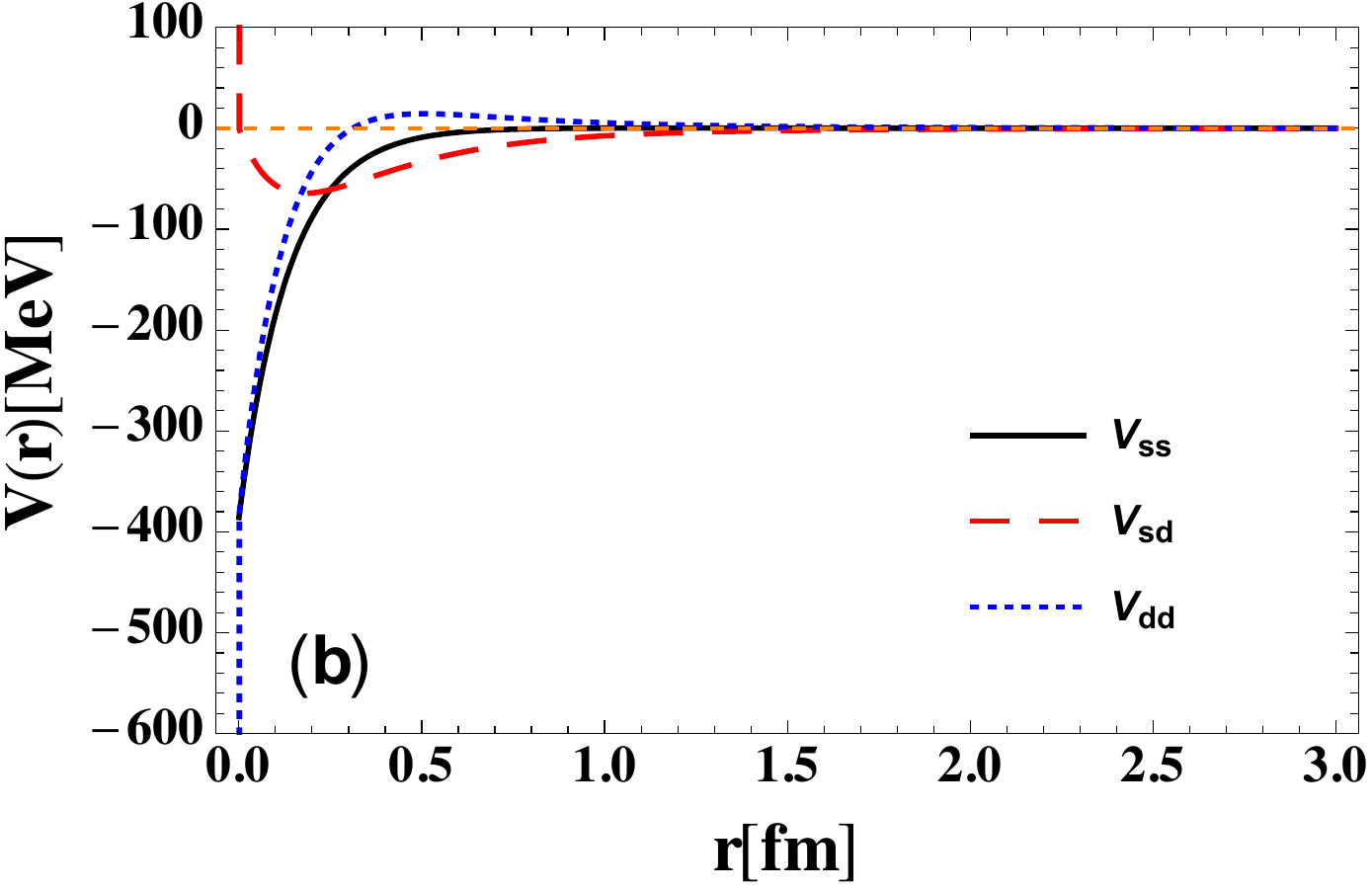}}
    \caption{The effective potentials $V_{BB^*\rightarrow BB^*}(\vec{r})$ of the $BB^*$ system
       with quantum number $J^P=1^+$, where (a) and (b)
        correspond to the isospin $I=0$ and $I=1$ cases, respectively.
        The $V_{ss}$ and $V_{dd}$ are the effective potentials for the S-wave and D-wave.
         The $V_{sd}$ represents the effective potential of S-D wave mixing. 
         For illustration, the value $1440~\mathrm{MeV}$ is used for the parameter $\Lambda$.}
    \label{BB-potential}
  \end{center}
\end{figure}

\begin{figure}[ht]
  \begin{center}
  \rotatebox{0}{\includegraphics*[width=0.45\textwidth]{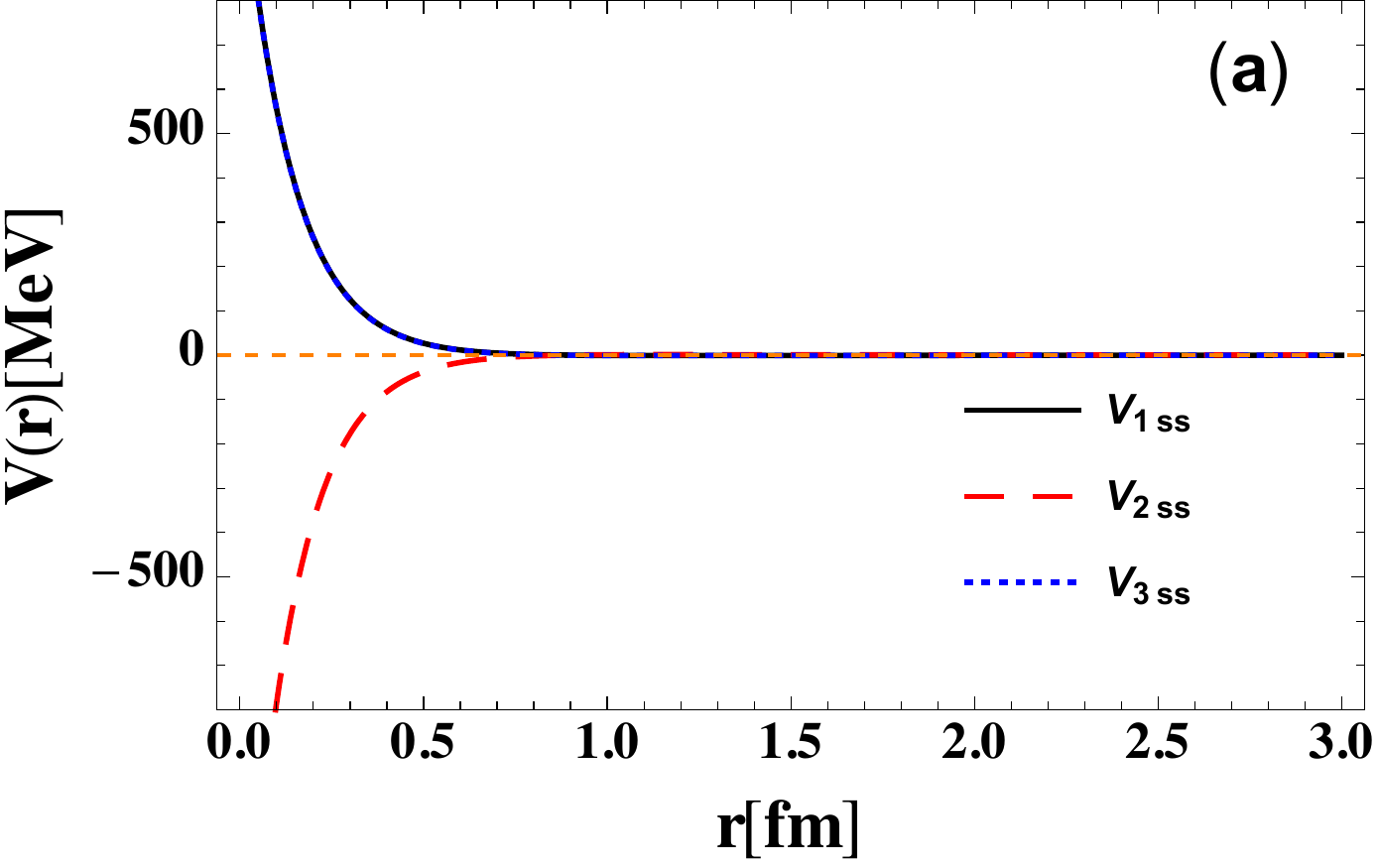}}
    \rotatebox{0}{\includegraphics*[width=0.45\textwidth]{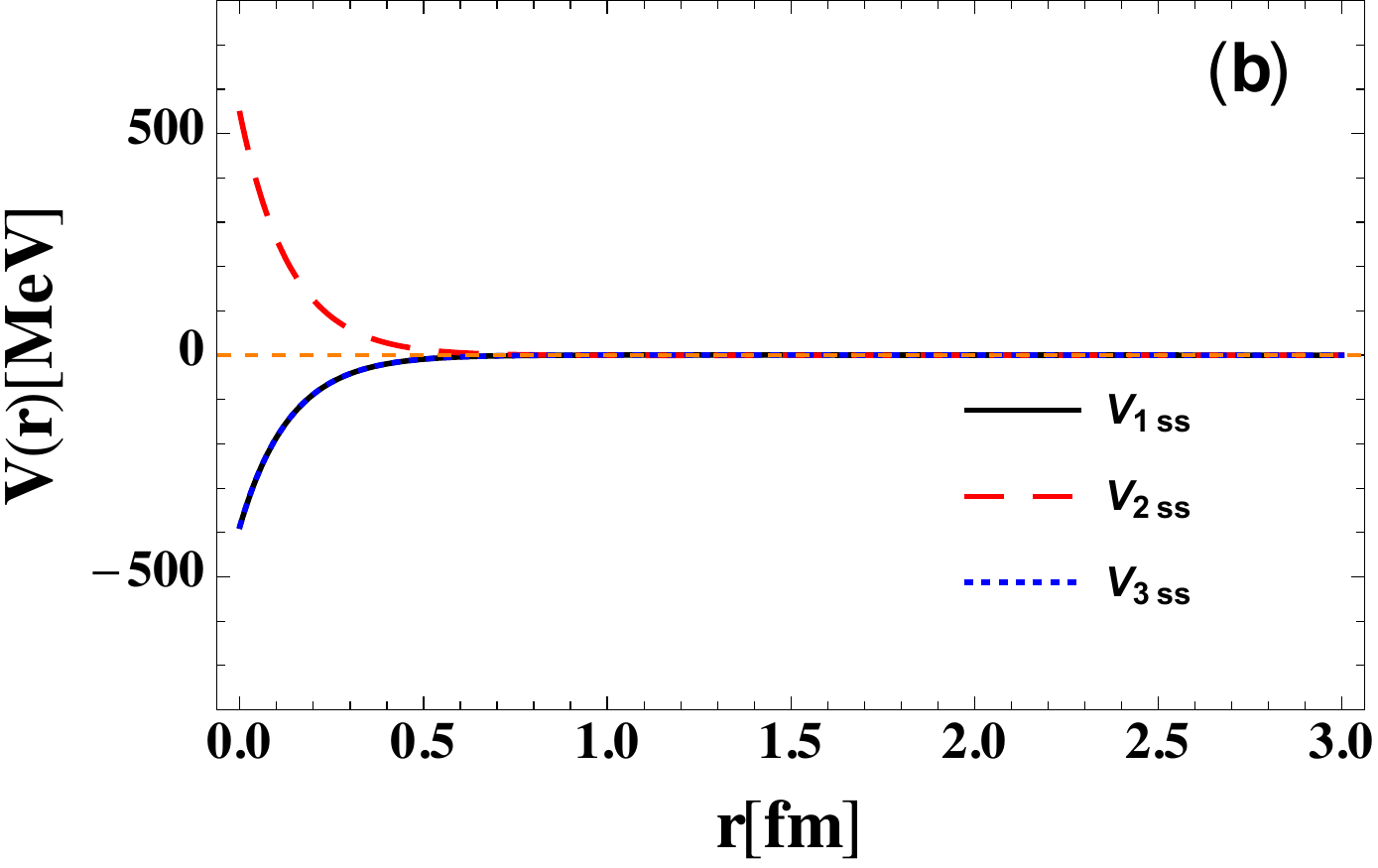}}
    \caption{The effective potentials for the S-wave of the $BB^*$ system with quantum number
      $J^P=1^+$, where (a) and (b)  correspond to the isospin $I=0$ and $I=1$ cases, respectively.
      The $V_{1ss}$,  $V_{2ss}$ and $V_{3ss}$ are the effective potentials
      $V_{BB^*\rightarrow BB^*}(\vec{r})$, $V_{BB^*\rightarrow B^*B^*}(\vec{r})$ and
      $V_{B^*B^*\rightarrow B^*B^*}(\vec{r})$ for the S-wave respectively. For illustration, the
      value $1440~\mathrm{MeV}$ is used for the parameter $\Lambda$.}
    \label{V1V2V3-potential}
  \end{center}
\end{figure}

The isosinglet and isotriplet $BB^*$ potentials $V_{BB^*\rightarrow BB^*}(\vec{r})$ in coordinate
space are shown in Fig.~\ref{BB-potential} (a)
and (b), respectively, with $\Lambda=1440~\mathrm{MeV}$. 
The isosinglet potential $V_{BB^*\rightarrow BB^*}(\vec{r})$ is repulsive which does not indicate a
bound solution. Nevertheless, the potential $V_{BB^*\rightarrow B^*B^*}(\vec{r})$ for the
isosinglet is attractive as shown in Fig.~\ref{V1V2V3-potential}. So there is still the
possibility of a binding solution. On the contrary, the isotriplet potential
$V_{BB^*\rightarrow BB^*}(\vec{r})$ is 
attractive, while its potential $V_{BB^*\rightarrow B^*B^*}(\vec{r})$ is repulsive. These
potentials in coordinate space can be expressed as  
\begin{eqnarray}
V_{BB^*\rightarrow BB^*}(\vec{r}) &=& -C_{\pi}(i,j)\frac{g^2}{12 \pi f^{2}_{\pi}} \{ \vec{\epsilon}\cdot\vec{\epsilon}^{\dag}  [~\tilde{m}_{\pi}^2\tilde{\Lambda} Y(\tilde{\Lambda} r)-\tilde{m}_{\pi}^3Y(\tilde{m}_{\pi}r)\nonumber\\
&+&(\Lambda^2-m_{\pi}^2)\tilde{\Lambda}\frac{e^{-\tilde{\Lambda} r}}{2}~]
+S_{T}(\vec{\epsilon}_3^{\dag}, \vec{\epsilon}_2)[-\tilde{m}_{\pi}^3 Z(\tilde{m}_{\pi}r)+ \tilde{\Lambda}^3 Z(\tilde{\Lambda} r)\nonumber\\
&+&(\Lambda^2-m_{\pi}^2)(1+\tilde{\Lambda} r)\frac{\tilde{\Lambda}}{2}Y(\tilde{\Lambda} r)~] \}, \label{VBB1}
\end{eqnarray}
\begin{eqnarray}
V_{BB^*\rightarrow B^*B^*}(\vec{r}) &=& C_{\pi}(i,j)\frac{g^2}{12 \pi f^{2}_{\pi}} \{ (\vec{\epsilon}_3\cdot i\vec{\epsilon}_4^{\dag}\times\vec{\epsilon}_2)  [~\tilde{m'}_{\pi}^2\tilde{\Lambda'} Y(\tilde{\Lambda'} r)-\tilde{m'}_{\pi}^3Y(\tilde{m'}_{\pi}r)\nonumber\\
&+&(\Lambda^2-m_{\pi}^2)\tilde{\Lambda'}\frac{e^{-\tilde{\Lambda'} r}}{2}~]
+S_{T}(\vec{\epsilon}_3, i\vec{\epsilon}_4^{\dag}\times\vec{\epsilon}_2)[-\tilde{m'}_{\pi}^3 Z(\tilde{m'}_{\pi}r)+ \tilde{\Lambda'}^3 Z(\tilde{\Lambda'} r)\nonumber\\
&+&(\Lambda^2-m_{\pi}^2)(1+\tilde{\Lambda'} r)\frac{\tilde{\Lambda'}}{2}Y(\tilde{\Lambda'} r)~] \}, \label{VBB2}
\end{eqnarray}
\begin{eqnarray}
V_{B^*B^*\rightarrow B^*B^*}(\vec{r}) &=& C_{\pi}(i,j)\frac{g^2}{12 \pi f^{2}_{\pi}} \{ (i\vec{\epsilon}_3^{\dag}\times\vec{\epsilon}_1\cdot i\vec{\epsilon}_4^{\dag}\times\vec{\epsilon}_2)  [~m_{\pi}^2\Lambda Y(\Lambda r)-m_{\pi}^3Y(m_{\pi}r)\nonumber\\
&+&(\Lambda^2-m_{\pi}^2)\Lambda\frac{e^{-\Lambda r}}{2}~]
+S_{T}(i\vec{\epsilon}_3^{\dag}\times\vec{\epsilon}_1,  i\vec{\epsilon}_4^{\dag}\times\vec{\epsilon}_2) [-m_{\pi}^3 Z(m_{\pi}r)+ \Lambda^3 Z(\Lambda r)\nonumber\\
&+&(\Lambda^2-m_{\pi}^2)(1+\Lambda r)\frac{\Lambda}{2}Y(\Lambda r)~] \}, \label{VBB3}
\end{eqnarray}
with $\tilde{m}^{(\prime)2}_{\pi}=m_{\pi}^2-\Delta M^{(\prime)2}$. The tensor operator $\hat{S}_{T}$ has the form
$
\hat{S}_{T}=3(\vec{r}\cdot \hat{\vec{\epsilon_b}})(\vec{r}\cdot
\hat{\vec{\epsilon_a}}^{\dag})-\hat{\vec{\epsilon_b}}\cdot
\hat{\vec{\epsilon_a}}^{\dag}
$ with $\epsilon$ the polarization vector of $B^{\ast}$.
The $C_{\pi}(i,j)$ are channel dependent coefficients, summarized in Table \ref{channel-coeff}.
The $c$ in Table~\ref{channel-coeff} represents the C-parity of the corresponding channel.

\begin{table}[htbp]
\caption{Channel dependent coefficients. Here, $c$ denotes the C-parity of the two-body system.}
\label{channel-coeff}
\begin{center}
\begin{tabular}{ c | c | c | c | c  }
\hline\hline 
{channel}  & {isospin}   & $C(i, j)$   & {channel} &  $C(i, j)$ \\
\hline
\multirow{2}{*}{$BB^{*}$} & $I=1$ &  $1/2$ &  $B^{+}B^{\ast +}$ & $1/2$    \\
\cline{2-5}
                                & $I=0$ &  $-3/2$  &  $B^{+}B^{\ast 0}$  & $-1/2$    \\
\hline
\multirow{2}{*}{$B\bar{B}^{*}$} & $I=1$ &  $c/2$ &  $B^{0}B^{\ast +}$ & $-1/2$  \\
\cline{2-5}
                                & $I=0$ &  $-3c/2$  & $B^{0}B^{\ast 0}$  & $1/2$   \\
\hline\hline
\end{tabular}
\end{center}
\end{table}


Since the tensor operator $\hat{S}_{T}$ leads to S-D wave mixing,
the contributions from D-wave should be taken into account. Thus the wave function $\Psi(\vec{r})$
has  two parts
\begin{equation}
\Psi(\vec{r})=\psi_S(\vec{r})+\psi_D(\vec{r}),
\end{equation}
with $\psi_S(\vec{r})$ and $\psi_D(\vec{r})$ the $S$-wave and
$D$-wave functions, respectively. In the matrix method, we use
Laguerre polynomials
\begin{equation}
\chi_{nl}(r)=\sqrt{\frac{(2\lambda)^{2l+3} n!}{\Gamma(2l+3+n)}}r^l
e^{-\lambda r}L^{2l+2}_n (2\lambda r), ~~~~n=1,2,3...  \label{Laguerre}
\end{equation}
as a set of orthogonal basis with the normalization condition
\begin{equation}
\int^\infty _0 \chi_{im}(r) \chi_{in}(r) r^2
dr=\delta_{ij}\delta_{mn}.
\end{equation}
Thus the total wave function can be expanded as
\begin{eqnarray*}
\psi(\vec{r}) &=& \sum^{n-1}_{i=0}a_i \chi_{i 0}(r)\phi_S +
\sum^{n-1}_{p=0}b_p \chi_{p 2}(r)\phi_D, \\
\psi'(\vec{r}) &=& \sum^{n-1}_{i=0}a'_i \chi_{i 0}(r)\phi_S +
\sum^{n-1}_{p=0}b'_p \chi_{p 2}(r)\phi_D.
\end{eqnarray*}
where $\phi_S$ and $\phi_D$ are the angular part of the spin and
orbital wave function for the $S$-wave ($^3S_1$) and $D$-wave ($^3D_1$) states, respectively.
$a^{(\prime)}_i$ and $b^{(\prime)}_i$ are the corresponding expansion coefficients of S-wave and
D-wave, respectively. After solving the coupled-channel Schr\"odinger equation with the S-D wave mixing,
we obtain the binding energy $E_b$ and its corresponding wave function 
$\Psi(\Lambda, \vec{r}_{ab})$ for a given  parameter $\Lambda$. Thus the wave function has the form
\begin{eqnarray}
\Psi(\Lambda, \vec{r}_{ab}) = \frac{1}{\sqrt{2}}\psi(\Lambda, \vec{r}_{ab}) |B_{a}B_{b}^{*}\rangle + \frac{1}{\sqrt{2}}\psi(\Lambda, \vec{r}_{ab}) |B_{a}^{*}B_{b}\rangle +\psi'(\Lambda, \vec{r}_{ab}) |B^{*}B^{*}\rangle.\label{WFBB}
\end{eqnarray}
Here, the wave function $\Psi(\Lambda, \vec{r}_{ab})$ is normalized. 
If we choose the value of the parameter $\Lambda=1440$ MeV for instance,
one finds a loosely bound state for the isospin triplet system with a binding energy of 5.08~MeV,
when the quantum number is $J^P=1^+$. There is also a loosely bound state for the isospin singlet
system, when the quantum number is $J^P=1^+$. If the value of the parameter is chosen at $\Lambda=1107.7$~MeV,
the isospin singlet and triplet systems have the same binding energy of 5.08~MeV. The
dependence of the binding energy on the parameter $\Lambda$ will
be given in Tables~\ref{Eb-lamd01}-\ref{Eb-lamd00} and discussed in Sec.~\ref{sec7}.

\section{Born-Oppenheimer potential}\label{sec4}

As discussed in Sec.~\ref{sec2}, the BO potential reflects the influence of  one of the mesons on
the dynamics of the other two.
For the $BBB^{\ast}$ (labeled as $a$, $b$ and $c$) system, 
one can derive the BO potential from $a$ for the $bc$ system.
The procedure is divided into the following three steps:
\begin{itemize}
\item Considering that the particle $b$ and $c$ are static with the separation $r_{bc}$, 
one can separate the degree of freedom of $a$ from the three-body system.
\item We assume the distance $r_{bc}$ is a parameter. The mesons $b$ and $c$ are static,
  and have one-pion interactions with meson $a$, which can be viewed as two static sources. 
\item We explore the dynamics for the meson $a$ in the limit $r_{bc}\to \infty$, and subtract the
  binding energy for the break-up state which is trivial for the three-body bound state.
\end{itemize}
Within this scheme, we divide the motion of the system into two parts, one is the motion of the meson
$a$ relative to the mesons $b$ and $c$. The other one is the relative motion between mesons $b$ and $c$
in the presence of the BO potential from $a$.

\begin{figure}[ht]
  \begin{center}
  \rotatebox{0}{\includegraphics*[width=0.33\textwidth]{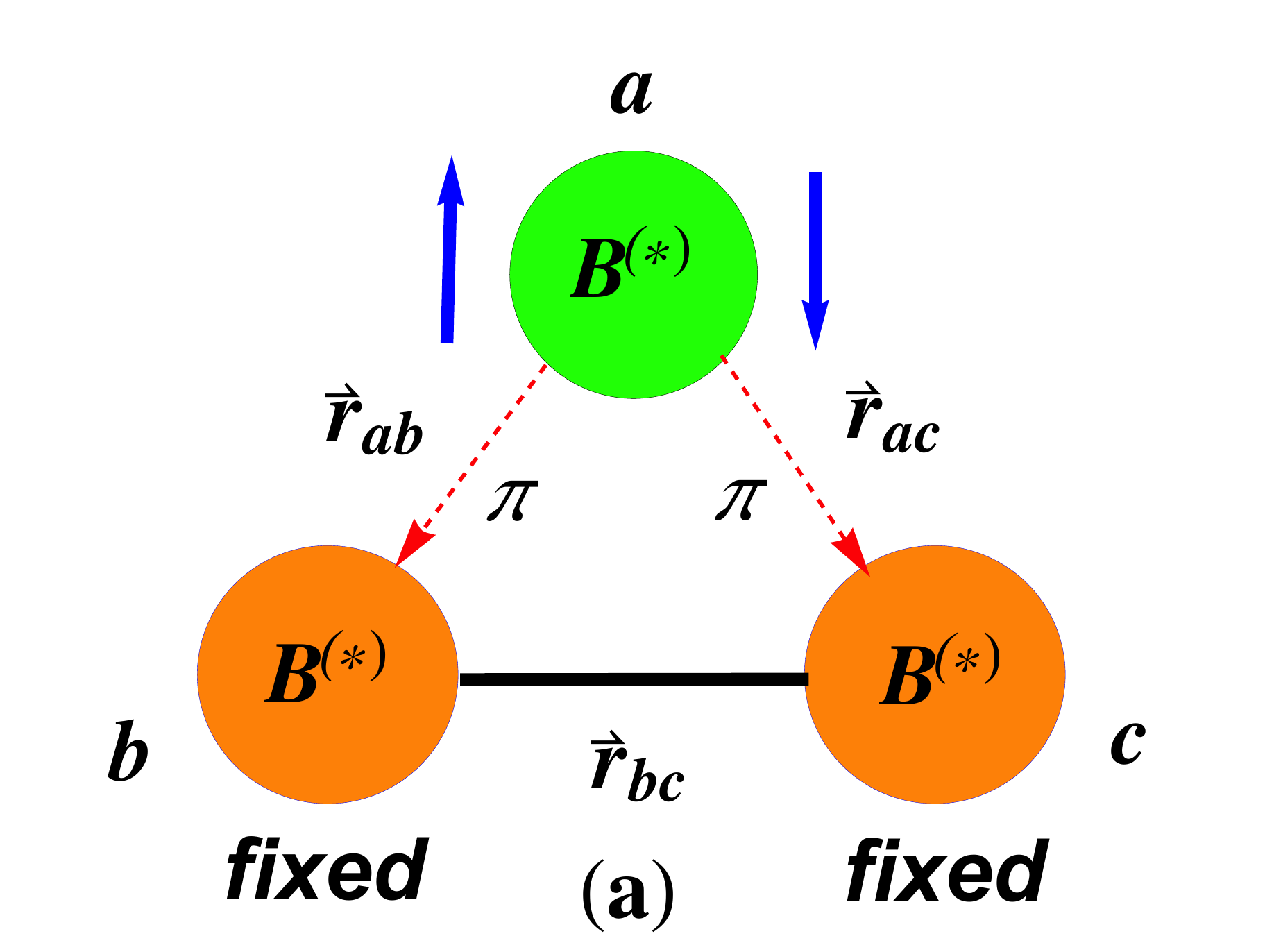}}
  \rotatebox{0}{\includegraphics*[width=0.33\textwidth]{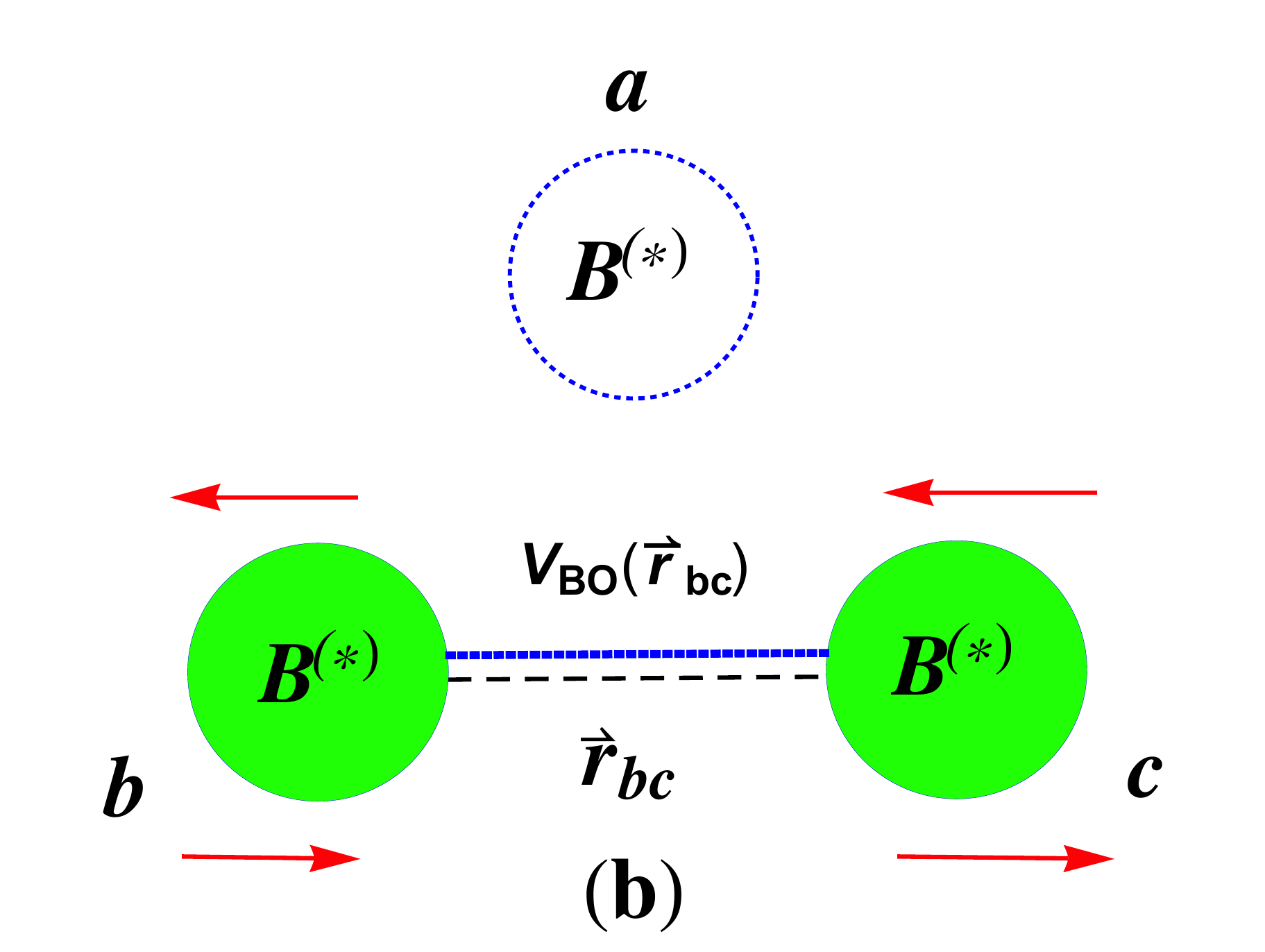}}
  \caption{Illustration of the BO potential. (a) illustrates the calculation procedure of
    the BO potential. (b) represents the role of the BO potential from the meson $a$ on
    the dynamics of the $bc$ two-body system.}
    \label{VBOS}
  \end{center}
\end{figure}

As illustrated in Figs.~\ref{VBOS}, we use $\vec{r}_{bc}$ to denote the relative displacement
 between $b$ and $c$. Further, $\vec{r}_{ab}$ and $\vec{r}_{ac}$ represent the displacement
  of the meson $a$ relative to the meson $b$ and $c$, respectively. 
  One can separate the effective potentials for the meson $a$
\begin{eqnarray}
V_{a}=\left(
         \begin{array}{cccccc}
           0 & V_1(\vec{r}_{ab}) &  V_1(\vec{r}_{ac})  &  V_2(\vec{r}_{ab}) & V_2(\vec{r}_{ac}) & 0 \\
          V_1(\vec{r}_{ba}) & 0  &  0  & V'_2(\vec{r}_{ba}) & 0 & 0 \\
          V_1(\vec{r}_{ca}) & 0 & 0 & 0 & V'_2(\vec{r}_{ac}) & 0 \\
          V_2(\vec{r}_{ba}) & V'_2(\vec{r}_{ab}) & 0 & V_3(\vec{r}_{ab}) & 0 & V_1(\vec{r}_{ac}) \\
          V_2(\vec{r}_{ca}) & 0 & V'_2(\vec{r}_{ca}) & 0 & V_3(\vec{r}_{ac}) & V_1(\vec{r}_{ab}) \\
          0 & 0 & 0 & V_1(\vec{r}_{ca}) & V_1(\vec{r}_{ba}) & 0 \\
         \end{array}
       \right). 
\end{eqnarray}
from the Eq.~(\ref{VBBB}).  
The remaining part
\begin{eqnarray}
V_{bc}=\left(
         \begin{array}{cccccc}
           0 & 0 &  0  &  0 & 0 & 0 \\
          0 & 0  &  V_1(\vec{r}_{bc})  & 0 & 0 & V_2(\vec{r}_{bc}) \\
          0 & V_1(\vec{r}_{cb}) & 0 & 0 & 0 & V'_2(\vec{r}_{bc}) \\
          0& 0 & 0 & 0 & V_1(\vec{r}_{bc}) & 0 \\
          0 & 0 & 0 & V_1(\vec{r}_{cb}) &0 & 0 \\
          0 & V_2(\vec{r}_{cb}) & V'_2(\vec{r}_{cb}) & 0 & 0 & V_3(\vec{r}_{bc}) \\
         \end{array}
       \right)
\end{eqnarray}
in Eq.~\eqref{VBBB} is the potential between $bc$.
As discussed in the previous section, one can obtain the two-body binding energy 
\begin{eqnarray}
  E_2 &=& \left( ~\frac{1}{\sqrt{2}}\psi(\vec{r}_{ab}), ~\frac{1}{\sqrt{2}}\psi(\vec{r}_{ab}),
  ~\psi'(\vec{r}_{ab})  \right)
\left(
        \begin{array}{ccc}
 T_{*} & V_1(\vec{r}_{ab})  & V_2(\vec{r}_{ab})  \\
 V_1(\vec{r}_{ba}) & T_{*}  & V'_2(\vec{r}_{ba})  \\
 V_2(\vec{r}_{ba}) & V'_2(\vec{r}_{ab})  & T_{**}+V_3(\vec{r}_{ab})  
       \end{array}
       \right)
      \left(
\begin{array}{c}
   \frac{1}{\sqrt{2}}\psi(\vec{r}_{ab})  \\
   \frac{1}{\sqrt{2}}\psi(\vec{r}_{ab})  \\
    \psi'(\vec{r}_{ab}) \\
\end{array}
\right) \nonumber\\ 
&=& \psi(\vec{r}_{ab}) T_{*} \psi(\vec{r}_{ab})+\psi'(\vec{r}_{ab}) T_{**} \psi'(\vec{r}_{ab}) + \psi(\vec{r}_{ab}) V_{1}^{ab} \psi(\vec{r}_{ab}) + 2\sqrt{2} \psi(\vec{r}_{ab}) V_{2}^{ab} \psi'(\vec{r}_{ab}) + \psi'(\vec{r}_{ab}) V_{3}^{ab} \psi'(\vec{r}_{ab}). \label{E2}
\end{eqnarray}
The $\psi(\vec{r}_{ab})$ and $\psi^\prime(\vec{r}_{ab})$ in the above equation are the eigenstate
wave functions in Eq.~(\ref{WFBB}). 

In OPE model, as the virtual pion can only be exchanged between two of the $BBB^*$ subsystems,
the wave function of $a$ can be either $\frac{1}{\sqrt{2}}\psi(\vec{r}_{ab}) |B_{a}^{*}B_{b}B_{c}\rangle
+ \frac{1}{\sqrt{2}}\psi(\vec{r}_{ab}) |B_{a}B_{b}^{*}B_{c}\rangle+\psi'(\vec{r}_{ab})
|B_{a}^{*}B_{b}^{*}B_{c}\rangle$ 
with pion exchanged between $a$ and $b$ or 
$\frac{1}{\sqrt{2}}\psi(\vec{r}_{ac}) |B_{a}^{*}B_{b}B_{c}\rangle+ \frac{1}{\sqrt{2}}\psi(\vec{r}_{ac})
|B_{a}B_{b}B_{c}^{*}\rangle+\psi'(\vec{r}_{ac}) |B_{a}^{*}B_{b}B_{c}^{*}\rangle $ with pion
exchanged between $a$ and $c$.
 The final wave function for the meson $a$ should be the superposition
of these two components
\begin{eqnarray}
\psi(\vec{r}_{ab}, \vec{r}_{ac}) &=& C \Big{\{} [\frac{1}{\sqrt{2}}\psi(\vec{r}_{ab})+\frac{1}{\sqrt{2}}\psi(\vec{r}_{ac})] |B_{a}^{*}B_{b}B_{c}\rangle + \frac{1}{\sqrt{2}}\psi(\vec{r}_{ab}) |B_{a}B_{b}^{*}B_{c}\rangle \nonumber\\
&+& \frac{1}{\sqrt{2}}\psi(\vec{r}_{ac}) |B_{a}B_{b}B_{c}^{*}\rangle +  \psi'(\vec{r}_{ab}) |B_{a}^{*}B_{b}^{*}B_{c}\rangle + \psi'(\vec{r}_{ac}) |B_{a}^{*}B_{b}B_{c}^{*}\rangle \Big{\}}. \label{WFa}
\end{eqnarray}

For simplicity, we neglect the mass difference for the $BB^*$ and $B^*B^*$ in the kinetic operator,
i.e. $T_{**}\approx T_{*}$. 
Then the Hamiltonian of the meson $a$ is 
\begin{eqnarray}
H_{a}\approx \left(
         \begin{array}{cccccc}
           T_{*} & V_1(\vec{r}_{ab}) &  V_1(\vec{r}_{ac})  &  V_2(\vec{r}_{ab}) & V_2(\vec{r}_{ac}) & 0 \\
          V_1(\vec{r}_{ba}) & T  &  0  & V'_2(\vec{r}_{ba}) & 0 & 0 \\
          V_1(\vec{r}_{ca}) & 0 & T & 0 & V'_2(\vec{r}_{ac}) & 0 \\
          V_2(\vec{r}_{ba}) & V'_2(\vec{r}_{ab}) & 0 & T_{*}+V_3(\vec{r}_{ab}) & 0 & V_1(\vec{r}_{ac}) \\
          V_2(\vec{r}_{ca}) & 0 & V'_2(\vec{r}_{ca}) & 0 & T_{*}+V_3(\vec{r}_{ac}) & V_1(\vec{r}_{ab}) \\
          0 & 0 & 0 & V_1(\vec{r}_{ca}) & V_1(\vec{r}_{ba}) & T_{*} \\
         \end{array}
       \right). 
\end{eqnarray}
Accordingly, one can obtain the energy eigenvalue of the meson $a$ 
\begin{eqnarray*}
E_a(\Lambda,~\vec{r}_{bc}) &=& \langle \psi(\vec{r}_{ab}, \vec{r}_{ac}) | H_a | \psi(\vec{r}_{ab}, \vec{r}_{ac})\rangle \\
&=& \frac{1}{1+\frac{1}{2}\langle \psi(\vec{r}_{ab})|\psi(\vec{r}_{ac})\rangle} \Big[ E_2+\frac{1}{2}\langle \psi(\vec{r}_{ab}) | T_{*} | \psi(\vec{r}_{ac})\rangle + \langle \psi(\vec{r}_{ab}) | V_{1}^{ba} | \psi(\vec{r}_{ac})\rangle + \sqrt{2} \langle \psi'(\vec{r}_{ab}) | V_{2}^{ba} | \psi(\vec{r}_{ac})\rangle \Big]
\end{eqnarray*}
where in the second step Eq.~(\ref{E2}) and the symmetry 
between $b$ and $c$ are used. Since both the two-body energy eigenvalue $E_2$ and the wave functions $\psi_b$,
$\psi_c$ depend on the parameter $\Lambda$, 
 $E_a$ is also a function of $\Lambda$. 
 
 \begin{figure}[ht]
  \begin{center}
  \rotatebox{0}{\includegraphics*[width=0.45\textwidth]{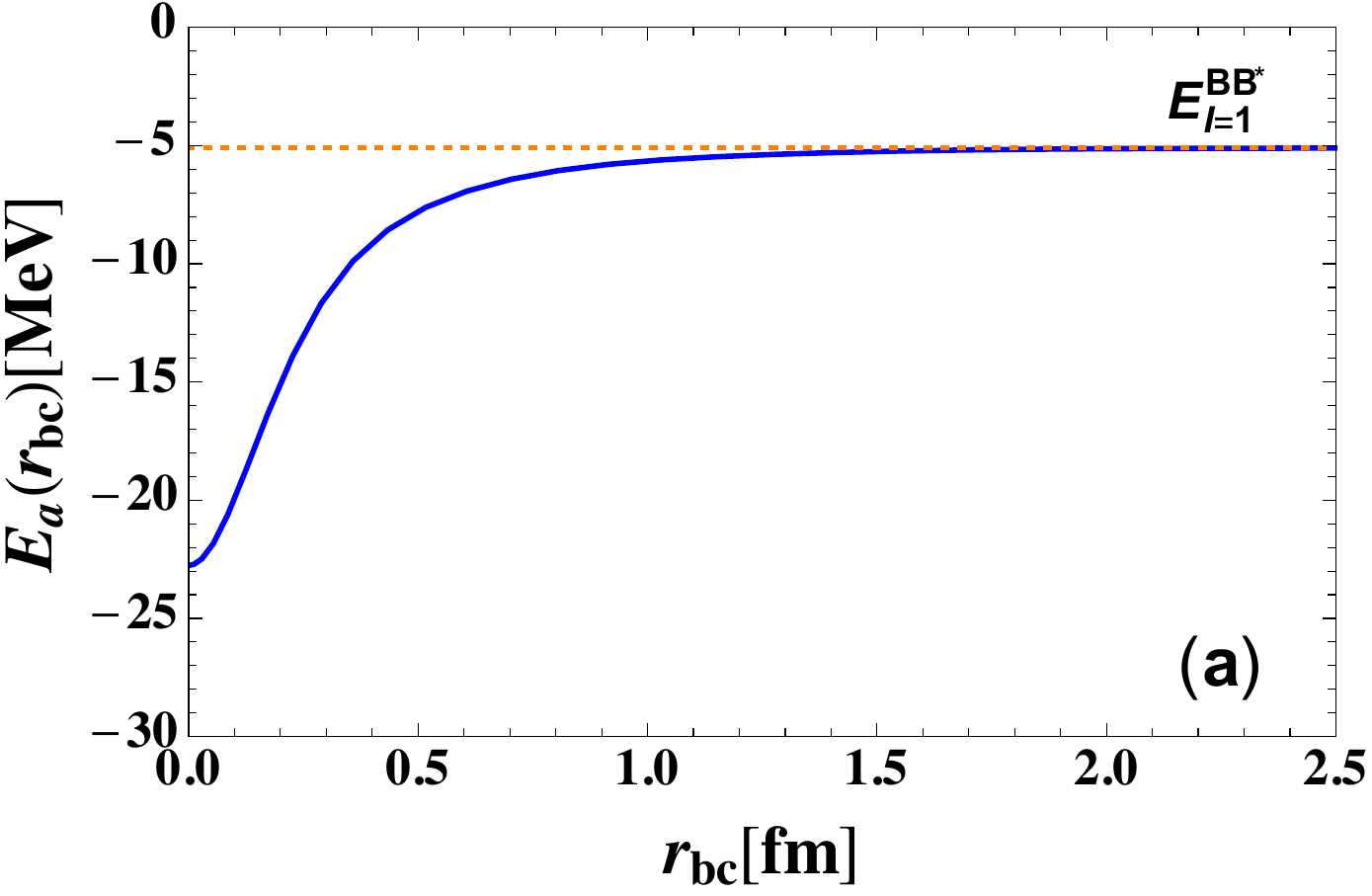}}
    \rotatebox{0}{\includegraphics*[width=0.45\textwidth]{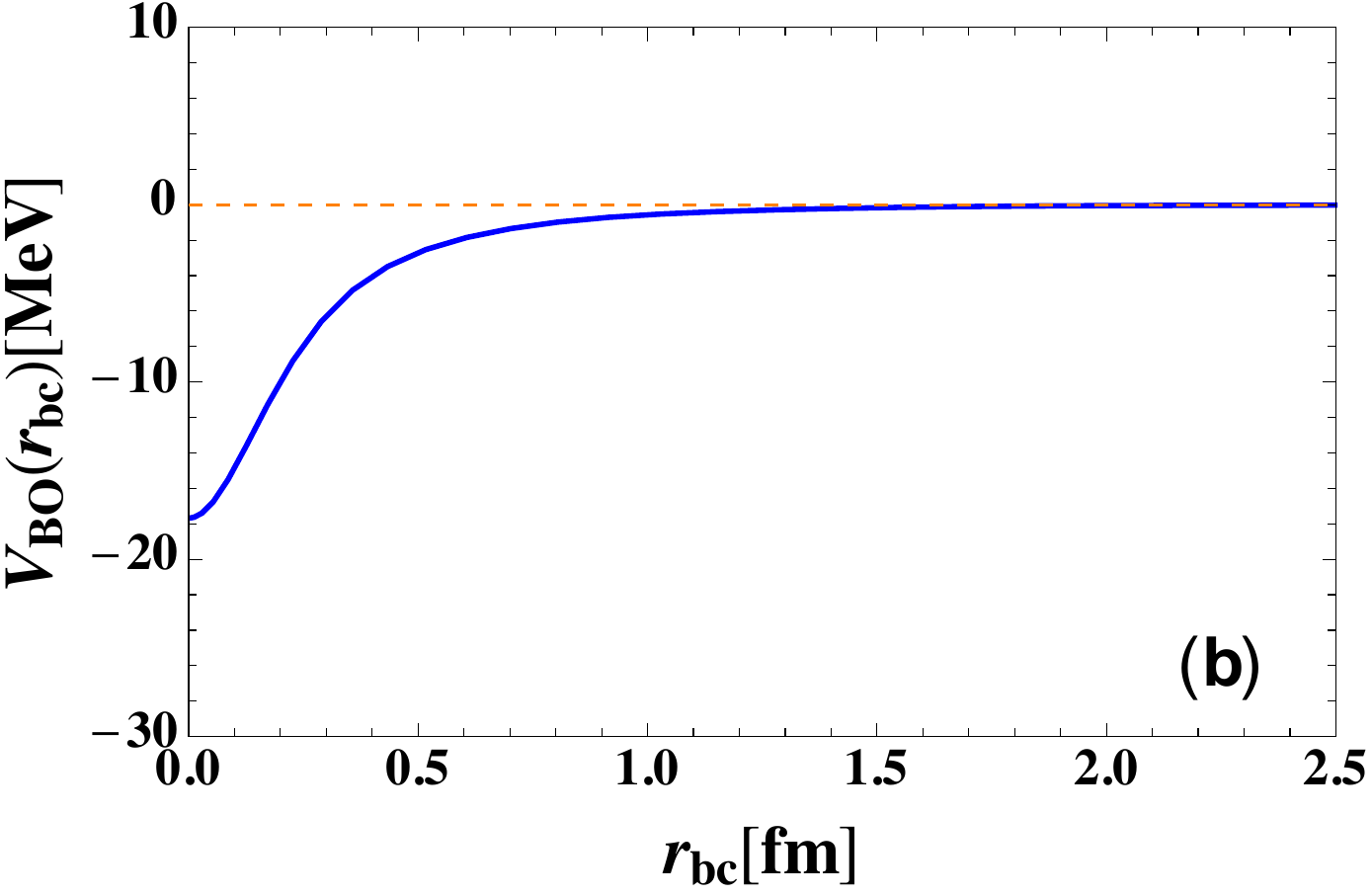}}
    \caption{The energy eigenvalue of the meson $a$ and its corresponding BO potential for the
      isospin triplet of the $BB^*$ system. (a) gives the energy eigenvalue of the meson $a$. When
      $r_{bc}\to \infty$,  $E_a$ tends to the two-body energy eigenvalue $E_2=E_{I=1}^{BB^*}$, i.e. the
      energy eigenvalue of the break-up state. The right panel gives the BO potential $V_{BO}$.
      Here we chose the parameter $\Lambda=1440$ MeV. }
    \label{VBO01}
  \end{center}
\end{figure}

\begin{figure}[ht]
  \begin{center}
  \rotatebox{0}{\includegraphics*[width=0.45\textwidth]{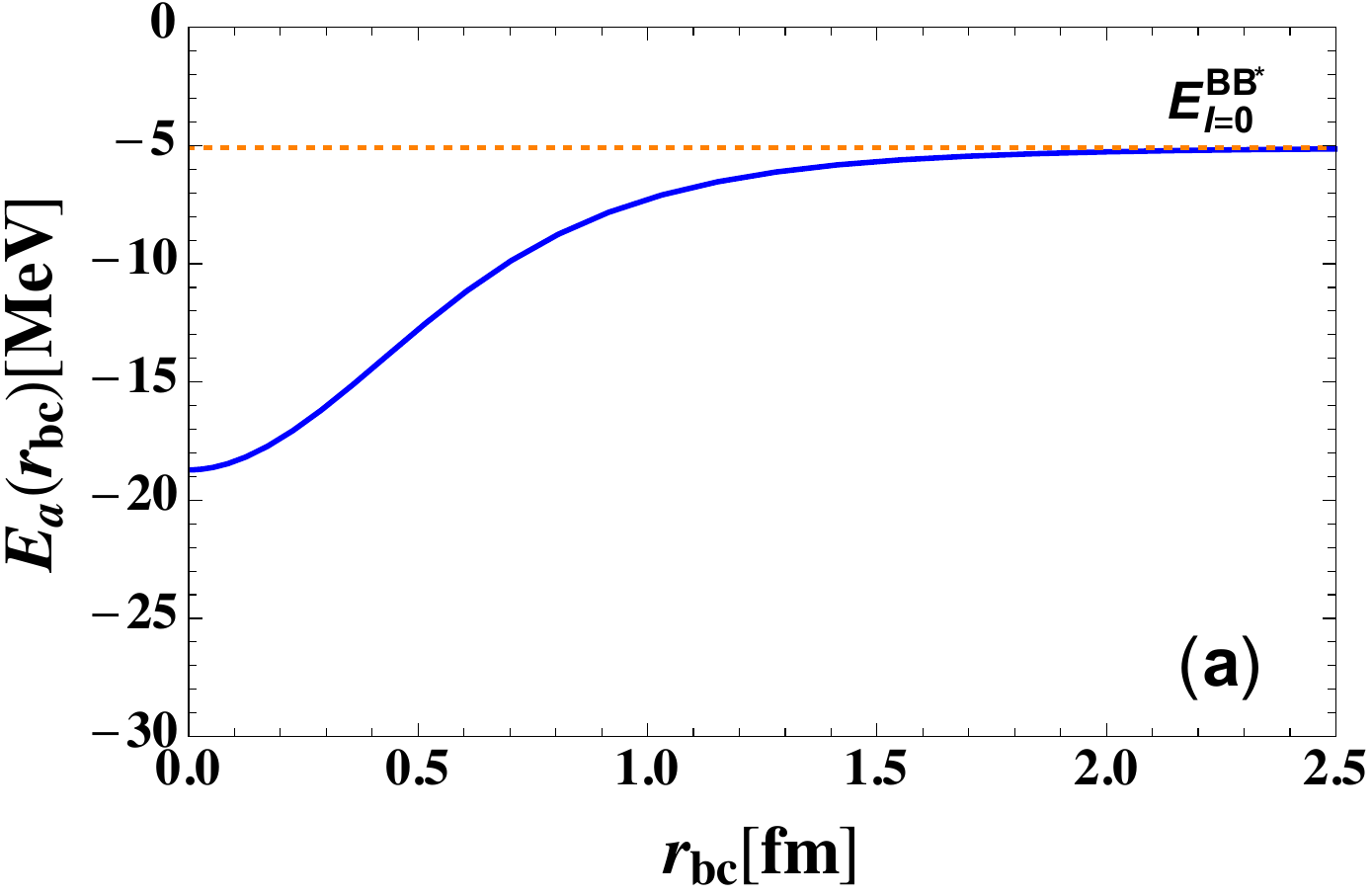}}
    \rotatebox{0}{\includegraphics*[width=0.45\textwidth]{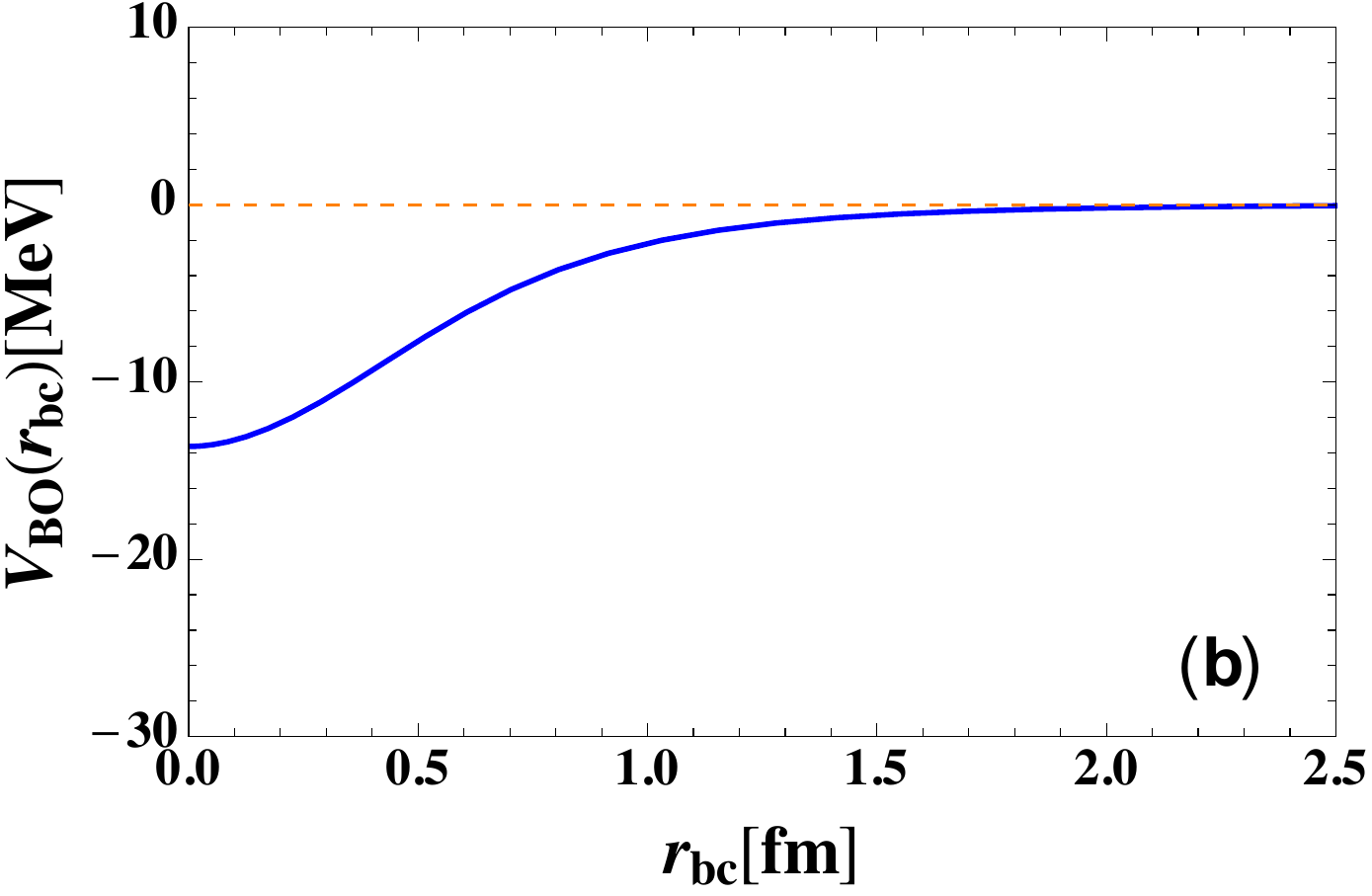}}
    \caption{The energy eigenvalue of the meson $a$ and its corrsponding BO potential for the
      isospin singlet of the $BB^*$ system. (a) is the energy eigenvalue of the meson $a$.
      When $r_{bc}\to \infty$, $E_a$ tends to the two-body energy eigenvalue $E_2=E_{I=0}^{BB^*}$, i.e.
      the energy eigenvalue of the break-up state. The right panel gives the BO potential $V_{BO}$.
      Here we chose the parameter $\Lambda=1107.7$ MeV. }
    \label{VBO00}
  \end{center}
\end{figure}

We take the parameter $\Lambda=1440$ MeV as an example and plot  $E_a$ for the isospin triplet
of the $BB^*$ system in Figs. \ref{VBO01}(a). As shown in the figure, 
the energy of the meson $a$ has a minimum $-17.65$~MeV when  $r_{bc}=0$,
which corresponds to the limit that the mesons $b$ and $c$ are on top of each other 
and the system is reduced to the $bc$-$a$ quasi-two-body system.
When  $r_{bc}\to \infty$, then $E_a$ tends to the two-body energy eigenvalue $E_2$, i.e.$-5.08$~MeV.
This corresponds to the situation that the meson $b$ is infinitely far away from the meson $c$.
Then the meson $a$ can only form a two-body bound state with either $b$ or $c$. 
It is not a three-body bound state anymore, but rather a two-body bound state plus a free meson state.
In fact, this is nothing but  the break-up state that we have discussed in the earlier sections.
We also plot  $E_a$ for the isospin triplet of the $BB^*$ system in Fig.~\ref{VBO00}(a), taking
the parameter $\Lambda=1107.7$~MeV. Similarly to the above,
$E_a$ tends to the two-body energy eigenvalue $-5.08$~MeV.  
Therefore, we should subtract the limiting value $E_2$ when investigating the three-body bound
state for the $BBB^*$ system. 
We define the BO potential as
\begin{equation}
V_{BO}(\Lambda,~\vec{r}_{bc})=E_a(\Lambda,~\vec{r}_{bc})- E_2(\Lambda). \label{DVBO}
\end{equation}
In other words, the BO potential between $b$ and $c$ is the energy eigenvalue
of the meson $a$ relative to that of the break-up state.

\section{The configurations of the three-body systems}\label{sec5}

In the OPE model, there is only one pion exchanged between any two constituents in the $BBB^{\ast}$ system.
The constituents will change themselves from vector mesons into pseudo-scalar mesons or vice verse 
when they exchange one pion. Each constituent has the same probability to be a vector meson or a
pseudo-scalar meson. 
Thus, the symbol $\ast$ is shared among them. Since only one virtual pion occurs in the
$BBB^{\ast}$ molecule, the virtual pion also be shared by the three mesons. 
We can thus write the $BBB^{\ast}$ as $B_{a}^{(\ast)}B_{b}^{(\ast)}B_{c}^{(\ast)}$. 

\begin{figure}[ht]
  \begin{center}
  \rotatebox{0}{\includegraphics*[width=0.33\textwidth]{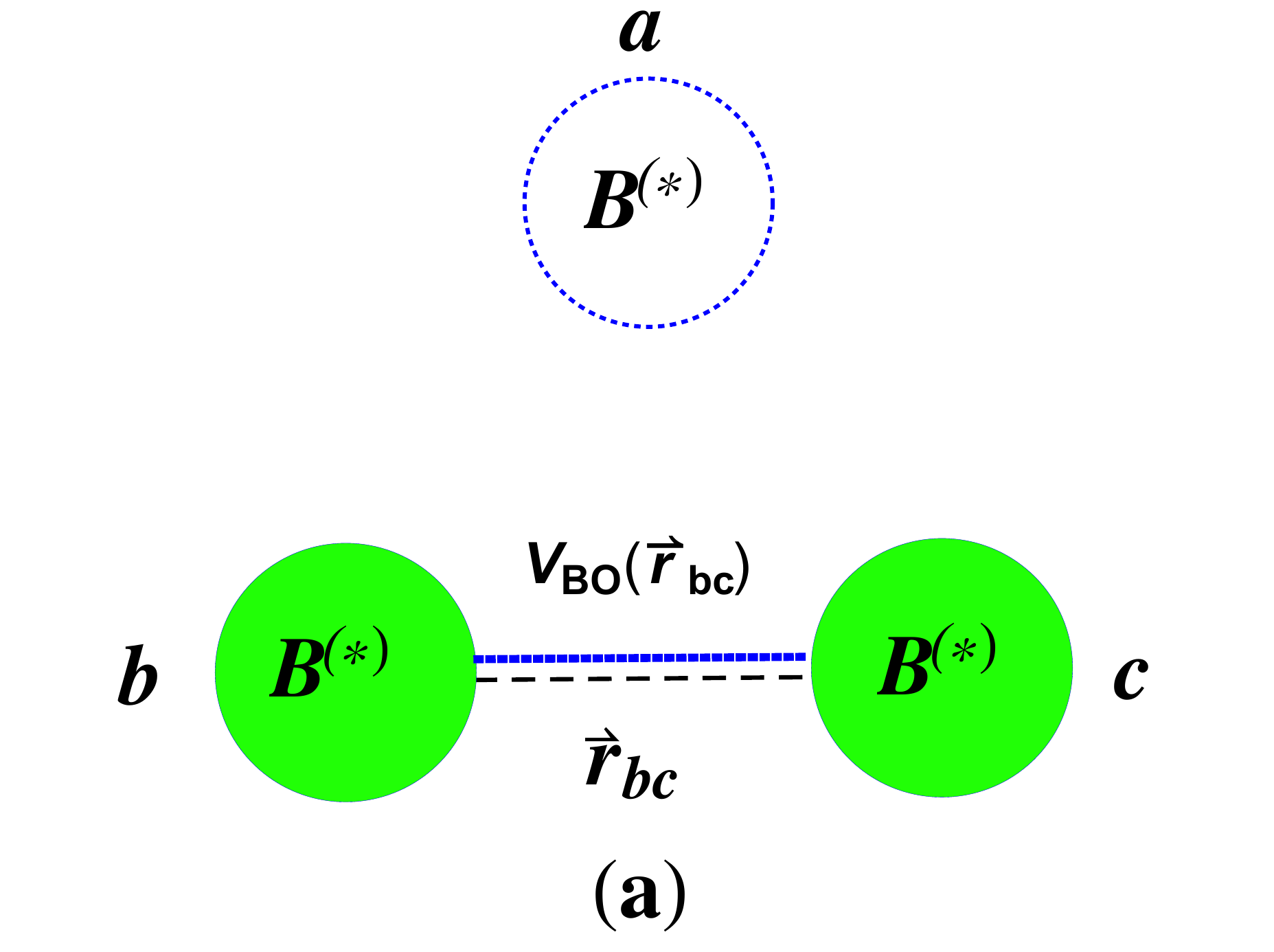}}
    \rotatebox{0}{\includegraphics*[width=0.33\textwidth]{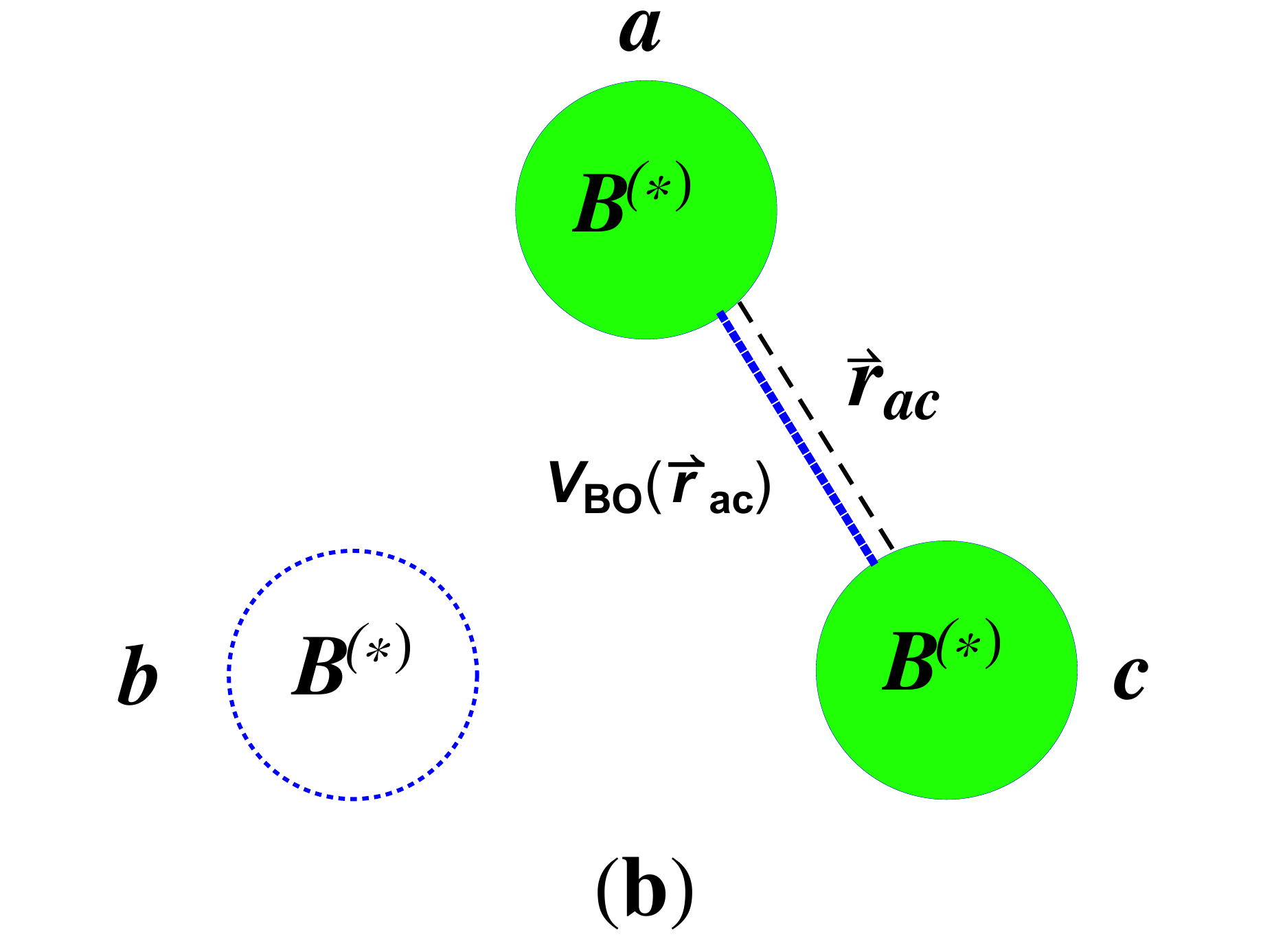}}
      \rotatebox{0}{\includegraphics*[width=0.33\textwidth]{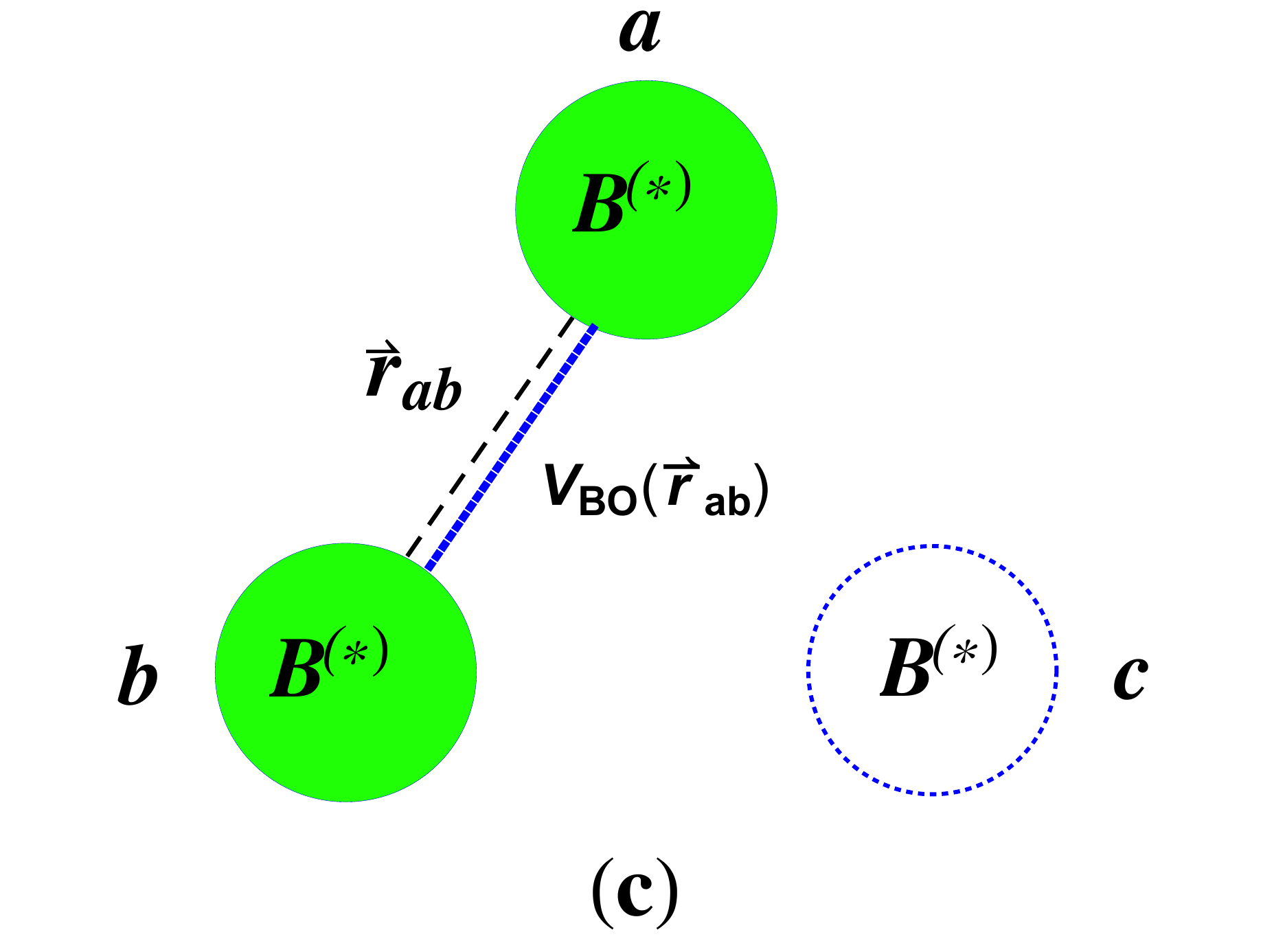}}
      \caption{ Three configurations of the $BBB^{\ast}$ system. (a), (b) and (c) correspond to the wave
        functions $\psi_{\slashed{a}}$, $\psi_{\slashed{b}}$ and $\psi_{\slashed{c}}$, respectively. }
    \label{BOPM}
  \end{center}
\end{figure}

The BO potential can describe the contribution for the one meson on the dynamics of the two remaining
mesons as we have discussed in the last section. Assuming that the meson $b$ and $c$ are much heavier
than the meson $a$, then we can use the Born-Oppenheimer approximation to separate the degree of freedom
of $a$ from the three-body system. In other words, it is a kind of an  adiabatic approximation that
we divide the degrees of freedom of the three-body system into a light one and a heavy one. The motion
of the light degree of freedom is the motion of meson $a$ relative to the three-body centre of mass.
The motion of the heavy degree of freedom is the relative  motion between meson $b$ and $c$. When
exploring the dynamics for the meson $a$, we can assume the meson $b$ and $c$ are static with the
distance $r_{bc}$. Then the three-body system can be simplified as a two-body system consisting of mesons
$b$ and $c$ but with an additional BO potential generated by the meson $a$. Overall, only the meson $a$
can be separated from the system due to the fact that  this meson is much lighter than the other ones.
A separation in this way can be a good approximation for this system.
With the same procedure that we derived Eq.~(\ref{WFa}), we obtain the wave functions
$\psi(\vec{r_{ab}},\vec{r_{ac}})$ for the meson $a$. The remaining degree of freedom is the
relative motion between meson $b$ and $c$ that can be described by a wave function assumed as
$\Phi(\vec{r_{be}})$, to be determined from the Schr\"{o}dinger equation. Then the total wave function of
the  system then has the form        
\begin{eqnarray*}
\Psi_T=\Phi(\vec{r}_{bc})\psi(\vec{r}_{ab},\vec{r}_{ac}). 
\end{eqnarray*}
Nevertheless, the true system $B_{a}^{(\ast)}B_{b}^{(\ast)}B_{c}^{(\ast)}$ is that the three mesons have
little mass difference. Every meson can be considered to be a lighter one and separated from the
three-body system. Thus, the system has the three basic simplification schemes. That is
we can divide the system $B_{a}^{(\ast)}B_{b}^{(\ast)}B_{c}^{(\ast)}$ into three kinds of two-body
subsystems, i.e., $B_{a}^{(\ast)}B_{b}^{(\ast)}$ with the BO potential created by the meson $c$,
$B_{b}^{(\ast)}B_{c}^{(\ast)}$ with the BO potential created by the meson $a$ and
$B_{a}^{(\ast)}B_{c}^{(\ast)}$ with the BO potential created by the meson $b$ as shown in Figs.~\ref{BOPM}.
These three simplification schemes can be regarded as three kinds of basic configurations.
The eigenstates of the three-body system should be combinations of them. As the most simplest
combination, one might expect the three-body eigenstate should be the superposition of the
three kinds of basic configurations. 
We use the $\psi_{\slashed{a}}$, $\psi_{\slashed{b}}$, $\psi_{\slashed{c}}$ to denote these
three configurations. The configuration wave function $\psi_{\slashed{a}}$ represent the configuration
that we omit the meson $B_a^{(\ast)}$ and add the corresponding BO potential instead.
Similarly, $\psi_{\slashed{b}}$, $\psi_{\slashed{c}}$ denote the configurations
with the BO potentials provided by the mesons $B_b^{(\ast)}$ and $B_c^{(\ast)}$, respectively.  

Taking the configuration function $\psi_{\slashed{a}}$ as an example,
we separate the motion of the $B_a^{(\ast)}$ relative to the other mesons  
$B_b^{(\ast)}$ and $B_c^{(\ast)}$ where their relative displacement $r_{bc}$ is regarded as
a parameter as shown in Fig.~\ref{BOPM}(a). 
The wave function of the $B_a^{(\ast)}$ has been discussed in the
last section and can be written as $\psi(\vec{r}_{ab}, \vec{r}_{ac})$. 
The remaining degree of freedom is the relative motion between $B_b^{(\ast)}$ and $B_c^{(\ast)}$, which can
be taken as $\Phi(\vec{r}_{bc})$. Thus we have the configuration function $\psi_{\slashed{a}}
=\Phi(\vec{r}_{bc})\psi(\vec{r}_{ab},~\vec{r}_{ac})$.    
The other two wave functions $\psi_{\slashed{b}}$ and $\psi_{\slashed{c}}$ can be obtained analogously,
i.e. $\psi_{\slashed{b}}=\Phi(\vec{r}_{ac})\psi(\vec{r}_{ab},~\vec{r}_{bc})$, $\psi_{\slashed{c}}=
\Phi(\vec{r}_{ab})\psi(\vec{r}_{bc},~\vec{r}_{ac})$ which correspond to the Fig.~\ref{BOPM}(b)
and Fig.~\ref{BOPM}(c), respectively. 
If we regard the three configuration functions as a set of  basis states, then the basis
constitute a configuration space $\{ \psi_{\slashed{a}},~\psi_{\slashed{b}},~\psi_{\slashed{c}} \}$.
The three-body eigenstate expressed as a superposition of the three kinds of basic configurations can
be described as a state vector in this configuration space. 
Thus, as an interpolating wave function, the three-body wave functions can be written as
\begin{eqnarray}
\Psi_T &=& \alpha\Phi(\vec{r}_{bc})\psi(\vec{r}_{ab},~\vec{r}_{ac})+\beta\Phi(\vec{r}_{ac})\psi(\vec{r}_{ab},~\vec{r}_{bc})+\gamma\Phi(\vec{r}_{ab})\psi(\vec{r}_{bc},~\vec{r}_{ac}) \nonumber\\
&=& \alpha\psi_{\slashed{a}}+\beta\psi_{\slashed{b}}+\gamma\psi_{\slashed{c}}  
 = \left(
\begin{array}{c}
   \alpha  \\
   \beta  \\
   \gamma 
\end{array}
\right), \label{Basis0}
\end{eqnarray}
where the $\Phi(\vec{r}_{bc})$, $\Phi(\vec{r}_{ac})$ and $\Phi(\vec{r}_{ab})$ are undetermined
functions that need to be solved. The $\alpha$, $\beta$ and $\gamma$ are the expansion  coefficients. 
According to Eq.~(\ref{WFa}), we rewrite the three basic configuration functions in the channel
space $\{ B_a^{\ast}B_bB_c, B_aB_b^{\ast}B_c, B_aB_bB_c^{\ast}, B_a^{\ast}B_b^{\ast}B_c, B_a^{\ast}B_bB_c^{\ast},
B_aB_b^{\ast}B_c^{\ast} \}$ as 
\begin{eqnarray}
\psi_{\slashed{a}} = C \Phi(\vec{r}_{bc}) \left(
\begin{array}{c}
   \frac{1}{\sqrt{2}}[\psi(\vec{r}_{ab})+\psi(\vec{r}_{ac})] \\
   \frac{1}{\sqrt{2}}\psi(\vec{r}_{ab})  \\
   \frac{1}{\sqrt{2}}\psi(\vec{r}_{ac})  \\
   \psi'(\vec{r}_{ab})  \\
   \psi'(\vec{r}_{ac})  \\
   0
\end{array}
\right), 
~
\psi_{\slashed{b}} = C \Phi(\vec{r}_{ac}) \left(
\begin{array}{c}
   \frac{1}{\sqrt{2}}\psi(\vec{r}_{ab})  \\
   \frac{1}{\sqrt{2}}[\psi(\vec{r}_{ab})+\psi(\vec{r}_{bc})] \\
   \frac{1}{\sqrt{2}}\psi(\vec{r}_{bc})  \\
   \psi'(\vec{r}_{ab})  \\
    0 \\
   \psi'(\vec{r}_{bc})
\end{array}
\right), 
~
\psi_{\slashed{c}} = C \Phi(\vec{r}_{ab}) \left(
\begin{array}{c}
   \frac{1}{\sqrt{2}}\psi(\vec{r}_{ac})  \\
   \frac{1}{\sqrt{2}}\psi(\vec{r}_{bc})  \\
   \frac{1}{\sqrt{2}}[\psi(\vec{r}_{bc})+\psi(\vec{r}_{ac})] \\
   0 \\
   \psi'(\vec{r}_{ac})  \\
   \psi'(\vec{r}_{bc})
\end{array}
\right), 
\end{eqnarray}
which can be expanded as a set of Laguerre polynomials
\begin{eqnarray*}
\psi_{\slashed{a}} &=& \sum_{i} \phi_i(\vec{r}_{bc})\psi(\vec{r}_{ab},~\vec{r}_{ac}), \\
\psi_{\slashed{b}} &=& \sum_{i} \phi_i(\vec{r}_{ac})\psi(\vec{r}_{ab},~\vec{r}_{bc}), \\
\psi_{\slashed{c}} &=& \sum_{i} \phi_i(\vec{r}_{ab})\psi(\vec{r}_{bc},~\vec{r}_{ac}). 
\end{eqnarray*}
Here the subscript $i$ is the order of Laguerre polynomials. We define the $i^{\rm th}$ order of the
configuration functions as $\psi_{\slashed{a}}^i = \phi_i(\vec{r}_{bc})\psi(\vec{r}_{ab},~\vec{r}_{ac})$,
$\psi_{\slashed{b}}^i = \phi_i(\vec{r}_{ac})\psi(\vec{r}_{ab},~\vec{r}_{bc})$ and $\psi_{\slashed{c}}^i
= \phi_i(\vec{r}_{ab})\psi(\vec{r}_{bc},~\vec{r}_{ac})$. Further,  $C$ is a normalisation constant. 

We expect the three-body bound state that we seek for can be expressed as a state vector in the
configuration space $\{ \psi_{\slashed{a}},~\psi_{\slashed{b}},~\psi_{\slashed{c}} \}$. 
However, the configuration functions in Eq.~(\ref{Basis0}) are not an orthogonal basis.  
Thus we orthonormalize the $\{ \psi_{\slashed{a}},~\psi_{\slashed{b}},~\psi_{\slashed{c}} \}$
into a new basis $\{ \tilde{\psi}_{\slashed{a}},~\tilde{\psi}_{\slashed{b}},~\tilde{\psi}_{\slashed{c}} \}$.
We use $\tilde{\psi}_{\slashed{a}}^i$, $\tilde{\psi}_{\slashed{b}}^i$ and $\tilde{\psi}_{\slashed{c}}^i$ to
denote the $i^{\rm th}$th order of the new configuration functions $\tilde{\psi}_{\slashed{a}}$,
$\tilde{\psi}_{\slashed{b}}$ and $\tilde{\psi}_{\slashed{c}}$, respectively. Then we have
\begin{eqnarray*}
  \tilde{\psi}_{\slashed{a}}^i &=& \frac{1}{N_i}\big[ (\psi_{\slashed{a}}^i+\psi_{\slashed{b}}^i+\psi_{\slashed{c}}^i)-\sum_{i} x_{ij}\psi_{\slashed{a}}^j \big], \\
  \tilde{\psi}_{\slashed{b}}^i &=& \frac{1}{N_i}\big[ (\psi_{\slashed{a}}^i+\psi_{\slashed{b}}^i+\psi_{\slashed{c}}^i)-\sum_{i} x_{ij}\psi_{\slashed{b}}^j \big], \\
  \tilde{\psi}_{\slashed{c}}^i &=& \frac{1}{N_i}\big[ (\psi_{\slashed{a}}^i+\psi_{\slashed{b}}^i+\psi_{\slashed{c}}^i)-\sum_{i} x_{ij}\psi_{\slashed{c}}^j \big], 
\end{eqnarray*}
where the $x_{ij}$ is a parameter matrix which will be determined later. The $N_i$ are normalization
coefficients. The parameter matrix $x_{ij}$ in the three configuration functions are the same due
to the interchange symmetry for the $B_a^{(\ast)}B_b^{(\ast)}B_c^{(\ast)}$ system.

Since the $i^{\rm th}$ order configuration function $\tilde{\psi}_{\slashed{a}}^i$ should be orthogonal
with the any order of the other configuration function $\tilde{\psi}_{\slashed{a}}^j$, one can gets
the orthogonalization condition
\begin{eqnarray*}
\langle \tilde{\psi}_{\slashed{a}}^i | \tilde{\psi}_{\slashed{b}}^j \rangle= \Big{\langle} \frac{1}{N_i}\big[ (\psi_{\slashed{a}}^i+\psi_{\slashed{b}}^i+\psi_{\slashed{c}}^i)-\sum_{i} x_{ik}\psi_{\slashed{a}}^k \big] \Big{|} \frac{1}{N_j}\big[ (\psi_{\slashed{a}}^j+\psi_{\slashed{b}}^j+\psi_{\slashed{c}}^j)-\sum_{i} x_{jl}\psi_{\slashed{b}}^l \big] \Big{\rangle} = 0,
\end{eqnarray*}
which gives
\begin{eqnarray}
x_{ik}\langle \psi_{\slashed{a}}^k | \psi_{\slashed{b}}^l \rangle x_{lj} - x_{ik}(\delta_{kj}+2\langle \psi_{\slashed{a}}^k | \psi_{\slashed{b}}^j \rangle) - x_{jl}(\delta_{il}+2\langle \psi_{\slashed{a}}^i | \psi_{\slashed{b}}^l \rangle) + 3\delta_{ij}+6\langle \psi_{\slashed{a}}^i | \psi_{\slashed{b}}^j \rangle = 0~. \label{orth}
\end{eqnarray}
This equation will determine the parameter matrix $x_{ij}$. Considering the normalization of the
$i^{\rm th}$ order configuration function $\tilde{\psi}_{\slashed{a}}^i$
\begin{eqnarray*}
\langle \tilde{\psi}_{\slashed{a}}^i | \tilde{\psi}_{\slashed{a}}^j \rangle= \Big{\langle} \frac{1}{N_i}\big[ (\psi_{\slashed{a}}^i+\psi_{\slashed{b}}^i+\psi_{\slashed{c}}^i)-\sum_{i} x_{ik}\psi_{\slashed{a}}^k \big] \Big{|} \frac{1}{N_j}\big[ (\psi_{\slashed{a}}^j+\psi_{\slashed{b}}^j+\psi_{\slashed{c}}^j)-\sum_{i} x_{jl}\psi_{\slashed{a}}^l \big] \Big{\rangle} = \delta_{ij}.
\end{eqnarray*}
one can obtain the normalization equation for the $N_i$ as
\begin{eqnarray}
\frac{1}{N^{\ast}_i N_j} \big[  3\delta_{ij}+ 6 \langle \psi_{\slashed{a}}^i | \psi_{\slashed{b}}^j \rangle -2x_{ij}-4 \sum_{m} x_{im} \langle \psi_{\slashed{a}}^m | \psi_{\slashed{b}}^j \rangle + \sum_{n} x_{in}x_{nj}  \big]=\delta_{ij}~. \label{norm}
\end{eqnarray}
After solving the equations for $x_{ij}$ and $N_i$, we obtain an orthonormalized configuration basis.
This basis constitutes a orthonormalized configuration space.
 Then the eigenvector for the three-body system $B_a^{(\ast)}B_b^{(\ast)}B_c^{(\ast)}$ can
 be written as a vector in the configuration space $\{ \tilde{\psi}_{\slashed{a}},~\tilde{\psi}_{\slashed{b}},
 ~\tilde{\psi}_{\slashed{c}} \}$. Therefore, we have
\begin{eqnarray*}
\Psi_T= \sum_{i}\tilde{\alpha}_{i}\tilde{\psi}_{\slashed{a}}^i+\sum_{j}\tilde{\beta}_{j}\tilde{\psi}_{\slashed{b}}^j+\sum_{k}\tilde{\gamma}_{k}\tilde{\psi}_{\slashed{c}}^k,
\end{eqnarray*}
where the $\tilde{\alpha}_{i}$, $\tilde{\beta}_{i}$ and $\tilde{\gamma}_{i}$ are the $i^{\rm th}$ order
expansion coefficients.

\section{Three-body Schr\"{o}dinger equation}\label{sec6}

As we have discussed in previous sections, if the three-body binding energy is below
the break-up  threshold, the three-body system will disintegrate into a two-body system and a free meson.  
Since we only focus on the three-body bound state, we could make an energy shift and remove
the energy eigenvalue $E_2$ for the break-up state and define a reduced Hamiltonian for
the three-body system as
\begin{equation*}
\mathcal{H}=H-E_2. 
\end{equation*}
The explicit form of $H$ is
\begin{eqnarray}
H=\left(
         \begin{array}{cccccc}
           T_{*}+T'_{*} & V_1(\vec{r}_{ab}) &  V_1(\vec{r}_{ac})  &  V_2(\vec{r}_{ab}) & V_2(\vec{r}_{ac}) & 0 \\
          V_1(\vec{r}_{ba}) & T_{*}+T'_{*}  &  V_1(\vec{r}_{bc})  & V'_2(\vec{r}_{ba}) & 0 & V_2(\vec{r}_{bc}) \\
          V_1(\vec{r}_{ca}) & V_1(\vec{r}_{cb}) & T+T' & 0 & V'_2(\vec{r}_{ac}) & V'_2(\vec{r}_{bc}) \\
          V_2(\vec{r}_{ba}) & V'_2(\vec{r}_{ab}) &  0 & T_{**}+ T'_{**}+V_3(\vec{r}_{ab})+\delta M  & V_1(\vec{r}_{bc}) & V_1(\vec{r}_{ac}) \\
          V_2(\vec{r}_{ca}) & 0 & V'_2(\vec{r}_{ca}) & V_1(\vec{r}_{cb}) & T_{*}+T'_{*}+V_3(\vec{r}_{ac})+\delta M & V_1(\vec{r}_{ab}) \\
          0 & V_2(\vec{r}_{cb}) & V'_2(\vec{r}_{cb}) & V_1(\vec{r}_{ca}) & V_1(\vec{r}_{ba}) & T_{*}+T'_{*}+V_3(\vec{r}_{bc})+\delta M \\
         \end{array}
       \right), \label{HBBB}
\end{eqnarray}
where $T_{*}=-({1}/{2\mu_{*}})\nabla_{ab}^2$, $T=-({1}/{2\mu})\nabla_{ab}^2$, $T_{**}=-({1}/{2\mu_{**}})
\nabla_{ab}^2$, $T'_{*}=-({1}/{2\mu'_{*}})\nabla_{\xi}^2$, $T'=-({1}/{2\mu'})\nabla_{\xi}^2$,
$T'_{**}=-({1}/{2\mu'_{**}})\nabla_{\xi}^2$ are the kinetic energy operators and the corresponding reduced  masses are $\mu_{*}=({M_BM_{B^*}})/({M_B+M_{B^*}})$, $\mu={M_B}/{2}$,
$\mu_{**}={M_{B^*}}/{2}$, $\mu'_{*}=({(M_B+M_{B^*})M_B})/({2M_B+M_{B^*}})$, $\mu'=({2M_BM_{B^*}})/({2M_B+M_{B^*}})$,
$\mu'_{**}=({2M_{B^*}M_B})/({2M_{B^*}+M_B})$. 
Here $\nabla_{ab}^2=({1}/{r_{ab}})({d^2}/{dr_{ab}^2})r_{ab}-({\overrightarrow{L}_{ab}^2}/{r_{ab}^2})$ and
$\nabla_{\xi}^2=({1}/{\xi})({d^2}/{d\xi^2})\xi-({\overrightarrow{L}_{\xi}^2}/{\xi^2})$
with $\vec{\xi}={\vec{r}_{ab}}/{2}-\vec{r}_{bc}$. $\vec{r_{bc}}$ is the direction of the meson $b$ relative to the meson $c$.
$\overrightarrow{L}_{ab}$ is the angular momentum operator between mesons $a$ and $b$. 
$\overrightarrow{L}_{\xi}$ is the relative angular momentum operator between two-body centre of mass
for the meson $a~b$ and the meson $c$. 
The mass gap is $\delta M=M_{B^*}-M_B$. 

The total Hamiltonian for the three-body system in the configuration space $\{ \tilde{\psi}_{\slashed{a}},~\tilde{\psi}_{\slashed{b}},~\tilde{\psi}_{\slashed{c}} \}$ can be written as
\begin{eqnarray}
H_T = \left(
        \begin{array}{ccc}
 H_{\slashed{a}\slashed{a}} & H_{\slashed{a}\slashed{b}}  & H_{\slashed{a}\slashed{c}} \\
 H_{\slashed{b}\slashed{a}} & H_{\slashed{b}\slashed{b}}  & H_{\slashed{b}\slashed{c}}  \\
 H_{\slashed{c}\slashed{a}} & H_{\slashed{c}\slashed{b}}  & H_{\slashed{c}\slashed{c}}  
       \end{array}
       \right)=
 \left(
        \begin{array}{ccc}
 \mathcal{H}_{\slashed{a}\slashed{a}}+E_2 & \mathcal{H}_{\slashed{a}\slashed{b}}+E_2  & \mathcal{H}_{\slashed{a}\slashed{c}}+E_2 \\
 \mathcal{H}_{\slashed{b}\slashed{a}}+E_2 & \mathcal{H}_{\slashed{b}\slashed{b}}+E_2  & \mathcal{H}_{\slashed{b}\slashed{c}}+E_2  \\
 \mathcal{H}_{\slashed{c}\slashed{a}}+E_2 & \mathcal{H}_{\slashed{c}\slashed{b}}+E_2  & \mathcal{H}_{\slashed{c}\slashed{c}}+E_2  
       \end{array}
       \right)=
       \left(
        \begin{array}{ccc}
 \mathcal{H}_{\slashed{a}\slashed{a}} & \mathcal{H}_{\slashed{a}\slashed{b}}  & \mathcal{H}_{\slashed{a}\slashed{c}} \\
 \mathcal{H}_{\slashed{b}\slashed{a}} & \mathcal{H}_{\slashed{b}\slashed{b}}  & \mathcal{H}_{\slashed{b}\slashed{c}}  \\
 \mathcal{H}_{\slashed{c}\slashed{a}} & \mathcal{H}_{\slashed{c}\slashed{b}}  & \mathcal{H}_{\slashed{c}\slashed{c}}  
       \end{array}
       \right)+E_2
       \left(
        \begin{array}{ccc}
        1 & 0 & 0 \\
        0 & 1 & 0 \\
        0 & 0 & 1
        \end{array}
       \right),
\end{eqnarray}
with $H_{\slashed{m}\slashed{n}}=\langle \tilde{\psi}_{\slashed{m}}| H | \tilde{\psi}_{\slashed{m}}\rangle$
$(m,~n=a,~b,~c)$.

The total reduced Hamiltonian for the three-body system $B_a^{(\ast)}B_b^{(\ast)}B_c^{(\ast)}$ in
the configuration space $\{ \tilde{\psi}_{\slashed{a}},~\tilde{\psi}_{\slashed{b}},
~\tilde{\psi}_{\slashed{c}} \}$ can be expressed as
\begin{eqnarray}
\mathcal{H}_T = \left(
        \begin{array}{ccc}
 \mathcal{H}_{\slashed{a}\slashed{a}} & \mathcal{H}_{\slashed{a}\slashed{b}}  & \mathcal{H}_{\slashed{a}\slashed{c}} \\
 \mathcal{H}_{\slashed{b}\slashed{a}} & \mathcal{H}_{\slashed{b}\slashed{b}}  & \mathcal{H}_{\slashed{b}\slashed{c}}  \\
 \mathcal{H}_{\slashed{c}\slashed{a}} & \mathcal{H}_{\slashed{c}\slashed{b}}  & \mathcal{H}_{\slashed{c}\slashed{c}}  
       \end{array}
       \right),
\end{eqnarray}
with $\mathcal{H}_{\slashed{m}\slashed{n}}=\langle \tilde{\psi}_{\slashed{m}}| \mathcal{H} |
\tilde{\psi}_{\slashed{m}}\rangle$ $(m,~n=a,~b,~c)$. Thus we have
\begin{equation*}
H_T=\mathcal{H}_{T}+E_2. 
\end{equation*}

The matrix element of the $\mathcal{H}_{\slashed{a}\slashed{a}}$ can be written as  
\begin{eqnarray}
\mathcal{H}_{\slashed{a}\slashed{a}}^{ij} &=& \langle \tilde{\psi}_{\slashed{a}}^i | \mathcal{H} | \tilde{\psi}_{\slashed{a}}^j \rangle = \Big{\langle} \frac{1}{N_i}\big[ (\psi_{\slashed{a}}^i+\psi_{\slashed{b}}^i+\psi_{\slashed{c}}^i)-\sum_{i} x_{im}\psi_{\slashed{a}}^m \big] \Big{|} \mathcal{H} \Big{|} \frac{1}{N_j}\big[ (\psi_{\slashed{a}}^j+\psi_{\slashed{b}}^j+\psi_{\slashed{c}}^j)-\sum_{i} x_{jn}\psi_{\slashed{a}}^n \big] \Big{\rangle} \nonumber\\
&=& 3\frac{1}{N_i N_j} \langle \psi_{\slashed{a}}^i | \mathcal{H} | \psi_{\slashed{a}}^j \rangle + 6\frac{1}{N_i N_j} \langle \psi_{\slashed{b}}^i | \mathcal{H} | \psi_{\slashed{a}}^j \rangle -x_{im} \frac{1}{N_i N_j} \langle \psi_{\slashed{a}}^m | \mathcal{H} | \psi_{\slashed{a}}^j \rangle -x_{jn}\frac{1}{N_i N_j} \langle \psi_{\slashed{a}}^i | \mathcal{H} | \psi_{\slashed{a}}^n \rangle \nonumber\\
&-& 2 x_{im} \frac{1}{N_i N_j} \langle \psi_{\slashed{b}}^m | \mathcal{H} | \psi_{\slashed{a}}^j \rangle -2 x_{jn} \frac{1}{N_i N_j} \langle \psi_{\slashed{b}}^i | \mathcal{H} | \psi_{\slashed{a}}^n \rangle + x_{im}x_{jn} \frac{1}{N_i N_j} \langle \psi_{\slashed{a}}^m | \mathcal{H} | \psi_{\slashed{a}}^n \rangle. \label{Haa}
\end{eqnarray}
where, in the last step, the interchange symmetry in the $B_a^{(\ast)}B_b^{(\ast)}B_c^{(\ast)}$ system is used. 
Similarly, we also have 
\begin{eqnarray}
\mathcal{H}_{\slashed{b}\slashed{a}}^{ij} &=& \langle \tilde{\psi}_{\slashed{b}}^i | \mathcal{H} |\tilde{\psi}_{\slashed{a}}^j \rangle = \Big{\langle} \frac{1}{N_i}\big[ (\psi_{\slashed{a}}^i+\psi_{\slashed{b}}^i+\psi_{\slashed{c}}^i)-\sum_{i} x_{im}\psi_{\slashed{b}}^m \big] \Big{|} \mathcal{H} \Big{|} \frac{1}{N_j}\big[ (\psi_{\slashed{a}}^j+\psi_{\slashed{b}}^j+\psi_{\slashed{c}}^j)-\sum_{i} x_{jn}\psi_{\slashed{a}}^n \big] \Big{\rangle} \nonumber\\
&=& 3\frac{1}{N_i N_j} \langle \psi_{\slashed{a}}^i | \mathcal{H} | \psi_{\slashed{a}}^j \rangle + 6\frac{1}{N_i N_j} \langle \psi_{\slashed{b}}^i | \mathcal{H} | \psi_{\slashed{a}}^j \rangle -x_{im} \frac{1}{N_i N_j} \langle \psi_{\slashed{a}}^m | \mathcal{H} | \psi_{\slashed{a}}^j \rangle -x_{jn}\frac{1}{N_i N_j} \langle \psi_{\slashed{a}}^i | \mathcal{H} | \psi_{\slashed{a}}^n \rangle \nonumber\\
&-& 2 x_{im} \frac{1}{N_i N_j} \langle \psi_{\slashed{b}}^m | \mathcal{H} | \psi_{\slashed{a}}^j \rangle -2 x_{jn} \frac{1}{N_i N_j} \langle \psi_{\slashed{b}}^i | \mathcal{H} | \psi_{\slashed{a}}^n \rangle + x_{im}x_{jn} \frac{1}{N_i N_j} \langle \psi_{\slashed{b}}^m | \mathcal{H} | \psi_{\slashed{a}}^n \rangle. \label{Hba}
\end{eqnarray}

There are two independent matrices 
\begin{eqnarray*}
\langle \psi_{\slashed{a}}^i | \mathcal{H} | \psi_{\slashed{a}}^j \rangle &=& |C|^2  \int d\vec{r}_{bc}  \Big{\{}    
  (\langle \psi_{ab} | \psi_{ab} \rangle+ \langle \psi_{ab} | \psi_{ac} \rangle)[ \phi_{bc}^i (T'+V_{BO}^{bc}) \phi_{bc}^j ] + (1+\langle \psi'_{ab} | \psi'_{ab} \rangle)[ \phi_{bc}^i (T'_{*}+V_{BO}^{bc}) \phi_{bc}^j ] \\
  &+& (\langle \psi_{ab} | \psi_{ac} \rangle+2\langle \psi'_{ab} | \psi'_{ac} \rangle)[ \phi_{bc}^i V_{1}^{bc} \phi_{bc}^j ]  \Big{\}}.
\end{eqnarray*}
\begin{eqnarray*}
\langle \psi_{\slashed{b}}^i | \mathcal{H} | \psi_{\slashed{a}}^j \rangle &=&  |C|^2  \int d\vec{r}_{bc}  \Big{\{} 
\frac{1}{2}\langle \phi_{ac}^i \psi_{ab} | T'+V_{BO}^{bc} | \phi_{bc}^j (\psi_{ab}+\psi_{ac}) \rangle + \frac{1}{2}\langle \phi_{ac}^i (\psi_{ab}+\psi_{bc}) | T'_{*}+V_{BO}^{bc} | \phi_{bc}^j \psi_{ab} \rangle \\
&+& \frac{1}{2}\langle  \phi_{ac}^i \psi_{bc}  | T'_{*}+V_{BO}^{bc} | \phi_{bc}^j \psi_{ac}  \rangle+
\langle  \phi_{ac}^i \psi'_{ab}  | T'_{*}+V_{BO}^{bc} | \phi_{bc}^j \psi'_{ab}  \rangle \\
&+& \frac{1}{2} \langle  \phi_{ac}^i \psi_{bc}  | V_{1}^{cb} | \phi_{bc}^j \psi_{ab}  \rangle +
\frac{1}{\sqrt{2}} \langle  \phi_{ac}^i \psi'_{bc}  | V_{2}^{cb} | \phi_{bc}^j \psi_{ab}  \rangle+
\langle \phi_{ac}^i \psi'_{ab}  | V_{1}^{cb} | \phi_{bc}^j \psi'_{ac}  \rangle \\
&+& \frac{1}{2}\langle \phi_{ac}^i (\psi_{ab}+\psi_{bc}) | V_{1}^{bc} | \phi_{bc}^j \psi_{ac} \rangle+
\frac{1}{\sqrt{2}} \langle  \phi_{ac}^i \psi'_{bc}  | V_{2}^{cb} | \phi_{bc}^j \psi_{ac}  \rangle
\Big{\}},
\end{eqnarray*}
to be determined, 
where we have used the abbreviations $\phi_{ab}^i$, $\phi_{bc}^i$, $\phi_{ac}^i$, $\psi_{ab}$,
$\psi_{bc}$, $\psi_{ac}$ for $\phi(\vec{r}_{ab})^i$, $\phi(\vec{r}_{bc})^i$, $\phi(\vec{r}_{ac})^i$,
$\psi(\vec{r}_{ab})$, $\psi(\vec{r}_{bc})$, $\psi(\vec{r}_{ac})$ respectively. 
The expression for the $\mathcal{H}_{\slashed{c}\slashed{a}}$ can be easily obtained by the
replacement $c\rightarrow b, ~b\rightarrow c$ on the expression for the
$\mathcal{H}_{\slashed{b}\slashed{a}}$. Similarly, the expression for the
$\mathcal{H}_{\slashed{c}\slashed{b}}$ is obtained by the replacement $c\rightarrow b, ~b\rightarrow a,
~a\rightarrow c$ on the expression for the $\mathcal{H}_{\slashed{b}\slashed{a}}$. In fact,
interchange invariance for the $B_a^{(\ast)}B_b^{(\ast)}B_c^{(\ast)}$ system can simplify the calculation,
i.e. $\mathcal{H}_{\slashed{c}\slashed{a}}=\mathcal{H}_{\slashed{c}\slashed{b}}=\mathcal{H}_{\slashed{b}\slashed{a}}$
and $\mathcal{H}_{\slashed{a}\slashed{a}}=\mathcal{H}_{\slashed{b}\slashed{b}}=\mathcal{H}_{\slashed{c}\slashed{c}}$. 

Based on the above discussion, the three-body Schr\"{o}dinger equation can finally be written as
\begin{eqnarray}
 \left(
        \begin{array}{ccc}
 \mathcal{H}_{\slashed{a}\slashed{a}} & \mathcal{H}_{\slashed{a}\slashed{b}}  & \mathcal{H}_{\slashed{a}\slashed{c}} \\
 \mathcal{H}_{\slashed{b}\slashed{a}} & \mathcal{H}_{\slashed{b}\slashed{b}}  & \mathcal{H}_{\slashed{b}\slashed{c}}  \\
 \mathcal{H}_{\slashed{c}\slashed{a}} & \mathcal{H}_{\slashed{c}\slashed{b}}  & \mathcal{H}_{\slashed{c}\slashed{c}}  
       \end{array}
       \right)
       \left(
\begin{array}{c}
   \tilde{\alpha}  \\
   \tilde{\beta}  \\
   \tilde{\gamma} 
\end{array}
\right)
= E_3
\left(
\begin{array}{c}
   \tilde{\alpha}  \\
   \tilde{\beta}  \\
   \tilde{\gamma} 
\end{array}
\right),  \label{SDE}
\end{eqnarray}
where the energy eigenvalue $E_3$ is the reduced three-body energy eigenvalue. The total energy
eigenvalue relative to the $BBB^{\ast}$ mass threshold is $E_T=E_3+E_2$. 
Solving the three-body Schr\"{o}dinger equation may partly answer whether the
three-body system has a loosely bound state or not.

\section{Application to the $NNN$ system}\label{sec7}

In order to verify the feasibility of the Born-Oppenheimer potential method for the three heavy system, we apply it to the three nucleon system. 
Since there is sufficient experimental data for this 
system, we can apply the formalism introduced above 
to investigate its binding energy and illustrate the 
feasibility of our formalism.  
As we know, the triton and the helium-3 ($^3$He) nucleus are the two possible bound states of the $NNN$ system,
both of them have the quantum numbers $I(J^P)=\frac{1}{2}({\frac{1}{2}}^+)$ but have different isospin on its $z$ direction.
They have the same structure and the binding energy if isospin symmetry breaking effect is neglected. 
The calculation on the three-nucleon system is much more straightforward than the $BBB^*$ system, as there are no other
coupled channels. For simplicity, we only write down the isospin wave functions of the triton and helium-3 nuclei, which are 
\begin{eqnarray*}
\left|0, \frac{1}{2}, \frac{1}{2}\right\rangle &=& \frac{1}{\sqrt{2}}[| (pn)p \rangle-| (np)p \rangle],  \\
\left|0, \frac{1}{2}, -\frac{1}{2}\right\rangle &=& \frac{1}{\sqrt{2}}[| (pn)n \rangle-| (np)n \rangle]. 
\end{eqnarray*}
The Lagrangian reads
\begin{equation*}
\mathcal{L}_{N} = g_{N} \bar{N} i \gamma_{5} \vec{\tau} N \cdot \vec{\pi}, 
\end{equation*}
where the $g_N=14.70$ is the coupling constants (we use here the pseudoscalar coupling, which is fine to the order we are
working, see e.g. Ref.~\cite{Bernard:1989fe}). $N=(\psi_p, \psi_n)$ is the nucleon doublet. Further, $\vec{\tau}=\{ \tau_1, \tau_2,
\tau_3 \}$ are the Pauli matrices, and $\vec{\pi}=\{ \frac{1}{\sqrt{2}}(\pi^{+}+\pi^{-}), \frac{i}{\sqrt{2}}(\pi^{+}+\pi^{-}),
\pi^0 \}$ are the $\pi$ fields. 
With a procedure similar to the one  discussed in the Sec. \ref{sec2}-\ref{sec6}, we can investigate the properties of the
break-up state formed by a deuteron and a free nucleon as well as the three-body bound states.  
As discussed in the above sections, there is only one free parameter $\Lambda$ in the
monopole form factor introduced in Sec.~\ref{sec3}, 
which reflects, in a rough way, the internal structure of the interacting hadrons. In other words,
the size of hadron is proportional to $1/\Lambda$, which is still unknown from the fundamental theory. 
Thus the parameter $\Lambda=899.60~\mathrm{MeV}$
 is fixed by the binding energy $E_2 = 2.23$ MeV of 
 deuteron in our calculations. With the so determined parameter $\Lambda$, we can obtain the BO potentials for the
 $NNN$ system using the formalism in the Sec.~\ref{sec2}-\ref{sec6}, with just the replacement of the effective potential
 $V_{BB*}$ by $V_{NN}$ in the calculations. This potential reads
\begin{eqnarray*}
V_{NN\rightarrow NN}(\vec{r}) &=& -C_{\pi}^{NN}(i,j)\frac{g_N^2}{12 M_N^2} \{ \vec{\sigma}_1\cdot\vec{\sigma}_2  [~\tilde{m}_{\pi}^2\tilde{\Lambda} Y(\tilde{\Lambda} r)-\tilde{m}_{\pi}^3Y(\tilde{m}_{\pi}r)\nonumber\\
&+&(\Lambda^2-m_{\pi}^2)\tilde{\Lambda}\frac{e^{-\tilde{\Lambda} r}}{2}~]
+S_{T}(\vec{\sigma}_1, \vec{\sigma}_2)[-\tilde{m}_{\pi}^3 Z(\tilde{m}_{\pi}r)+ \tilde{\Lambda}^3 Z(\tilde{\Lambda} r)\nonumber\\
&+&(\Lambda^2-m_{\pi}^2)(1+\tilde{\Lambda} r)\frac{\tilde{\Lambda}}{2}Y(\tilde{\Lambda} r)~] \}, \label{VNN}
\end{eqnarray*}
where the $C_{\pi}^{NN}(i,j)$ is the channel-dependent coefficient for the two-nucleon system, $M_N$ is the mass of nucleon,
$g_N$ is the pion-nucleon coupling constant, and $\sigma_1$ and $\sigma_2$ are the spin Pauli matrix 
for the nucleon 1 and 2 in the scattering process $1+2\rightarrow 3+4$. 

\begin{table}[htbp]
\caption{Bound state solutions of the $NNN$ system with  isospin $I_3=1/2$. 
$E_2$ is the energy eigenvalue of its subsystem. 
 $E_3$ is the reduced three-body energy eigenvalue relative to the break-up state of the $NNN$ system.  
$E_T$ is the total three-body energy eigenvalue relative to the $NNN$ threshold. 
 $V_{BO}(0)$ is the minimum of the BO potential. 
$r_{rms}$ represents the root-mean-square radius of any two $N$ in the $NNN$ system. 
The $S$-wave and $D$-wave represent the probabilities for $S$-wave and
 $D$-wave components in any two $N$ in the $NNN$ system.}\label{E3-EbNN}
\begin{center}
\begin{tabular}{ c | c | c | c  | c  | c | c  | c  | c  }
\hline\hline 
  $\Lambda$(MeV)  &  $E_2$(MeV)  & $E_3$(MeV) & $E_T$(MeV)  & $V_{BO}(0)$(MeV)  & S wave(\%) & D wave(\%) & $r_{rms}$(fm)   \\
\hline
        830.00 & -0.18 & -1.93 & -2.11 & -4.54 & 94.01 & 5.99 & 4.21 \\
\hline       
       850.00 &  -0.67 & -2.71 & -3.38 & -5.36 & 93.36 & 6.64 & 4.00 \\
\hline       
       870.00 &  -1.23 & -3.65 & -4.88 & -6.32 & 92.68 & 7.32 & 3.78  \\
\hline       
       890.00 &  -1.88 & -4.77 & -6.66 & -7.42 & 91.99 & 8.01 & 3.54 \\
\hline       
       899.60 &  -2.23 & -5.38 & -7.62 & -8.00 & 91.66 & 8.34 & 3.42 \\
\hline
       900.00 &  -2.25 & -5.41 & -7.66 & -8.03 & 91.64 & 8.36 & 3.42 \\
\hline
       920.00 &  -3.05 & -6.85 & -9.90 & -9.35 & 90.97 & 9.03 & 3.18 \\
\hline
       940.00 &  -3.98 & -8.51 & -12.49 & -10.83 & 90.35 & 9.65 & 2.95 \\
\hline
       960.00 &  -5.03 & -10.42 & -15.45 & -12.46 & 89.76 & 10.24 & 2.74 \\
\hline
       980.00 &  -6.21 & -12.57 & -18.78 & -14.23 & 89.23 & 10.77 & 2.54 \\
\hline
      1000.00 &  -7.55 & -14.97 & -22.51 & -16.14 & 88.73 & 11.27 & 2.37 \\
\hline
      1020.00 &  -9.04 & -17.61 & -26.65 & -18.19 & 88.27 & 11.73 & 2.23 \\
\hline
      1040.00 & -10.69 & -20.51 & -31.20 & -20.37 & 87.84 & 12.16 & 2.10 \\
\hline\hline
\end{tabular}
\end{center}
\end{table}

After solving the three-body Schr\"{o}dinger equation, i.e.Eq. (\ref{SDE}), one can obtain
 the dependence of the binding of the three-nucleon system on the parameter
  $\Lambda$ (Table~\ref{E3-EbNN}). As shown in the table, there is a three-body bound 
  state with the reduced binding energy and the total three-body bound energy
   in the range of $1.93$-$20.51$ $\mathrm{MeV}$ and $2.11$-$31.20$ $\mathrm{MeV}$, respectively,
  when the parameter $\Lambda$ varies from $830~ \mathrm{MeV}$ to $1040~ \mathrm{MeV}$. 
  The corresponding isospin singlet two-body subsystem $NN$ has the
  binding enrgy in the range of $0.18$-$10.69$ $\mathrm{MeV}$.
  The root-mean-square of the system decreases from 
  4.21~$\mathrm{fm}$ to 2.10~$\mathrm{fm}$ when the parameter increases.
  Once the parameter $\Lambda=899.60~\mathrm{MeV}$ fixed by the deuteron 
  binding energy, the reduced three-body binding energy and the total binding
  energy relative to the three free nucleons are 5.38~$\mathrm{MeV}$ and 
  7.62~$\mathrm{MeV}$, respectively. The later is comparable with the 
  empirical binding energies of the triton (8.48~$\mathrm{MeV}$) and helium-3 (7.80~$\mathrm{MeV}$) nuclei. 
  Noet again that there is no numerical difference between the binding energies of triton 
  and helium-3 in the calculation, as the isospin breaking has not be considered.

\begin{figure}[ht]
  \begin{center}
     \rotatebox{0}{\includegraphics*[width=0.45\textwidth]{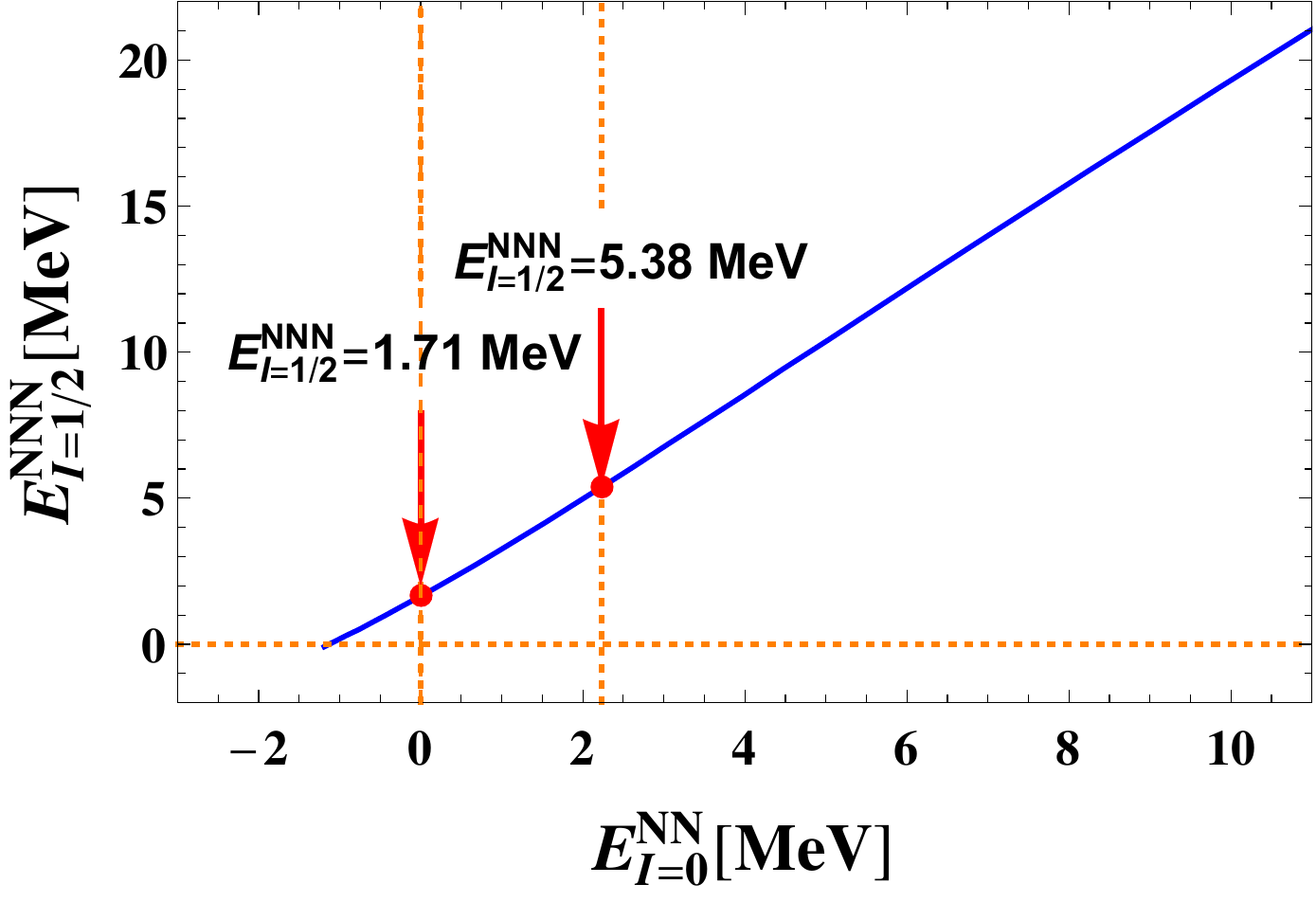}}
     \caption{Dependence of the reduced three-body binding energy on the binding energy of
       its two-body subsystem (the deuteron). 
       The left red point is the critical point which indicates the lower
       limit of the required binding energy of the deuteron to form a three-body bound state. The right
       one is our numerical result of the binding energy of the triton or the helium-3 nucleus.}
    \label{E3-EbNNN}
  \end{center}
\end{figure}

For a better illustration of the binding property,
 we plot the dependence of the reduced three-body binding energy
  on the two-body binding energy of its deuteron subsystem.
  As shown in Fig.~{\ref{E3-EbNNN}}, the binding energy of the three-nucleon system
   becomes larger when the binding energy of its subsystem $NN$ increases.
    There are two red points in the figure, 
    the left red point is the critical point which indicates
     the lower limit of the required binding energy of the deuteron
      to form a three-body bound state. It is very interesting that 
      even though the binding energy of its subsystem is zero, 
      there is a small binding energy of the three-nucleon system, which is 1.71 $\mathrm{MeV}$. 
    This is reminsicent of a Borromean state, 
    where a three-body system may have a bound state despite the fact that none of its subsystems forms a bound state.
    The other red point is our numerical result of the binding energy of triton or helium-3. It is a little below the experimental
    values since in our calculations we use the BOP method to construct our interpolating wave functions, which can be regarded as
    a version of the variational principle. As we know, this always give an upper limit of the energy of a system.

\section{Numerical results on the $BBB^*$ system}\label{sec8}

The application of the BOP method to the three-nucleon system has verified its feasibility to some extent. Now we return to the system mainly discussed in this paper, i.e. the three $B$ mesons system $BBB^*$. 
There is only one free parameter $\Lambda$ in the monopole form factor which is undetermined in our calculations. 
For the deuteron case, the parameter $\Lambda$ is within the range $0.8\sim 1.5~\mathrm{GeV}$.
One would expect that the size of heavier bottom system is smaller than the size of deuteron,
leading to a larger $\Lambda$. Thus, we vary the parameter $\Lambda$ from $0.9~\mathrm{GeV}$ 
to $1.6~\mathrm{GeV}$ to study whether the $BBB^{\ast}$ system is bound or not. 

In order to show the properties of the two-body interactions for the $BB^{\ast}$, we first present
the numerical results for the break-up state in Tables~\ref{Eb-lamd01}-\ref{Eb-lamd00}. We plot
the effective potentials for the $BB^{\ast}$ in Fig.~\ref{BB-potential} and Fig.~\ref{V1V2V3-potential},
where the regularization parameter is fixed at 1440~MeV. In these figures,  (a) and (b) correspond
to the isospin $I=0$ and $I=1$ cases, respectively. After carefully solving the coupled-channel
Schr\"{o}dinger equation with the treatment of the S-D wave mixing, we find  loosely bound
states for both cases.  

\begin{figure}[ht]
  \begin{center}
  \rotatebox{0}{\includegraphics*[width=0.45\textwidth]{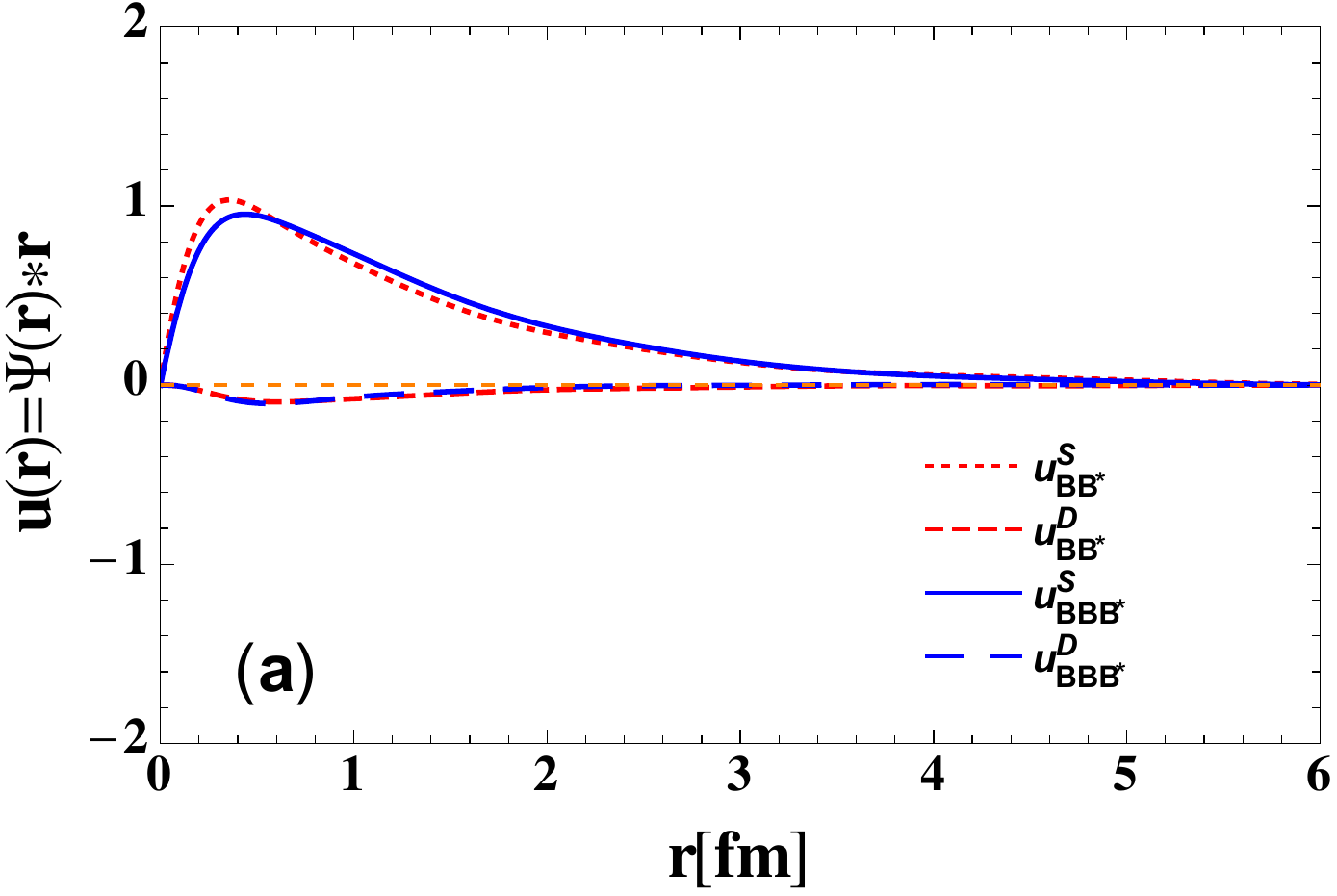}}
   \rotatebox{0}{\includegraphics*[width=0.45\textwidth]{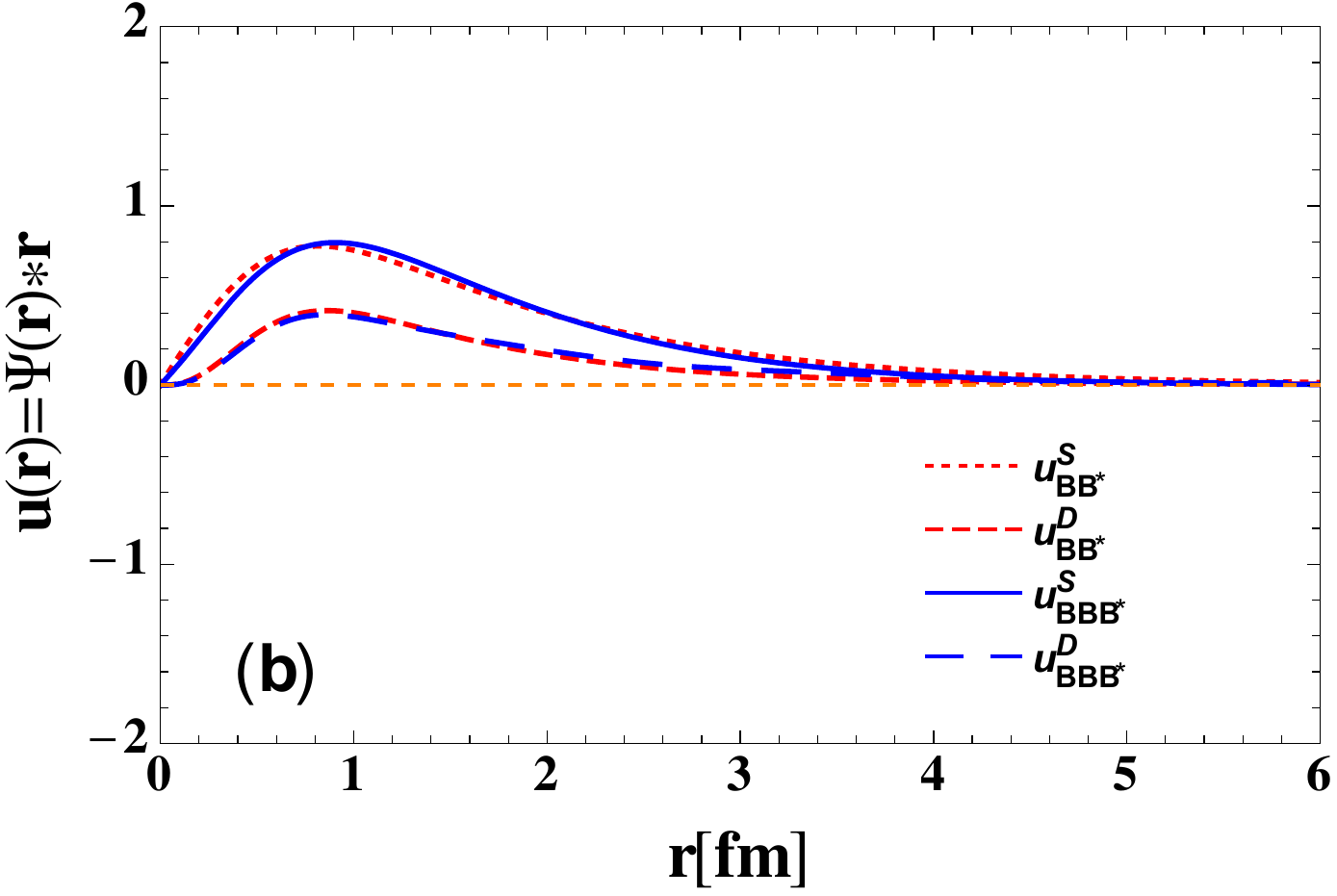}}
   \caption{Plot of various wave functions. The blue lines represent the wave functions for
     any two constituents in the $BBB^{\ast}$. The red lines denote the wave functions for its
     subsystem $BB^{\ast}$.  (a) corresponds to the isospin states $|1,\frac{3}{2},\pm\frac{1}{2}
     (\pm\frac{3}{2})\rangle$ and $|1,\frac{1}{2},\pm\frac{1}{2}\rangle$ cases. (b) corresponds to
     the isospin state $|0,\frac{1}{2},\pm\frac{1}{2}\rangle$ case. Here we chose the parameter
     $\Lambda=1440$~MeV in t (a) and $\Lambda=1107.7$~MeV in  (b) for a better comparison of
     all the cases, since they have the same two-body binding energy of 5.08~MeV. }
    \label{WF}
  \end{center}
\end{figure}

For the isospin triplet case, i.e. $I_2=1$, the dependence of
the binding energy of the two-body $BB^*$ system on the regularization parameter $\Lambda$
is shown in Table~\ref{Eb-lamd01}. The energy threshold of the break-up state
for the $BBB^{\ast}$ is just the two-body energy eigenvalue of the $BB^*$ plus the mass of the three
static free meson. We use $E_2$ denote the energy eigenvalue of the $BB^*$. 
When the parameter $\Lambda$ varies from $1380~\mathrm{MeV}$ to $1560~\mathrm{MeV}$,
there is a bound state solution with the binding energy $2.11\sim 15.59~\mathrm{MeV}$
and the root-mean-square radius $1.51\sim 0.67~\mathrm{fm}$. The $S$-wave component takes over
$99.13\%\sim 99.36\%$ comparing to the value $0.87\%\sim 0.64\%$ for $D$-wave.
The $BB^*$ and $B^*B^*$ channels have probabilities $91.91\%\sim 76.50\%$
and $8.09\%\sim 23.50\%$, respectively. The proportion of the $B^*B^*$ channel and D-wave component
are relatively small. However, as the value of the regularization parameter $\Lambda$ increases,
the $B$ and $B^*$ interacting with pion is more like a point particle, the proportion of
the $B^*B^*$ channel increases greatly, while the D-wave component decreases. 
We plot the radial wave functions of the $S$-wave and $D$-wave 
Fig.~\ref{WF}(a) for the system $BB^{\ast}$, obviously, the bound state we have
found is ground state.

\begin{table}[htbp]
\caption{Bound state solutions of the $BB^{\ast}$ system with the isospin $I_2=1$. $\Lambda$ is the
  parameter in the form factor. $E_2$ is the energy eigenvalue. The binding energy is $-E_2$.
  $r_{rms}$ is the root-mean-square radius. $\alpha$ and $\beta$ are the probabilities for the
  components $BB^{\ast}$ and $B^{\ast}B^{\ast}$, respectively. }
     \label{Eb-lamd01}
\begin{center}
\begin{tabular}{ c | c | c  c  | c  c  | c | c  c }
\hline\hline 
\multicolumn{1}{c}{~} & \multicolumn{1}{c}{~} & \multicolumn{2}{c}{$BB^{\ast}$} &  \multicolumn{2}{c}{$B^{\ast}B^{\ast}$}  &\multicolumn{1}{c}{~} & \multicolumn{2}{c}{ Proportion } \\
 \hline
  $\Lambda$(MeV) & $E_2$(MeV) & S wave(\%) & D wave(\%)  & S wave(\%) & D wave(\%) &  $r_{rms1}$(fm) & $\alpha$(\%) & $\beta$(\%) \\
\hline
   1380 & -2.11 & 99.13 & 0.87 &  99.21 & 0.79 & 1.51 & 91.91 & 8.09    \\
\hline
   1400 & -2.94 & 99.12 & 0.88 & 99.36 & 0.64 & 1.30 & 90.22 & 9.78 \\
\hline
   1420 & -3.93 & 99.14 & 0.86 & 99.49 & 0.51 & 1.15 & 88.47 & 11.53 \\
\hline
   1440 & -5.08 & 99.16 & 0.84 & 99.59 & 0.41 & 1.03 & 86.69 & 13.31 \\
\hline
   1460 & -6.40 & 99.19 & 0.81 & 99.68 & 0.33 & 0.94 & 84.91 & 15.09  \\
\hline
   1480 & -7.88 & 99.22 & 0.78 & 99.74 & 0.26 & 0.86 & 83.14 & 16.86 \\
\hline
   1500 & -9.54 & 99.25 & 0.75 & 99.80 & 0.20 & 0.80 & 81.40 & 18.60 \\
\hline
   1520 & -11.38 & 99.29 & 0.71 & 99.84 & 0.16 & 0.75 & 79.70 & 20.30 \\
\hline
   1540 & -13.39 & 99.32 & 0.68 & 99.88 & 0.12 & 0.71 & 78.07 & 21.93 \\
\hline
   1560 & -15.59 & 99.36 & 0.64 & 99.91 & 0.09 & 0.67 & 76.50 & 23.50 \\
\hline\hline
\end{tabular}
\end{center}
\end{table}

For the isospin singlet case, i.e. $I_2=0$, the dependence of
the binding energy of the two-body $BB^*$ system on the regularization parameter $\Lambda$
is shown in Table~\ref{Eb-lamd00}. We also use $E_2$ to denote the energy eigenvalue of the $BB^*$. 
When the parameter varies from $1040~\mathrm{MeV}$ to $1220~\mathrm{MeV}$,
there is a bound state solution with  binding energy $1.88\sim 14.78~\mathrm{MeV}$
and the root-mean-square radius $2.03\sim 0.95~\mathrm{fm}$. The $S$-wave component is $86.09\%\sim 77.21\%$
compared to the value $13.91\%\sim 22.79\%$ for $D$-wave.
The $BB^*$ and $B^*B^*$ channels have probabilities $92.05\%\sim 75.09\%$
and $7.95\%\sim 24.91\%$, respectively. The proportion of the $B^*B^*$ channel and D-wave component are
relatively small which is similar with the case of isospin triplet. As the value of the
regularization parameter $\Lambda$ increases, the proportion of the $B^*B^*$ channel increases greatly.
Different from the case of isospin triplet the S-wave component decrease and D-wave increase as
$\Lambda$ increases. As the parameter $\Lambda$ increases, all of the effective potentials become
stronger. The S-wave potential increases faster than the D-wave potential for the isospin triplet case,
while it is reverse for the isospin singlet case.
In order to check whether the bound state we have found is the ground state, we also plot the radial
wave functions of the $S$-wave and $D$-wave Fig.~\ref{WF}(b) for the system $BB^{\ast}$.

\begin{table}[htbp]
\caption{Bound state solutios of the $BB^{\ast}$ with the isospin $I_2=0$. $\Lambda$ is the
  parameter in the form factor. $E_2$ is the energy eigenvalue. The binding energy is $-E_2$.
  $r_{rms}$ is the root-mean-square radius. $\alpha$ and $\beta$ are the probabilities for the
  component $BB^{\ast}$ and $B^{\ast}B^{\ast}$, respectively. }
     \label{Eb-lamd00}
\begin{center}
\begin{tabular}{ c | c | c  c  | c  c  | c | c  c }
\hline\hline 
\multicolumn{1}{c}{~} & \multicolumn{1}{c}{~} & \multicolumn{2}{c}{$BB^{\ast}$} &  \multicolumn{2}{c}{$B^{\ast}B^{\ast}$}  &\multicolumn{1}{c}{~} & \multicolumn{2}{c}{ Proportion } \\
 \hline
  $\Lambda$(MeV) & $E_2$(MeV) & S wave(\%) & D wave(\%)  & S wave(\%) & D wave(\%) &  $r_{rms1}$(fm) & $\alpha$(\%) & $\beta$(\%) \\
\hline
   1040 & -1.88 & 86.09 & 13.91 & 68.84 & 31.16 & 2.03 & 92.05 & 7.95    \\
\hline
   1060 & -2.63 & 84.59 & 15.41 & 69.41 & 30.59 & 1.79 &  90.18 & 9.82 \\
\hline
   1080 & -3.54 & 83.26 & 16.74 & 69.90 & 30.10 & 1.60 & 88.22 & 11.78 \\
\hline
   1100 & -4.62 & 82.07 & 17.93 & 70.33 & 29.67 & 1.45 & 86.22 & 13.78 \\
\hline
   1120 & -5.87 & 81.01 & 18.99 & 70.69 & 29.31 & 1.33 & 84.22 & 15.78  \\
\hline
   1140 & -7.29 & 80.07 & 19.93 & 70.99 & 29.01 & 1.23 & 82.25 & 17.75 \\
\hline
   1160 & -8.89 & 79.23 & 20.77 & 71.25 & 28.75 & 1.14 & 80.34 & 19.66 \\
\hline
   1180 & -10.67 & 78.49 & 21.51 & 71.46 & 28.54 & 1.07 & 78.50 & 21.50 \\
\hline
   1200 & -12.63 & 77.81 & 22.19 & 71.63 & 28.37 & 1.01 & 76.75 & 23.25 \\
\hline
   1220 & -14.78 & 77.21 & 22.79 & 71.77 & 28.23 & 0.95 & 75.09 & 24.91 \\
\hline\hline
\end{tabular}
\end{center}
\end{table}

In Sec.~\ref{sec2}, we have listed the isospin wave functions of the $BBB^*$ which are expressed as
$|I_2,I_3,I_{3z}\rangle$. After solving the three-body Schr\"odinger equation via the method
of Sec.~\ref{sec5}, we find that all of these isospin eigenstates have bound state solutions.
As long as the two-body system $BB^*$ has a loosely bound state, the three-body system $BBB^*$ is most
likely to have a loosely bound state, too. We have collected the dependence of the three-body bound state
solutions on the  two-body binding energy in Tables~\ref{E3-Eb01}-\ref{E3-Eb00}.   

The bound state solutions for the state $|1,\frac{3}{2},\pm\frac{1}{2}(\pm\frac{3}{2})\rangle$ are
shown in Table~\ref{E3-Eb01}. The three-body binding energy relative to their break-up states is 5.67~MeV,
when the parameter $\Lambda$ is chosen at 1440~MeV and the two-body binding energy of their
subsystems $BB^{\ast}$ is 5.08~MeV. To search for the dependence on the binding energy of the two-body
system $E_2$, we change the parameter $\Lambda$. It turns out that if the value of the $E_2$ varies
from $-0.18~\mathrm{MeV}$ to $-17.97~\mathrm{MeV}$, then the reduced three-body energy eigenvalue
$E_3$ decreases from $-0.19~\mathrm{MeV}$ to $-18.99~\mathrm{MeV}$ and the total three-body energy
eigenvalue $E_T$ decreases from $-0.38~\mathrm{MeV}$ to $-36.95~\mathrm{MeV}$. The structure of the
three-body bound state is a regular triangle with the root-mean-square length of one side decreasing
from 3.98~fm to 0.65~fm. In order to illustrate the strength of the BO potential, we also collect
its minimum $V_{BO}(0)$ in the table within the range of -3.43$\sim$-37.24 MeV as the $E_2$ increases.
As  $E_2$ increases, the effective attraction between $B$ and $B^*$ becomes stronger, the BO potential
becomed deeper, so then the three-body system becomes tighter and has a larger binding energy. From
the results in the table, we can also see that the dominant wave between any two $B^{(*)}$ in
the $BBB^{\ast}$ is $S$ wave and the dominant channel is the $BBB^{\ast}$ instead of the $BB^{\ast}B^{\ast}$
channel. For  comparison, we plot the wave functions for any two $B^{(*)}$ in the $BBB^{\ast}$ system
and that for the two-body $BB^*$ system in Fig.~\ref{WF}(a) with $\Lambda=1440$~MeV. The shapes of
these exhibit  little difference. From another perspective, one more $B$ meson has little effect
on the size of the system but greatly increases the binding energy.

\begin{table}[htbp]
\caption{Bound state solutions of the $BBB^{\ast}$ with the isospin $I_3=3/2$. 
$E_2$ is the energy eigenvalue of its subsystem $BB^{\ast}$ with the isospin $I_2=1$. 
 $E_3$ is the reduced three-body energy eigenvalue relative to the break-up state of the $BBB^{\ast}$ system.  
$E_T$ is the total three-body energy eigenvalue relative to the $BBB^{\ast}$ threshold. 
 $V_{BO}(0)$ is the minimum of the BO potential. 
$r_{rms}$ represents the root-mean-square radius of any two $B$ in the $BBB^{\ast}$ system. 
The $S$-wave and $D$-wave represent the probabilities for $S$-wave and
 $D$-wave components in any two $B$ in the $BBB^{\ast}$. 
The $\alpha$ and $\beta$ denote the probabilities for the $BBB^{\ast}$ and $BB^{\ast}B^{\ast}$ components,
respectively. 
 }\label{E3-Eb01}
\begin{center}
\begin{tabular}{ c | c | c | c  | c  | c | c  | c  | c  }
\hline\hline 
  $E_2$(MeV)  & $E_3$(MeV) & $E_T$(MeV)  & $V_{BO}(0)$(MeV)  & S wave(\%) & D wave(\%) & $r_{rms}$(fm)  & $\alpha$(\%) & $\beta$(\%)  \\
\hline
        -0.18 & -0.19 & -0.38 & -3.43 & 99.76 & 0.24 & 3.98 & 97.50 & 2.50 \\
\hline       
        -0.48 & -0.45 & -0.93 & -4.88 & 99.68 & 0.32 & 3.34 & 96.39 & 3.61 \\
\hline       
       -0.89 & -0.85 & -1.74 & -6.62 & 99.59 & 0.41 & 2.67 & 95.02 & 4.98  \\
\hline       
       -1.43 & -1.42 & -2.85 & -8.56 & 99.49 & 0.51 & 2.11 & 93.78 & 6.22 \\
\hline       
       -2.11 & -2.20 & -4.31 & -10.65 & 99.41 & 0.59 & 1.71 & 91.91 & 8.09 \\
\hline
       -2.94 & -3.17 & -6.11 & -12.87 & 99.34 & 0.66 & 1.43 & 90.22 & 9.78  \\
\hline
       -3.93 & -4.33 & -8.26 & -15.21 & 99.29 & 0.71 & 1.24 & 88.47 & 11.53 \\
\hline
       -5.08 & -5.67 & -10.75 & -17.65 & 99.25 & 0.75 & 1.09 & 86.69 & 13.31 \\
\hline
       -6.40 & -7.18 & -13.58 & -20.19 & 99.22 & 0.78 & 0.98 & 84.91 & 15.09 \\
\hline
       -7.88 & -8.83 & -16.71 & -22.83 & 99.20 & 0.80 & 0.90 & 83.14 & 16.86 \\
\hline
      -9.54 & -10.61 & -20.16 & -25.55 & 99.18 & 0.82 & 0.83 & 81.40 & 18.60 \\
\hline
      -11.38 & -12.51 & -23.89 & -28.36 & 99.17 & 0.83 & 0.77 & 79.70 & 20.30 \\
\hline
      -13.39 & -14.62 & -28.01 & -31.24 & 99.16 & 0.84 & 0.72 & 78.07 & 21.93 \\
\hline
      -15.59 & -16.75 & -32.34 & -34.20 & 99.15 & 0.85 & 0.68 & 76.50 & 23.50 \\
\hline
      -17.97 & -18.99 & -36.95 & -37.24 & 99.14 & 0.86 & 0.65 & 75.01 & 24.99 \\
\hline\hline
\end{tabular}
\end{center}
\end{table}

For the state $|0,\frac{1}{2},\pm\frac{1}{2}\rangle$, we also find a loosely bound solution, which
is  shown in Table~\ref{E3-Eb00}. The three-body binding energy relative to their break-up states is
7.18~MeV, when the parameter $\Lambda$ is chosen at 1107.7~MeV and the two-body binding energy of their
subsystems $BB^{\ast}$ is 5.08~MeV. In order to show the dependence on the binding energy of the
two-body system $E_2$, we also change the parameter $\Lambda$. We find that if the value of the $E_2$
varies from $-0.19~\mathrm{MeV}$ to $-17.10~\mathrm{MeV}$, then the reduced three-body energy
eigenvalue  $E_3$ decreases from $-0.32~\mathrm{MeV}$ to $-20.96~\mathrm{MeV}$ and the total
three-body energy eigenvalue $E_T$ decreases from $-0.51~\mathrm{MeV}$ to $-38.06~\mathrm{MeV}$.
The structure of the three-body bound state is a regular triangle with the root-mean-square length
of one side decreasing from 3.89~fm to 0.93~fm. As an illustration for the strength of the BO potential,
we also list its minimum $V_{BO}(0)$ in the table within the range of $-$2.15$\sim -30.15$~MeV as
$E_2$ increases. Similar to the $I_3=\frac{3}{2}$ case, the dominant wave between any two $B^{(*)}$
in the $BBB^{\ast}$ is $S$ wave and the dominant channel is the $BBB^{\ast}$ one. In order to show
that one more $B$ meson has little effect on the size of the system, we also plot the wave
functions for any two $B^{(*)}$ in the $BBB^{\ast}$ system and that for the two-body $BB^*$ system
in Fig.~\ref{WF}(b) with $\Lambda=1107.7$ MeV. Here we chose the parameter $\Lambda=1107.70$ MeV
for a better comparison with the $|1,\frac{3}{2},\pm\frac{1}{2}(\pm\frac{3}{2})\rangle$ case,
since both cases have the same two-body binding energy 5.08~MeV. 

\begin{table}[htbp]
\caption{Bound state solutions of the $BBB^{\ast}$ with  isospin $I=1/2$. 
$E_2$ is the energy eigenvalue of its subsystem $BB^{\ast}$ with the isospin $I=0$. 
 $E_3$ is the reduced three-body energy eigenvalue relative to the break-up state of the $BBB^{\ast}$ system.  
$E_T$ is the total three-body energy eigenvalue relative to the $BBB^{\ast}$ threshold. 
 $V_{BO}(0)$ is the minimum of the BO potential. 
$r_{rms}$ represents the root-mean-square radius of any two $B$ in the $BBB^{\ast}$ system. 
The $S$-wave and $D$-wave represent the probabilities for $S$-wave and
 $D$-wave components in any two $B$ in the $BBB^{\ast}$. 
The $\alpha$ and $\beta$ denote the probabilities for the $BBB^{\ast}$ and $BB^{\ast}B^{\ast}$ components,
respectively. 
 }\label{E3-Eb00}
\begin{center}
\begin{tabular}{ c | c | c | c  | c  | c | c  | c  | c  }
\hline\hline 
  $E_2$(MeV)  & $E_3$(MeV) & $E_T$(MeV)  & $V_{BO}(0)$(MeV)  & S wave(\%) & D wave(\%) & $r_{rms}$(fm)  & $\alpha$(\%) & $\beta$(\%)  \\
\hline
       -0.19 & -0.32 & -0.51 & -2.15 & 94.66 & 5.34 & 3.89 & 97.68 & 2.32 \\
\hline       
       -0.44 & -0.64 & -1.08 & -3.04 & 92.56 & 7.44 & 3.27 & 96.66 & 3.34 \\
\hline       
       -0.80 & -1.13 & -1.93 & -4.19 & 90.43 & 9.57 & 2.69 & 95.36 & 4.64 \\
\hline
       -1.27 & -1.82 & -3.09 & -5.57 & 88.49 & 11.51 & 2.24 & 93.80 & 6.20 \\
\hline       
       -1.88 & -2.72 & -4.60 & -7.14 & 86.72 & 13.28 & 1.93 & 92.05 & 7.95 \\
\hline
       -2.63 & -3.82 & -6.45 & -8.89 & 85.09 & 14.91 & 1.70 & 90.18 & 9.82  \\
\hline
       -3.54 & -5.11 & -8.65 & -10.78 & 83.60 & 16.40 & 1.53 & 88.22 & 11.78 \\
\hline
       -4.62 & -6.57 & -11.20 & -12.82 & 82.27 & 17.73 & 1.40 & 86.22 & 13.78 \\
\hline
       -5.04 & -7.13 & -12.17 & -13.56 & 81.84 & 18.16 & 1.36 & 85.52 & 14.48 \\
\hline
       -7.29 & -10.00 & -17.29 & -17.25 & 80.06 & 19.94 & 1.21 & 82.25 & 17.75 \\
\hline
      -8.89 & -11.93 & -20.83 & -19.63 & 79.15 & 20.85 & 1.13 & 80.34 & 19.66  \\
\hline
      -10.67 & -14.00 & -24.68 & -22.12 & 78.36 & 21.64 & 1.07 & 78.50 & 21.50 \\
\hline
      -12.63 & -16.20 & -28.84 & -24.70 & 77.66 & 22.34 & 1.02 & 76.75 & 23.25 \\
\hline
      -14.78 & -18.52 & -33.30 & -27.38 & 77.03 & 22.97 & 0.97 & 75.09 & 24.91 \\
\hline
      -17.10 & -20.96 & -38.06 & -30.15 & 76.46 & 23.54 & 0.93 & 73.52 & 26.48 \\
\hline\hline
\end{tabular}
\end{center}
\end{table}

The state $|1,\frac{1}{2},\pm\frac{1}{2}\rangle$ also has a loosely bound solution. However, in
our calculation the states $|1,\frac{1}{2},\pm\frac{1}{2}\rangle$ and $|1,\frac{3}{2},
\pm\frac{1}{2}(\pm\frac{3}{2})\rangle$ are degenerate. This is due to the fact that in the OPE model
only two-body interactions are considered, and the two-body interaction is only depends on the
total isospin of the two interacting mesons. The states $|1,\frac{1}{2},\pm\frac{1}{2}\rangle$ and
$|1,\frac{3}{2},\pm\frac{1}{2}(\pm\frac{3}{2})\rangle$ have the same two-body interaction but
may have different three-body interaction. If we further consider the calculation to the
next-to-next leading order, this degeneracy may disappear. The calculation that contains three-body
interactions via pion exchange is quite complicated, which is left for the further work.   

\begin{figure}[ht]
  \begin{center}
     \rotatebox{0}{\includegraphics*[width=0.45\textwidth]{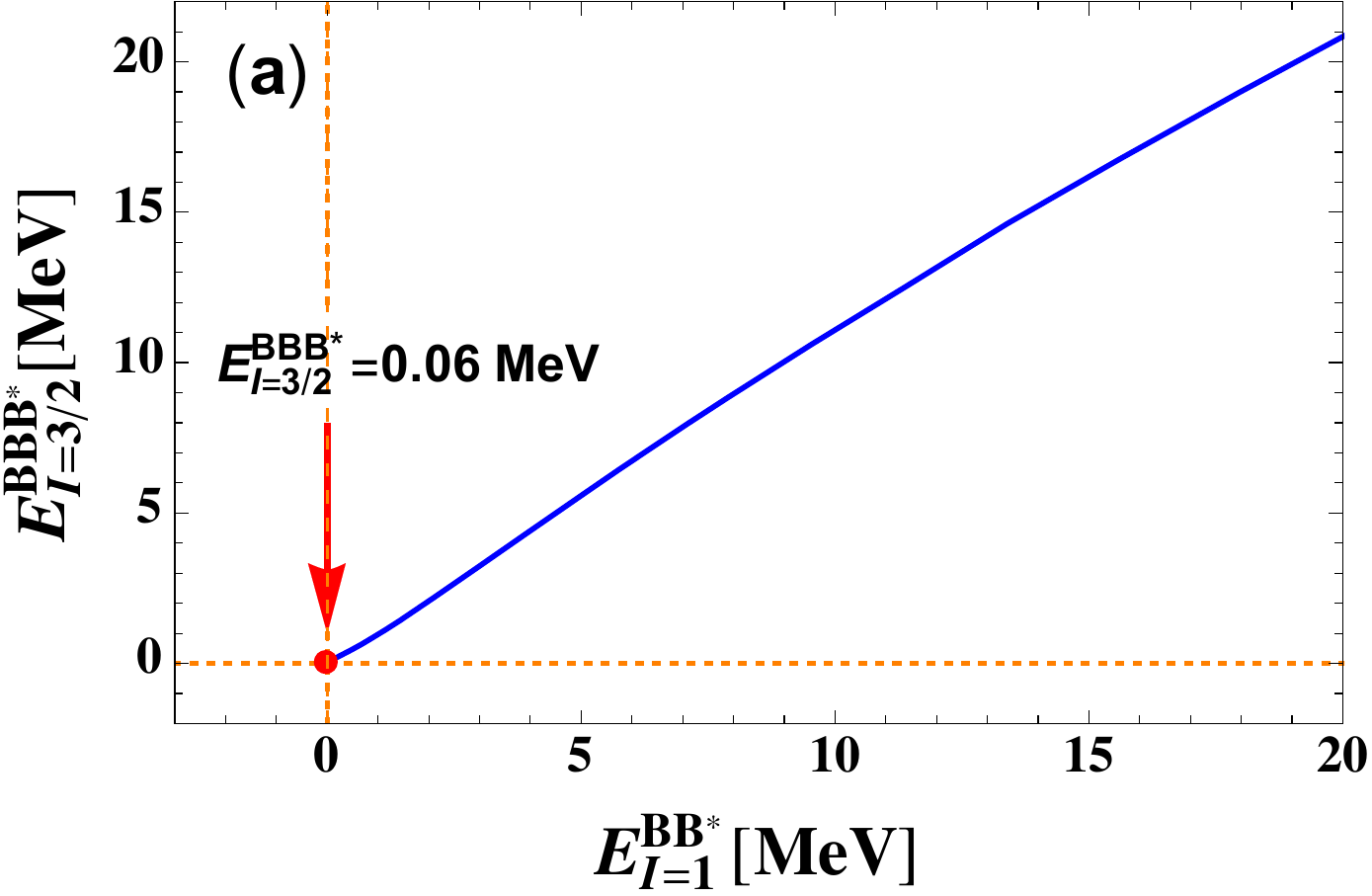}}
     \rotatebox{0}{\includegraphics*[width=0.45\textwidth]{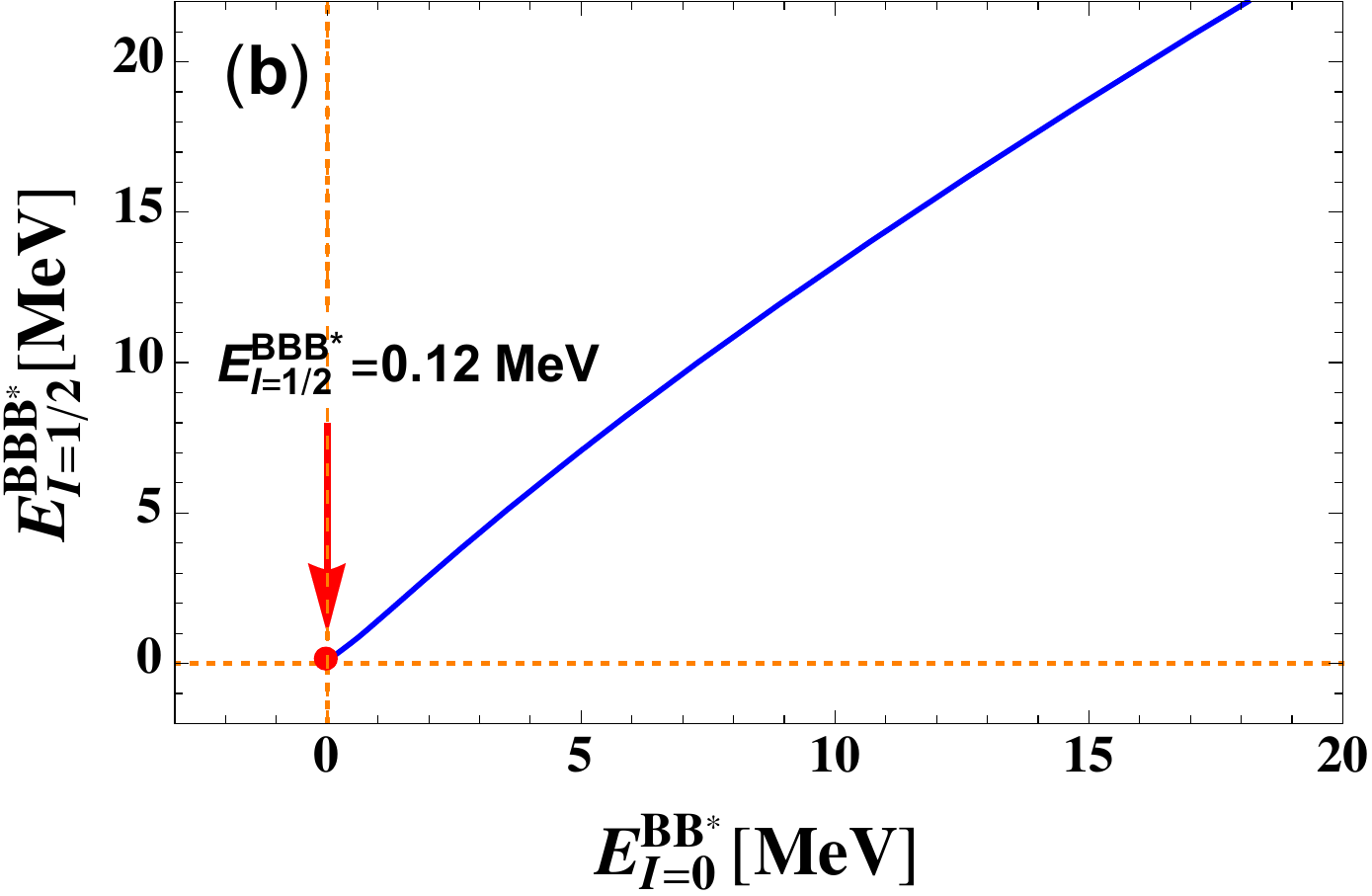}}
     \caption{Dependence of the reduced three-body binding energy on the two-body binding energy of
       its subsystem $BB^{\ast}$. The red point is the critical point which indicates the lower
       limit of the required binding energy of the isotriplet $BB^{\ast}$ to form a three-body bound state.
       (a) corresponds to the isospin states $|1,\frac{3}{2},\pm\frac{1}{2}(\pm\frac{3}{2})\rangle$
       and $|1,\frac{1}{2},\pm\frac{1}{2}\rangle$ cases, while (b) corresponds to the isospin state
       $|0,\frac{1}{2},\pm\frac{1}{2}\rangle$ case.}
    \label{E3-Eb}
  \end{center}
\end{figure}

The numerical results show  that the three-body binding energy $|E_3|$ increases as the two-body
binding energy $|E_2|$ increases. One may wonder whether there is a critical value of the $|E_2|$, below which
the three-body system has no bound state solution.
After lots of calculations, it turns out that all of the isospin eigenstates have no such critical value.
That is to say, no matter what a small value for the two-body binding energy $|E_2|$,
as long as the two-body system $BB^*$ has a loosely bound state, the three-body system $BBB^*$
probably has a loosely bound state. To show this conclusion explicitly, we plot the dependence of
the three-body binding energy on the variety of two-body binding energy in Fig.~\ref{E3-Eb}.
The isospin eigenstate $|1,\frac{3}{2},\pm\frac{1}{2}(\pm\frac{3}{2})\rangle$ and $|1,\frac{1}{2},
\pm\frac{1}{2}\rangle$ cases are shown in Fig.~\ref{E3-Eb}(a), where  $E_{I=3/2}^{BBB^*}$ and
$E_{I=1}^{BB^*}$ denote the reduced three-body binding energy and two-body binding energy, respectively.
When the two-body binding energy $E_{I=1}^{BB^*}$ approaches 0~MeV, the reduced three-body binding
energy $E_{I=3/2}^{BBB^*}$ approaches a small value of about 0.06~MeV. Similarly, we also plot the
dependence curve for the $|0,\frac{1}{2},\pm\frac{1}{2}\rangle$ case in Fig.~\ref{E3-Eb}(b),
where  $E_{I=1/2}^{BBB^*}$ and $E_{I=0}^{BB^*}$ denote the reduced three-body binding energy and
two-body binding energy in this case, respectively. As shown in the figure, the reduced
three-body binding energy $E_{I=1/2}^{BBB^*}$ also has a small value of about 0.12~MeV,
when the two-body binding energy $E_{I=0}^{BB^*}$ approaches zero.

\section{Summary and Discussion}\label{sec9}

In the present paper, we have performed an extensive study on the possibility of the $BBB^{\ast}$ system to
form  tri-meson molecules. Based on the Born-Oppenheimer potential method as well as the OPE scheme,
we derived the three-body Schr\"odinger equation for the system $BBB^{\ast}$. Since the
regularization parameter $\Lambda$ is difficult to be pinned down, we choose the parameter in the range of
0.9$\sim$1.6~GeV and show various bound state solutions of the $BBB^{\ast}$ system. After
careful treatments of the S-D wave mixing and the coupled-channel $BB^{\ast}B^{\ast}$ effects, we
found that all of the isospin eigenstates expressed by the $|I_2,I_3,I_{3z}\rangle$ have bound state solutions.
For instance, in the states $|1,\frac{3}{2},\pm\frac{1}{2}(\pm\frac{3}{2})\rangle$ and
$|1,\frac{1}{2},\pm\frac{1}{2}\rangle$, the three-body binding energy relative to their break-up
states is 5.67~MeV, when the parameter $\Lambda$ is chosen at 1440~MeV and the two-body binding
energy of their subsystems $BB^{\ast}$ is 5.08~MeV. In the state $|0,\frac{1}{2},\pm\frac{1}{2}\rangle$,
the three-body binding energy relative to their break-up states is 7.18~MeV, when the parameter
$\Lambda$ is chosen at 1107.7~MeV and the two-body binding energy of their subsystems $BB^{\ast}$ is
also 5.08~MeV.  After careful calculations, we find no critical value for the two-body binding energy,
which indicates the lower limit of the required binding energy of their subsystem $BB^{\ast}$ to form
a three-body bound state. That is to say, no matter how small the two-body binding energy is,
as long as the two-body subsystem $BB^*$ has a loosely bound state, the three-body system
$BBB^*$ is most likely to have a loosely bound state, too.  

The BO potential method we have used in this paper is an adiabatic approximation that divide
the degrees of freedom of the motion for the three-body system into a light one and a heavy one.
Then we simplify the three-body system into a two-body system only with heavy degrees of freedom
but with an additional BO potential generated by the relative light meson. Since the system
$B_{a}^{(\ast)}B_{b}^{(\ast)}B_{c}^{(\ast)}$ has little mass difference on its constituents,
the motion of every constituent can be regarded as a light degree of freedom. Therefore, the
eigenstates of the three-body system should be the combinations of all of the possible cases.
As the most simplest combination, one might expect the three-body eigenstate should be a
superposition all of the possible cases. In other words, the three-body bound state solutions we
have listed in the last section are approximate solutions. It may be that the strict solutions will
be a more complicated combinations. To answer this question requires further study.   

Our calculations are based on the OPE scheme, which is leading order in the chiral power counting
(neglecting contact interactions). Since only one virtual pion occurs in the $BBB^{\ast}$ molecule,
the virtual pion is also shared by the three mesons. Therefore, the three constituents in the
$BBB^{\ast}$ system share one virtual pion which corresponds to a delocalized pion bond.
It is attractive and strong enough to make them form a three-body molecular state.   

To summarize briefly, with the delicate efforts of the long-range one-pion exchange, the S-D wave mixing
and coupled-channel effects, we have investigated the existence of the loosely bound
tri-meson molecules $BBB^{\ast}$ and find that it is very easy to form a tri-meson molecular state
as long as its two-body subsystem $BB^*$ has a molecular state. Hopefully, the present extensive
investigations will be useful to the understanding of the few-body hadronic systems and the
future well-developed experiments on hadron collisions will provide us with a platform to seek
out the  tri-meson molecules.  

\section*{Acknowledgements}

This work is
supported in part by the DFG (Grant No. TRR110) and the NSFC (Grant No. 11621131001) through funds 
provided to
the Sino-German CRC 110 ``Symmetries and the Emergence of Structure
in QCD''. The work of UGM was also supported by the Chinese Academy 
of Sciences (CAS) President's International Fellowship Initiative (PIFI) 
(Grant No. 2018DM0034) and by VolkswagenStiftung (Grant No. 93562).
QW is also supported by the Thousand Talents Plan for Young Professionals and 
research startup funding at SCNU.

\section*{Appendix: Some helpful functions and Fourier transformations }

The functions $Y(\tilde{m}_{\pi}r)$ and $Z(\tilde{m}_{\pi}r)$ in Eqs.~(\ref{VBB1})-(\ref{VBB3}) are defined as
\begin{equation*}
Y(\tilde{m}_{\pi}r)=\frac{\exp(\tilde{m}_{\pi}r)}{\tilde{m}_{\pi}r},
\end{equation*}
\begin{equation*}
Z(\tilde{m}_{\pi}r)=(1+\frac{3}{\tilde{m}_{\pi}r}+\frac{3}{(\tilde{m}_{\pi}r)^2})Y(\tilde{m}_{\pi}r),
\end{equation*}
where
\begin{equation*}
\tilde{m}^2_{\pi}=m^2_{\pi}-(\Delta M)^2.
\end{equation*}

Fourier transformation formulas read
\begin{eqnarray*}
&&4\pi~(\frac{\Lambda^2-m_{\pi}^2}{\tilde{\Lambda}^2+\vec{q}^2})^2
\frac{1}{\vec{q}^2+\tilde{m}_{\pi}^2}
\rightarrow \tilde{m}_{\pi}Y(\tilde{m}_{\pi}r)-\tilde{\Lambda}
Y(\tilde{\Lambda}
r)-(\Lambda^2-m_{\pi}^2)\frac{e^{-\tilde{\Lambda}
r}}{2\tilde{\Lambda}},
\end{eqnarray*}
\begin{eqnarray*}
&&4\pi~(\frac{\Lambda^2-m_{\pi}^2}{{\tilde{\Lambda}}^2+\vec{q}^2})^2
\frac{\vec{q}^2}{\vec{q}^2+\tilde{m}_{\pi}^2}
\rightarrow \tilde{m}_{\pi}^2[\tilde{\Lambda} Y(\tilde{\Lambda} r)-\tilde{m}_{\pi}Y(\tilde{m}_{\pi}r)]
+(\Lambda^2-m_{\pi}^2)\tilde{\Lambda}\frac{e^{-\tilde{\Lambda}
r}}{2},
\end{eqnarray*}
\begin{eqnarray*}
&&4\pi~(\frac{\Lambda^2-m_{\pi}^2}{{\tilde{\Lambda}}^2+\vec{q}^2})^2
\frac{(\vec{\epsilon_1}\cdot\vec{q})(\vec{\epsilon_2}\cdot\vec{q})}{\vec{q}^2+\tilde{m}_{\pi}^2} 
\rightarrow \frac{1}{3}\vec{\epsilon_1}\cdot\vec{\epsilon_2}[\tilde{m}_{\pi}^2\tilde{\Lambda} Y(\tilde{\Lambda} r)-\tilde{m}_{\pi}^3Y(\tilde{m}_{\pi}r)\nonumber\\
&+&(\Lambda^2-m_{\pi}^2)\tilde{\Lambda}\frac{e^{-\tilde{\Lambda} r}}{2}]
+\frac{1}{3}S_{T}[-\tilde{m_{\pi}}^3 Z(\tilde{m_{\pi}}r)+ \tilde{\Lambda}^3 Z(\tilde{\Lambda} r) 
+(\Lambda^2-m_{\pi}^2)(1+\tilde{\Lambda} r)\frac{\tilde{\Lambda}}{2}Y(\tilde{\Lambda} r)~],  
\end{eqnarray*}
where $
\hat{S}_{T}=3(\vec{r}\cdot \hat{\vec{\epsilon_b}})(\vec{r}\cdot
\hat{\vec{\epsilon_a}}^{\dag})-\hat{\vec{\epsilon_b}}\cdot
\hat{\vec{\epsilon_a}}^{\dag}
$. 

The polarization vector is the S-D wave space have the following substitution
\begin{eqnarray*}
\vec{\epsilon_b}\cdot \vec{\epsilon_a}^{\dag}\rightarrow\left(
                                                \begin{array}{cc}
                                                  1 & 0 \\
                                                  0 & 1 \\
                                                \end{array}
                                              \right),
\end{eqnarray*}
\begin{eqnarray*}
S_{T}\rightarrow\left(
         \begin{array}{cc}
           0 & -\sqrt{2} \\
           -\sqrt{2} & 1 \\
         \end{array}
       \right),
\end{eqnarray*}
\begin{eqnarray*}
i\vec{\epsilon}_3^{\dag}\times\vec{\epsilon}_1\cdot i\vec{\epsilon}_4^{\dag}\times\vec{\epsilon}_2\rightarrow\left(
                                                \begin{array}{cc}
                                                  -1 & 0 \\
                                                  0 & -1 \\
                                                \end{array}
                                              \right),
\end{eqnarray*}
\begin{eqnarray*}
S_{T}(i\vec{\epsilon}_3^{\dag}\times\vec{\epsilon}_1,  i\vec{\epsilon}_4^{\dag}\times\vec{\epsilon}_2)\rightarrow\left(
         \begin{array}{cc}
           0 & \sqrt{2} \\
           \sqrt{2} & -1 \\
         \end{array}
       \right),
\end{eqnarray*}
\begin{eqnarray*}
\vec{\epsilon}_3\cdot i\vec{\epsilon}_4^{\dag}\times\vec{\epsilon}_2\rightarrow\left(
                                                \begin{array}{cc}
                                                  \sqrt{2} & 0 \\
                                                  0 & \sqrt{2} \\
                                                \end{array}
                                              \right),
\end{eqnarray*}
\begin{eqnarray*}
S_{T}(\vec{\epsilon}_3, i\vec{\epsilon}_4^{\dag}\times\vec{\epsilon}_2)\rightarrow\left(
         \begin{array}{cc}
           0 & 1 \\
           1 & -\frac{1}{\sqrt{2}} \\
         \end{array}
       \right). 
\end{eqnarray*}


\begin{thebibliography}{}

\bibitem{Klempt:2007cp} 
  E.~Klempt and A.~Zaitsev,
  Phys.\ Rept.\  {\bf 454}, 1 (2007)
  doi:10.1016/j.physrep.2007.07.006
  [arXiv:0708.4016 [hep-ph]].


\bibitem{Klempt:2009pi} 
  E.~Klempt and J.~M.~Richard,
  Rev.\ Mod.\ Phys.\  {\bf 82}, 1095 (2010)
  doi:10.1103/RevModPhys.82.1095
  [arXiv:0901.2055 [hep-ph]].


\bibitem{Brambilla:2010cs} 
  N.~Brambilla {\it et al.},
  Eur.\ Phys.\ J.\ C {\bf 71}, 1534 (2011)
  doi:10.1140/epjc/s10052-010-1534-9
  [arXiv:1010.5827 [hep-ph]].


\bibitem{Olsen:2014qna} 
  S.~L.~Olsen,
  Front.\ Phys.\ (Beijing) {\bf 10}, no. 2, 121 (2015)
  doi:10.1007/S11467-014-0449-6
  [arXiv:1411.7738 [hep-ex]].


\bibitem{Oset:2016lyh} 
  E.~Oset {\it et al.},
  Int.\ J.\ Mod.\ Phys.\ E {\bf 25}, 1630001 (2016)
  doi:10.1142/S0218301316300010
  [arXiv:1601.03972 [hep-ph]].


\bibitem{Chen:2016qju} 
  H.~X.~Chen, W.~Chen, X.~Liu and S.~L.~Zhu,
  Phys.\ Rept.\  {\bf 639}, 1 (2016)
  doi:10.1016/j.physrep.2016.05.004
  [arXiv:1601.02092 [hep-ph]].


\bibitem{Chen:2016spr} 
  H.~X.~Chen, W.~Chen, X.~Liu, Y.~R.~Liu and S.~L.~Zhu,
  Rept.\ Prog.\ Phys.\  {\bf 80}, no. 7, 076201 (2017)
  doi:10.1088/1361-6633/aa6420
  [arXiv:1609.08928 [hep-ph]].


\bibitem{Esposito:2016noz} 
  A.~Esposito, A.~Pilloni and A.~D.~Polosa,
  Phys.\ Rept.\  {\bf 668}, 1 (2017)
  doi:10.1016/j.physrep.2016.11.002
  [arXiv:1611.07920 [hep-ph]].


\bibitem{Lebed:2016hpi} 
  R.~F.~Lebed, R.~E.~Mitchell and E.~S.~Swanson,
  Prog.\ Part.\ Nucl.\ Phys.\  {\bf 93}, 143 (2017)
  doi:10.1016/j.ppnp.2016.11.003
  [arXiv:1610.04528 [hep-ph]].


\bibitem{Hosaka:2016pey} 
  A.~Hosaka, T.~Iijima, K.~Miyabayashi, Y.~Sakai and S.~Yasui,
  PTEP {\bf 2016}, no. 6, 062C01 (2016)
  doi:10.1093/ptep/ptw045
  [arXiv:1603.09229 [hep-ph]].


\bibitem{Dong:2017gaw} 
  Y.~Dong, A.~Faessler and V.~E.~Lyubovitskij,
  Prog.\ Part.\ Nucl.\ Phys.\  {\bf 94}, 282 (2017).
  doi:10.1016/j.ppnp.2017.01.002


\bibitem{Guo:2017jvc} 
  F.~K.~Guo, C.~Hanhart, U.-G.~Mei{\ss}ner, Q.~Wang, Q.~Zhao and B.~S.~Zou,
  Rev.\ Mod.\ Phys.\  {\bf 90}, no. 1, 015004 (2018)
  doi:10.1103/RevModPhys.90.015004
  [arXiv:1705.00141 [hep-ph]].


\bibitem{Olsen:2017bmm} 
  S.~L.~Olsen, T.~Skwarnicki and D.~Zieminska,
  Rev.\ Mod.\ Phys.\  {\bf 90}, no. 1, 015003 (2018)
  doi:10.1103/RevModPhys.90.015003
  [arXiv:1708.04012 [hep-ph]].



\bibitem{Francis:2016hui} 
  A.~Francis, R.~J.~Hudspith, R.~Lewis and K.~Maltman,
  Phys.\ Rev.\ Lett.\  {\bf 118}, no. 14, 142001 (2017)
  doi:10.1103/PhysRevLett.118.142001
  [arXiv:1607.05214 [hep-lat]].

\bibitem{Karliner:2017qjm} 
  M.~Karliner and J.~L.~Rosner,
  Phys.\ Rev.\ Lett.\  {\bf 119}, no. 20, 202001 (2017)
  doi:10.1103/PhysRevLett.119.202001
  [arXiv:1707.07666 [hep-ph]].
  
\bibitem{Eichten:2017ffp} 
  E.~J.~Eichten and C.~Quigg,
  Phys.\ Rev.\ Lett.\  {\bf 119}, no. 20, 202002 (2017)
  doi:10.1103/PhysRevLett.119.202002
  [arXiv:1707.09575 [hep-ph]].


\bibitem{Machleidt:1987hj} 
  R.~Machleidt, K.~Holinde and C.~Elster,
  Phys.\ Rept.\  {\bf 149}, 1 (1987).
  doi:10.1016/S0370-1573(87)80002-9


\bibitem{Tornqvist:2003na} 
  N.~A.~Tornqvist,
  hep-ph/0308277.


\bibitem{Malfliet:1968tj} 
  R.~A.~Malfliet and J.~A.~Tjon,
  Nucl.\ Phys.\ A {\bf 127}, 161 (1969).
  doi:10.1016/0375-9474(69)90775-1


\bibitem{Eichmann:2009qa} 
  G.~Eichmann, R.~Alkofer, A.~Krassnigg and D.~Nicmorus,
  Phys.\ Rev.\ Lett.\  {\bf 104}, 201601 (2010)
  doi:10.1103/PhysRevLett.104.201601
  [arXiv:0912.2246 [hep-ph]].


\bibitem{Ishii:1995bu} 
  N.~Ishii, W.~Bentz and K.~Yazaki,
  Nucl.\ Phys.\ A {\bf 587}, 617 (1995).
  doi:10.1016/0375-9474(95)00032-V


\bibitem{Eichmann:2011vu} 
  G.~Eichmann,
  Phys.\ Rev.\ D {\bf 84}, 014014 (2011)
  doi:10.1103/PhysRevD.84.014014
  [arXiv:1104.4505 [hep-ph]].


\bibitem{Huang:1993yd} 
  S.~z.~Huang and J.~Tjon,
  Phys.\ Rev.\ C {\bf 49}, 1702 (1994)
  doi:10.1103/PhysRevC.49.1702
  [hep-ph/9308362].


\bibitem{Ishii:1993np} 
  N.~Ishii, W.~Bentz and K.~Yazaki,
  Phys.\ Lett.\ B {\bf 301}, 165 (1993).
  doi:10.1016/0370-2693(93)90683-9


\bibitem{Ishikawa:2002ti} 
  S.~Ishikawa,
  Few Body Syst.\  {\bf 32}, 229 (2003)
  doi:10.1007/s00601-003-0001-7
  [nucl-th/0206064].


\bibitem{SanchisAlepuz:2011jn} 
  H.~Sanchis-Alepuz, G.~Eichmann, S.~Villalba-Chavez and R.~Alkofer,
  Phys.\ Rev.\ D {\bf 84}, 096003 (2011)
  doi:10.1103/PhysRevD.84.096003
  [arXiv:1109.0199 [hep-ph]].


\bibitem{Elster:2008hn} 
  C.~Elster, W.~Gl\"{o}ckle and H.~Witala,
  Few Body Syst.\  {\bf 45}, 1 (2009)
  doi:10.1007/s00601-008-0003-6
  [arXiv:0807.1421 [nucl-th]].


\bibitem{Eichmann:2008ef} 
  G.~Eichmann, I.~C.~Cloet, R.~Alkofer, A.~Krassnigg and C.~D.~Roberts,
  Phys.\ Rev.\ C {\bf 79}, 012202 (2009)
  doi:10.1103/PhysRevC.79.012202
  [arXiv:0810.1222 [nucl-th]].


\bibitem{Popovici:2010ph} 
  C.~Popovici, P.~Watson and H.~Reinhardt,
  Phys.\ Rev.\ D {\bf 83}, 025013 (2011)
  doi:10.1103/PhysRevD.83.025013
  [arXiv:1010.4254 [hep-ph]].


\bibitem{Fujiwara:2003wr} 
  Y.~Fujiwara, M.~Kohno and Y.~Suzuki,
  Few Body Syst.\  {\bf 34}, 237 (2004)
  doi:10.1007/s00601-004-0021-y
  [nucl-th/0310028].


\bibitem{Stadler:1991zz} 
  A.~Stadler, W.~Gl\"{o}ckle and P.~U.~Sauer,
  Phys.\ Rev.\ C {\bf 44}, 2319 (1991).
  doi:10.1103/PhysRevC.44.2319


\bibitem{MartinezTorres:2008gy} 
  A.~Martinez Torres, K.~P.~Khemchandani, L.~S.~Geng, M.~Napsuciale and E.~Oset,
  Phys.\ Rev.\ D {\bf 78}, 074031 (2008)
  doi:10.1103/PhysRevD.78.074031
  [arXiv:0801.3635 [nucl-th]].


\bibitem{MartinezTorres:2011vh} 
  A.~Martinez Torres, K.~P.~Khemchandani, D.~Jido and A.~Hosaka,
  Phys.\ Rev.\ D {\bf 84}, 074027 (2011)
  doi:10.1103/PhysRevD.84.074027
  [arXiv:1106.6101 [nucl-th]].


\bibitem{Torres:2011jt} 
  A.~Martinez Torres, D.~Jido and Y.~Kanada-En'yo,
  Phys.\ Rev.\ C {\bf 83}, 065205 (2011)
  doi:10.1103/PhysRevC.83.065205
  [arXiv:1102.1505 [nucl-th]].


\bibitem{Ren:2018pcd} 
  X.~L.~Ren, B.~B.~Malabarba, L.~S.~Geng, K.~P.~Khemchandani and A.~Martínez Torres,
  Phys.\ Lett.\ B {\bf 785}, 112 (2018)
  doi:10.1016/j.physletb.2018.08.034
  [arXiv:1805.08330 [hep-ph]].


\bibitem{MartinezTorres:2009xb} 
  A.~Martinez Torres, K.~P.~Khemchandani, D.~Gamermann and E.~Oset,
  Phys.\ Rev.\ D {\bf 80}, 094012 (2009)
  doi:10.1103/PhysRevD.80.094012
  [arXiv:0906.5333 [nucl-th]].


\bibitem{Xiao:2011rc} 
  C.~W.~Xiao, M.~Bayar and E.~Oset,
  Phys.\ Rev.\ D {\bf 84}, 034037 (2011)
  doi:10.1103/PhysRevD.84.034037
  [arXiv:1106.0459 [hep-ph]].


\bibitem{Xie:2010ig} 
  J.~J.~Xie, A.~Martinez Torres and E.~Oset,
  Phys.\ Rev.\ C {\bf 83}, 065207 (2011)
  doi:10.1103/PhysRevC.83.065207
  [arXiv:1010.6164 [nucl-th]].


\bibitem{Dias:2017miz} 
  J.~M.~Dias, V.~R.~Debastiani, L.~Roca, S.~Sakai and E.~Oset,
  Phys.\ Rev.\ D {\bf 96}, no. 9, 094007 (2017)
  doi:10.1103/PhysRevD.96.094007
  [arXiv:1709.01372 [hep-ph]].


\bibitem{Jido:2008kp} 
  D.~Jido and Y.~Kanada-En'yo,
  Phys.\ Rev.\ C {\bf 78}, 035203 (2008)
  doi:10.1103/PhysRevC.78.035203
  [arXiv:0806.3601 [nucl-th]].


\bibitem{MartinezTorres:2008kh} 
  A.~Martinez Torres, K.~P.~Khemchandani and E.~Oset,
  Phys.\ Rev.\ C {\bf 79}, 065207 (2009)
  doi:10.1103/PhysRevC.79.065207
  [arXiv:0812.2235 [nucl-th]].


\bibitem{Bayar:2012rk} 
  M.~Bayar and E.~Oset,
  Nucl.\ Phys.\ A {\bf 883}, 57 (2012)
  doi:10.1016/j.nuclphysa.2012.04.005
  [arXiv:1203.5313 [nucl-th]].


\bibitem{Oset:2012gi} 
  E.~Oset, D.~Jido, T.~Sekihara, A.~Martinez Torres, K.~P.~Khemchandani, M.~Bayar and J.~Yamagata-Sekihara,
  Nucl.\ Phys.\ A {\bf 881}, 127 (2012)
  doi:10.1016/j.nuclphysa.2012.02.005
  [arXiv:1203.4798 [hep-ph]].


\bibitem{MartinezTorres:2010ax} 
  A.~Martinez Torres, E.~J.~Garzon, E.~Oset and L.~R.~Dai,
  Phys.\ Rev.\ D {\bf 83}, 116002 (2011)
  doi:10.1103/PhysRevD.83.116002
  [arXiv:1012.2708 [hep-ph]].


\bibitem{Liang:2013yta} 
  W.~Liang, C.~W.~Xiao and E.~Oset,
  Phys.\ Rev.\ D {\bf 88}, no. 11, 114024 (2013)
  doi:10.1103/PhysRevD.88.114024
  [arXiv:1309.7310 [hep-ph]].


\bibitem{Bayar:2015oea} 
  M.~Bayar, X.~L.~Ren and E.~Oset,
  Eur.\ Phys.\ J.\ A {\bf 51}, no. 5, 61 (2015)
  doi:10.1140/epja/i2015-15061-8
  [arXiv:1501.02962 [hep-ph]].


\bibitem{YamagataSekihara:2010qk} 
  J.~Yamagata-Sekihara, L.~Roca and E.~Oset,
  Phys.\ Rev.\ D {\bf 82}, 094017 (2010)
  Erratum: [Phys.\ Rev.\ D {\bf 85}, 119905 (2012)]
  doi:10.1103/PhysRevD.82.094017, 10.1103/PhysRevD.85.119905
  [arXiv:1010.0525 [hep-ph]].


\bibitem{Roca:2010tf} 
  L.~Roca and E.~Oset,
  Phys.\ Rev.\ D {\bf 82}, 054013 (2010)
  doi:10.1103/PhysRevD.82.054013
  [arXiv:1005.0283 [hep-ph]].


\bibitem{Xie:2011uw} 
  J.~J.~Xie, A.~Martinez Torres, E.~Oset and P.~Gonzalez,
  Phys.\ Rev.\ C {\bf 83}, 055204 (2011)
  doi:10.1103/PhysRevC.83.055204
  [arXiv:1101.1722 [nucl-th]].


\bibitem{Debastiani:2017vhv} 
  V.~R.~Debastiani, J.~M.~Dias and E.~Oset,
  Phys.\ Rev.\ D {\bf 96}, no. 1, 016014 (2017)
  doi:10.1103/PhysRevD.96.016014
  [arXiv:1705.09257 [hep-ph]].


\bibitem{Ikeda:2007nz} 
  Y.~Ikeda and T.~Sato,
  Phys.\ Rev.\ C {\bf 76}, 035203 (2007)
  doi:10.1103/PhysRevC.76.035203
  [arXiv:0704.1978 [nucl-th]].


\bibitem{Hajduk:1979yn} 
  C.~Hajduk and P.~U.~Sauer,
  Nucl.\ Phys.\ A {\bf 322}, 329 (1979).
  doi:10.1016/0375-9474(79)90429-9


\bibitem{Ikeda:2008ub} 
  Y.~Ikeda and T.~Sato,
  Phys.\ Rev.\ C {\bf 79}, 035201 (2009)
  doi:10.1103/PhysRevC.79.035201
  [arXiv:0809.1285 [nucl-th]].


\bibitem{Gal:2013dca} 
  A.~Gal and H.~Garcilazo,
  Phys.\ Rev.\ Lett.\  {\bf 111}, 172301 (2013)
  doi:10.1103/PhysRevLett.111.172301
  [arXiv:1308.2112 [nucl-th]].


\bibitem{Dreissigacker:1981az} 
  K.~Dreissigacker, S.~Furui, C.~Hajduk, P.~U.~Sauer and R.~Machleidt,
  Nucl.\ Phys.\ A {\bf 375}, 334 (1981).
  doi:10.1016/0375-9474(82)90018-5


\bibitem{Konig:2015aka} 
  S.~K\"{o}nig, H.~W.~Grie{\ss}hammer, H.~W.~Hammer and U.~van Kolck,
  J.\ Phys.\ G {\bf 43}, no. 5, 055106 (2016)
  doi:10.1088/0954-3899/43/5/055106
  [arXiv:1508.05085 [nucl-th]].


\bibitem{Konig:2016yka} 
  S.~K\"{o}nig and H.~W.~Hammer,
  EPJ Web Conf.\  {\bf 113}, 04011 (2016).
  doi:10.1051/epjconf/201611304011


\bibitem{Wilbring:2017fwy} 
  E.~Wilbring, H.-W.~Hammer and U.-G.~Mei{\ss}ner,
  arXiv:1705.06176 [hep-ph].


\bibitem{Schmidt:2018vvl} 
  M.~Schmidt, M.~Jansen and H.-W.~Hammer,
  Phys.\ Rev.\ D {\bf 98}, no. 1, 014032 (2018)
  doi:10.1103/PhysRevD.98.014032
  [arXiv:1804.00375 [hep-ph]].


\bibitem{Hammer:2017uqm} 
  H.~W.~Hammer, J.~Y.~Pang and A.~Rusetsky,
  JHEP {\bf 1709}, 109 (2017)
  doi:10.1007/JHEP09(2017)109
  [arXiv:1706.07700 [hep-lat]].


\bibitem{Hammer:2017kms} 
  H.-W.~Hammer, J.-Y.~Pang and A.~Rusetsky,
  JHEP {\bf 1710}, 115 (2017)
  doi:10.1007/JHEP10(2017)115
  [arXiv:1707.02176 [hep-lat]].


\bibitem{Meng:2017jgx} 
  Y.~Meng, C.~Liu, U.-G.~Mei{\ss}ner and A.~Rusetsky,
  Phys.\ Rev.\ D {\bf 98}, no. 1, 014508 (2018)
  doi:10.1103/PhysRevD.98.014508
  [arXiv:1712.08464 [hep-lat]].



\bibitem{Garcilazo:2018rwu} 
  H.~Garcilazo and A.~Valcarce,
  Phys.\ Lett.\ B {\bf 784}, 169 (2018)
  doi:10.1016/j.physletb.2018.07.055
  [arXiv:1808.00226 [hep-ph]].


\bibitem{Garcilazo:2019igo} 
  H.~Garcilazo and A.~Valcarce,
  Phys.\ Rev.\ C {\bf 99}, no. 1, 014001 (2019)
  doi:10.1103/PhysRevC.99.014001
  [arXiv:1901.05678 [hep-ph]].



\bibitem{Ma:2017ery} 
  L.~Ma, Q.~Wang and U.-G.~Mei{\ss}ner,
  Chin.\ Phys.\ C {\bf 43}, no. 1, 014102 (2019)
  doi:10.1088/1674-1137/43/1/014102
  [arXiv:1711.06143 [hep-ph]].


\bibitem{Moroz:2014eba} 
  S.~Moroz and Y.~Nishida,
  Phys.\ Rev.\ A {\bf 90}, no. 6, 063631 (2014)
  doi:10.1103/PhysRevA.90.063631
  [arXiv:1407.7664 [cond-mat.quant-gas]].


\bibitem{Braaten:2014qka} 
  E.~Braaten, C.~Langmack and D.~H.~Smith,
  Phys.\ Rev.\ D {\bf 90}, no. 1, 014044 (2014)
  doi:10.1103/PhysRevD.90.014044
  [arXiv:1402.0438 [hep-ph]].


\bibitem{Zhao:2014gqa} 
  L.~Zhao, L.~Ma and S.~L.~Zhu,
  Phys.\ Rev.\ D {\bf 89}, no. 9, 094026 (2014)
  doi:10.1103/PhysRevD.89.094026
  [arXiv:1403.4043 [hep-ph]].


\bibitem{Patrignani:2016xqp} 
  C.~Patrignani {\it et al.} [Particle Data Group],
  Chin.\ Phys.\ C {\bf 40}, no. 10, 100001 (2016).
  doi:10.1088/1674-1137/40/10/100001


\bibitem{Bernard:1989fe} 
  V.~Bernard and U.-G.~Mei{\ss}ner,
  Phys.\ Rev.\ C {\bf 39}, 2054 (1989).
  doi:10.1103/PhysRevC.39.2054
   
\end{thebibliography}
\end{document}